\documentclass[opre,sglanonrev]{informs4}
\OneAndAHalfSpacedXI
\usepackage{anyfontsize,newtxtext,newtxmath}
\usepackage{float}
\usepackage{natbib}
 \bibpunct[, ]{(}{)}{,}{a}{}{,}%
 \def\bibfont{\footnotesize}%

\usepackage[hidelinks]{hyperref}
\renewcommand{\EMAIL}[1]{\href{mailto:#1}{#1}}
\hypersetup{
  pdftitle={Large Deviations in Ranking and Selection:
Limits and Value of Targeting Rate-Optimal
Sample Allocations},
  pdfauthor={Zaile Li and Yuchen Wan},
  pdfsubject={Optimal allocation ratios and the probability of incorrect selection in fixed-budget ranking and selection},
  pdfkeywords={ranking and selection, best-arm identification, large deviations, rate optimality, probability of incorrect selection},
  pdfdisplaydoctitle=true
}

\TheoremsNumberedThrough
\EquationsNumberedThrough
\newcommand{\E}{{\rm E}}
\newcommand{\R}{\mathbb R}
\newcommand{\N}{\mathcal N}
\newcommand{\sbest}{w^{\star}}
\newcommand{\kl}{\operatorname{kl}}
\newcommand{\KL}{\operatorname{KL}}

\usepackage{xcolor}
\definecolor{linkblue}{RGB}{17,14,178}
\usepackage{hyperref}

\hypersetup{
  colorlinks=true,
  linkcolor=linkblue,
  citecolor=linkblue,
  filecolor=linkblue,
  urlcolor=linkblue
}

\begin{document}

\TITLE{Large Deviations in Ranking and Selection: Limits and Value of Targeting Rate-Optimal Sample Allocations}
\RUNAUTHOR{Li and Wan}
\RUNTITLE{Limits and Value of Targeting Rate-Optimal Allocations}
\ARTICLEAUTHORS{%
  \AUTHOR{Zaile Li}
  \AFF{Department of Industrial Systems Engineering and Management, National University of Singapore, Singapore\\
       \EMAIL{zaile.li@nus.edu.sg}}
  \AUTHOR{Yuchen Wan}
  \AFF{Department of Intelligent Business and Management, Soochow University, China\\
       \EMAIL{ycwan@suda.edu.cn}}
}
\ABSTRACT{Fixed-budget ranking and selection (best-arm identification) seeks to identify the best alternative using a given sampling budget. Large deviations theory identifies the static sample allocation ratios that maximize the exponential decay rate (exponent) of the probability of incorrect selection (PICS), motivating many adaptive procedures that learn and track these ratios or attain them asymptotically. We study the performance of procedures that attain rate-optimal sample allocations and identify both the limits and the value of targeting them. We show that, in many scenarios, large deviations theory itself prevents such procedures from achieving exponential decay of PICS, let alone the optimal rate. Specifically, we find that for independent normal alternatives with unknown variances, every consistent procedure that attains the problem-specific rate-optimal allocation ratios has zero PICS exponent for any problem with a unique best alternative; consequently, PICS does not decay exponentially. Motivated by this finding, we establish a general necessary and sufficient condition under which attaining predefined allocation ratios forces a zero exponent. Using this condition, we then obtain a complete characterization when all alternatives belong to a common one-parameter natural exponential family. Whether exponential decay of PICS is possible depends on the number of alternatives, the distributional family, and whether there are tied inferior alternatives. These limits do not make rate-optimal allocations uninformative. We show that by reserving a nonzero fraction of the budget for uniform exploration, allocation-tracking procedures can preserve exponential convergence of PICS and strictly improve on uniform exploration alone.
The results may clarify what achieving a rate-optimal sample allocation can and cannot imply for adaptive selection performance and offer insights for procedure design.
}

\HISTORY{Working draft, \today.}
\maketitle

\section{Introduction}

Ranking and selection (R\&S) is a classical problem in stochastic simulation. The goal is to identify the alternative with the largest mean performance from a finite set. Since the means are unknown, an R\&S procedure must use observations from the alternatives to make its selection. In machine learning, this problem is studied under the name of \emph{best-arm identification}. It is a form of \emph{pure exploration} because the observations are collected for the purpose of making a final selection \citep{audibert2010,kaufmann2016}. Applications of R\&S range from comparing treatments in clinical trials and tuning algorithm hyperparameters to designing neural network architectures.

Three formulations of R\&S in the literature reflect different ways of accounting for sampling effort and the quality of the final selection \citep{hong2021review,chicklihong2026}. In fixed-precision R\&S, the experimenter specifies the required selection accuracy, and the procedure seeks to attain it with as few observations as possible. In fixed-budget R\&S, the procedure is given a budget of observations and allocates them to maximize selection accuracy. A third formulation, the economics of selection, weighs the cost of additional observations against the expected value of a better final selection \citep{chicklihong2026}.

We study fixed-budget R\&S procedures that allocate the budget adaptively. At each sampling stage, the procedure uses the observations collected so far to choose which alternative to sample next. When the budget is exhausted, it makes its final selection based on all the information collected. We measure selection accuracy by the probability of correct selection (PCS), the probability that the procedure selects the alternative with the largest true mean. A fundamental property is consistency: whenever one alternative has the largest true mean, the probability of selecting it approaches one as the sampling budget grows. Two consistent procedures can nevertheless require very different budgets to attain the same accuracy. Large deviations theory provides a natural framework for studying this difference: it describes how the probability of incorrect selection (PICS), the complement of PCS, decays as the budget grows. We refer to the exponential decay rate of PICS as its \emph{exponent}. For a consistent procedure, this exponent can be zero, meaning that PICS decays subexponentially (for example, polynomially) rather than exponentially fast.

The seminal work of \citet{glynn2004} establishes this rate perspective, which has become a standard theoretical foundation for fixed-budget R\&S. In their formulation, consider an oracle that knows the distribution of each alternative and plans the experiment before sampling begins by choosing an \emph{allocation ratio} for each alternative, the fraction of the total budget assigned to it. Under static sampling with these ratios and selection of the largest sample mean, \citet{glynn2004} characterize the exact PICS exponent via large deviations theory and formulate the optimization problem of maximizing this exponent to obtain an \emph{optimal static allocation}, also called a \emph{rate-optimal allocation}. They further derive the optimality conditions characterizing these ratios, which depend on the underlying problem configuration, that is, on the distributions of the alternatives. For independent normal observations with positive variances and a unique best alternative, the optimal ratios are unique and positive \citep[Proposition~2]{shin2018}; with two alternatives, for example, they are proportional to the true standard deviations.

In practice, the distributions are unknown. Choosing where to sample next directly on the basis of PCS is difficult because the value of an observation depends on how the procedure will use it in later sampling decisions and in the final selection. The optimal static allocation offers a more tractable target. For normal observations, for example, a procedure can substitute the current sample means and variances into the static allocation problem to estimate the optimal ratios. It can then sample an alternative whose observed allocation ratio is below its estimated target and update the estimates as new observations arrive. Tracking the optimal ratios therefore serves as a \emph{surrogate objective} for improving PCS.

This idea underlies many existing fixed-budget R\&S approaches. For example, the optimal computing budget allocation (OCBA) approach of \citet{chen2000} uses estimates of the means and variances in an approximate static allocation problem. Computing the exact optimal allocation poses a difficulty of its own, because repeatedly solving the allocation problem as the estimates change can be costly. The Balancing Optimal Large Deviations (BOLD) procedure of \citet{chenryzhov2023} instead uses the conditions that characterize the optimal static allocation to decide where to sample next. The Welch divergence procedure of \citet{shin2018} also uses estimated means and variances to guide this decision. More broadly, the rate-optimal allocation perspective has also been widely adopted across other formulations, including stochastic control (e.g., \citealt{peng2018}), constrained simulation optimization (e.g., \citealt{hunter2013}), subset and top-$m$ selection (e.g., \citealt{gao2015,zhang2023}), and simulation optimization under input uncertainty (e.g., \citealt{wangzhou2025,songkim2026}), to name a few.

These procedures are often supported by asymptotic guarantees on their observed allocation ratios. When those ratios converge to the optimal ratios as the budget grows, we say that the procedure \emph{tracks} the optimal allocation. We use \emph{rate-optimal} only to describe an allocation that maximizes the static exponent. The target is problem-specific. The same procedure, given none of the true parameters, may therefore track different allocation ratios at different problem configurations.

The optimal static allocation also provides a way to evaluate procedures developed on other grounds. Expected-improvement procedures, for example, assess the value of taking one more observation. For normal observations with known variances, \citet{ryzhov2016} relates the relative allocation ratios among inferior alternatives to the OCBA ratios. \citet{chenryzhov2019cei} show that a modified complete expected-improvement procedure tracks the optimal allocation ratios for all alternatives. These results give a reason to expect strong selection performance. Among static allocations, the rate-optimal ratios give the fastest exponential decay of PICS. It is then natural to expect an adaptive procedure that tracks those ratios to achieve exponential decay as well.

The static and adaptive statements are nevertheless different. Static optimality describes a procedure that knows the distributions before sampling begins. Tracking describes the typical limiting allocation of a procedure that must learn those distributions from the observations. Earlier work cautions that an allocation guarantee need not imply an exponential PICS guarantee: \citet[Section~4]{chenryzhov2019} observe that a procedure tracking the optimal static allocation may fail to achieve exponential decay, and \citet{wuzhou2018} prove subexponential decay for OCBA and two variants under normal sampling when the initial sample size does not grow with the total budget. Beyond these important early warnings, however, there remains limited theoretical analysis (and, surprisingly, little direct numerical evidence) on how tracking a rate-optimal allocation relates to, and differs from, achieving (optimal) exponential decay of PICS.

In this paper, we explore both the connection and the gap between tracking and selection accuracy, establishing when tracking fails to deliver exponential decay and how the same allocation information can still improve selection accuracy. We summarize our main findings below.

\subsection{Limits under Normal Settings}

Our first result (Theorem~\ref{thm:zero}, Section~\ref{sec:gauss}) establishes a striking negative finding in the normal setting with unknown variances. Consider independent normal alternatives with unrestricted means and arbitrary unknown positive variances. If a single procedure is consistent and tracks the problem-specific rate-optimal allocation throughout this distributional setting, then its PICS has zero exponent at any configuration with a unique best alternative. Thus PICS converges to zero, but not exponentially fast, even though the corresponding static oracle has a positive exponent. The conclusion holds for any number of alternatives, including two, and for any final selection rule.

Perhaps surprisingly, this impossibility arises from the very balance conditions that define the rate-optimal allocation. Under Cram\'er's large-deviations rate functions for the sample means, the rate-optimal allocation ratios in the normal setting satisfy two balance conditions: one equalizes the exponential decay rate of each inferior alternative's sample mean overtaking the best alternative, and the other balances the allocation to the best alternative against those to all other alternatives (see \citealt{chenryzhov2023} for a good exposition). Our proof combines these balance conditions with a change-of-measure lemma (Lemma~\ref{lem:comparison}) linking the PICS exponent at one configuration to the Kullback--Leibler (KL) divergences and tracked allocation ratios at a different-best configuration, that is, one with a different best alternative (Section~\ref{sec:gauss}).

Unknown variances are essential to this phenomenon. When the variances are known (Theorem~\ref{kv:theorem}, Section~\ref{kv:section}), some consistent procedure that tracks the rate-optimal allocation can achieve a positive lower PICS exponent (ensuring exponential decay) at every configuration with two alternatives, and with three or more alternatives whenever all means are distinct; by contrast, at least two tied inferior means still force a zero exponent. These known-variance conclusions follow from the general characterization developed in the next subsection. Table~\ref{tab:normal-tracking} summarizes this distinction between known and unknown variances.

\begin{table}[!ht]
\caption{Possibility of exponential PICS decay under normal alternatives with a unique best alternative (Theorems~\ref{thm:zero} and~\ref{kv:theorem}). A $+$ indicates that some consistent procedure tracking the rate-optimal allocation achieves a positive lower PICS exponent at that configuration, whereas $-$ indicates that every such procedure has zero exponent.}
\label{tab:normal-tracking}
\centering
\small
\begin{tabular}{@{}p{.30\linewidth}ccc@{}}
\hline
& $k=2$ & \multicolumn{2}{c}{$k\ge3$} \\
\cline{3-4}
Setting & Distinct means & All means distinct & \begin{tabular}[b]{@{}c@{}}At least two\\[-2pt]tied inferior means\end{tabular} \\
\hline
Normal (known variances) & $+$ & $+$ & $-$ \\
Normal (unknown variances) & $-$ & $-$ & $-$ \\
\hline
\end{tabular}
\end{table}

Section~\ref{sec:numerical} evaluates several classical adaptive procedures motivated by rate-optimal allocations, namely OCBA, BOLD, and modified complete expected improvement (mCEI), across all six settings in Table~\ref{tab:normal-tracking}. Their estimated PICS exponent behavior aligns closely with this theoretical classification.

\subsection{General Setting and One-Parameter Natural Exponential Families}

We next ask when tracking any predefined, problem-specific allocation vector forces a zero exponent. Extending the change-of-measure comparison between a given configuration and different-best configurations, we establish a general \emph{necessary and sufficient} condition in terms of KL divergence (Theorem~\ref{thm:zero-characterization}, Section~\ref{sec:information}): tracking forces a zero exponent at a given configuration if and only if the KL divergence between a different-best configuration and the given one, weighted by the ratios tracked at the different-best configuration, can be made arbitrarily small. Thus, the entire question reduces to evaluating a single weighted KL quantity, providing a general tool for determining when tracking predefined allocation ratios can achieve exponential PICS decay across different distributional settings.

We then specialize this characterization to rate-optimal allocations when all alternatives belong to a common one-parameter natural exponential family, which covers normal observations with a common known variance as well as Bernoulli, Poisson, and exponential observations (Theorem~\ref{thm:nef-tangent}, Section~\ref{sec:ld-oracle}). In this family, Cram\'er's sample-mean rate function coincides directly with the KL divergence between distributions in the family. Exploiting this link, we develop a geometric tangent representation of the rate function that directly visualizes the rate-optimal allocation and the resulting weighted KL divergence. Together with the pattern of ties among the means, this geometry yields a complete classification.

Whether exponential decay remains possible depends on the number of alternatives, whether any inferior means tie, and how the rate function behaves at extreme mean values of the alternatives. In particular, with three or more alternatives, a zero exponent (subexponential decay of PICS) is forced whenever at least two inferior means tie or when the distributional family fails the boundary conditions of Theorem~\ref{thm:nef-tangent}, which covers exponential, Poisson, and Bernoulli observations; by contrast, normal observations with a common known variance allow a positive lower exponent whenever all means are distinct. With two alternatives, normal and exponential observations still allow a positive lower exponent, whereas Bernoulli and Poisson observations force a zero exponent. Table~\ref{tab:nef-tracking} summarizes these conclusions.

\begin{table}[!ht]
\caption{Possibility of exponential PICS decay for common one-parameter natural exponential families with a unique best alternative (Theorem~\ref{thm:nef-tangent}). A $+$ indicates that some consistent procedure tracking the rate-optimal allocation achieves a positive lower PICS exponent at that configuration, $-$ indicates that every such procedure has zero exponent, and $+$ or $-$ indicates that either outcome can occur depending on the distributional family.}
\label{tab:nef-tracking}
\centering
\small
\begin{tabular}{@{}lccc@{}}
\hline
& $k=2$ & \multicolumn{2}{c}{$k\ge3$} \\
\cline{3-4}
Distributional family & Distinct means & All means distinct & \begin{tabular}[b]{@{}c@{}}At least two\\[-2pt]tied inferior means\end{tabular} \\
\hline
Normal with common known variance & $+$ & $+$ & $-$ \\
Exponential & $+$ & $-$ & $-$ \\
Poisson & $-$ & $-$ & $-$ \\
Bernoulli & $-$ & $-$ & $-$ \\
\hline
Any common family & $+$ or $-$ & $+$ or $-$ & $-$ \\
\hline
\end{tabular}
\end{table}

Section~\ref{sec:numerical} also evaluates OCBA and BOLD under exponential observations, as they provide convenient, general templates for tracking rate-optimal allocations across different distributional families. Their estimated PICS exponent behavior aligns closely with the exponential results in Table~\ref{tab:nef-tracking} for both $k=2$ and $k\ge3$.

\subsection{The Value of Tracking Rate-Optimal Sample Allocations}

The preceding limits concern \emph{pure} tracking of the rate-optimal allocation. However, they do not mean that rate-optimal allocations are uninformative. If the immediate goal is simply exponential convergence of PICS (setting aside achieving the unattainable static optimum), one straightforward baseline is uniform exploration, or equal allocation, which assigns an equal fraction $1/k$ of the budget to every alternative. Whenever the alternative distributions have exponential tails, equal allocation readily guarantees exponential decay of PICS with a strictly positive exponent, even though the resulting exponent can be suboptimal.

We find that deviating from rate-optimal allocations by combining equal allocation with rate-optimal tracking produces a strict synergy: the resulting hybrid procedures achieve a PICS exponent strictly larger than \emph{both} equal allocation using the full budget and pure rate-optimal tracking (which has zero exponent in all the collapsed regimes above). From our perspective, this outcome is unexpected: since pure tracking alone yields a zero exponent, one might expect mixing it with equal allocation to dilute equal allocation rather than strengthen it. 
We study two mechanisms to implement this combination: (1) A two-stage pilot-and-tracking mechanism that reserves a nonzero fraction of the budget for initial equal-allocation sampling and then applies a rate-optimal ratio tracking procedure over the remaining budget, in the spirit of \citet{wuzhou2018} (Theorems~\ref{pv:gauss} and~\ref{dt:theorem}); (2) An online mixture mechanism that adjusts the empirical target sampling ratios on the fly at every update toward a convex combination of equal-allocation ratios and the estimated rate-optimal allocation (Theorem~\ref{dt:theorem}). Both mechanisms may strictly beat equal allocation whenever $k\ge3$, even at configurations where pure tracking forces a zero exponent. Section~\ref{sec:numerical} illustrates the finite-budget gains of the two-stage pilot-and-tracking mechanism under unknown-variance normal sampling.

This synergy shows that although pure rate-optimal tracking may fail to sustain exponential decay on its own, pairing it with systematic uniform exploration restores exponential decay and strictly improves the PICS exponent over uniform exploration alone. Overall, our paper demonstrates that in many settings, large-deviations rate-optimal allocation ratios do not by themselves guarantee exponential decay of PICS for adaptive procedures that track them or attain them asymptotically, and that incorporating systematic uniform exploration alleviates this limitation while strictly improving on equal allocation alone. At the same time, our analysis is asymptotic; the strong finite-budget performance often observed in practice for rate-optimal tracking procedures merits deeper study from a finite-budget perspective, which we leave for future work.

\subsection{Related Literature}\label{sec:literature}

As noted above, \citet[Section~4]{chenryzhov2019} and \citet{wuzhou2018} provide important early warnings that an allocation guarantee need not imply an exponential PICS guarantee. Specifically, \citet{wuzhou2018} prove subexponential decay for OCBA and two variants under normal sampling with unknown variances when the initial sample size does not grow linearly with the total budget, and show that an initial sample proportional to the budget can restore exponential decay for those procedures. Our work differs from and extends these findings in both negative and positive directions. On the negative side, our impossibility results are procedure-independent: they attribute the zero PICS exponent directly to the joint requirements of consistency and tracking rate-optimal allocations, extending beyond unknown-variance normal alternatives (Theorem~\ref{thm:zero}) to known-variance normal configurations with tied inferior means (Theorem~\ref{kv:theorem}) and to broad classes of common one-parameter natural exponential families (Theorem~\ref{thm:nef-tangent}). On the positive side, whereas \citet{wuzhou2018} restore exponential decay for OCBA rules without establishing gains over equal allocation, we analyze both two-stage pilot-and-tracking rules and an anytime online mixture rule (Theorems~\ref{pv:gauss} and~\ref{dt:theorem}) and prove that combining equal allocation with rate-optimal tracking strictly improves the PICS exponent over equal allocation.

A related line of work in best-arm identification asks whether one procedure can attain the optimal static exponent at every configuration. \citet{degenne2023} proves that this is impossible for two Bernoulli alternatives, and \citet{goldberger2026} quantifies the shortfall at some configuration for more than two alternatives in a regular one-parameter natural exponential family. Weighted KL quantities evaluated at different-best configurations also appear in the allocation game of \citet{komiyama2022}. For constant allocation ratios, \citet[Theorem~1]{degenne2023} gives an exact exponent in one-parameter exponential families, while \citet{komiyama2026} study attainable fractions of a related information bound under normal sampling. Complementing these impossibility and constant-ratio results, our characterization (Theorem~\ref{thm:zero-characterization}) identifies exactly when tracking any prescribed problem-dependent allocation forces a zero exponent at a given configuration, and shows precisely how mixing uniform exploration with tracking recovers exponential gains.

The remainder of the paper is organized as follows. Section~\ref{sec:problem} introduces the fixed-budget problem, static rate-optimal allocation, consistency, allocation tracking, and PICS exponents. Section~\ref{sec:gauss} proves the zero-exponent result for normal observations with unknown variances. Section~\ref{sec:information} establishes the general weighted-KL characterization. Section~\ref{sec:ld-oracle} applies it to common one-parameter natural exponential families and develops the complete characterization and geometric insights. Section~\ref{sec:applications} evaluates the classification on common distributional examples and analyzes normal observations with known variances. Section~\ref{dt:section} studies mixtures of equal allocation and tracking. Section~\ref{sec:numerical} illustrates our theoretical findings through numerical experiments. Section~\ref{sec:conclusion} concludes the paper, and the electronic companion contains the proofs.

\section{Problem Formulation and Rate-Optimal Sample Allocations}\label{sec:problem}

Consider a fixed-budget R\&S problem with $k\ge2$ alternatives, where $k$ remains fixed as the sampling budget grows. A \emph{problem configuration} $P=(P_1,\ldots,P_k)$ specifies the distribution of every alternative. Observations from alternative $i$ are independent and identically distributed according to $P_i$, which has finite mean $\mu_i(P)$, and observations from different alternatives are independent. The distributions are unknown to the procedure, and we assume $P$ has a unique best alternative $b(P)=\argmax_i\mu_i(P)$, although the other means may tie. When no confusion arises, we write $\mu_i$ and $b$ for $\mu_i(P)$ and $b(P)$.

A fixed-budget procedure $\pi$ receives a budget of $T$ observations. At each sampling stage, it chooses an alternative using the observations collected so far and, possibly, auxiliary randomization independent of all observation streams. It then collects one new observation from that alternative. Let $N_i(T)$ be the number of observations collected from alternative $i$, so $\sum_iN_i(T)=T$. The vector of observed allocation ratios is $\left(N_1(T)/T,\ldots,N_k(T)/T\right)$. After the final observation, the procedure selects an alternative $\widehat b_T$ using the information it has collected. Let $\Pr_P$ denote the probability when the observations are generated under $P$ and procedure $\pi$ is used. Its PICS and PCS are, respectively, $\Pr_P\{\widehat b_T\ne b(P)\}$ and $\Pr_P\{\widehat b_T=b(P)\}$. Here, $\Pr_P$ is taken over both the observations and any auxiliary randomization used by the procedure.

\subsection{Static Rate-Optimal Sample Allocation}\label{sec:static-allocation}

A static procedure fixes its allocation before observing any data. Write $\mathcal S_k=\{w\in[0,1]^k:\sum_iw_i=1\}$ for the allocation simplex. For $w\in\mathcal S_k$ with positive coordinates, assign alternative $i$ a deterministic number $n_i(T)$ of observations such that $\sum_i n_i(T)=T$ and $\lim_{T\to\infty}n_i(T)/T=w_i$. The procedure then selects the alternative with the largest sample mean, using a tie-breaking rule fixed before sampling. Denote its PICS by $p_T^{\rm stat}(w,P)$. When the limit exists, its \emph{static exponent} is
\[
  \Gamma(w,P)
  =\lim_{T\to\infty}\frac{1}{T}
    \log\frac{1}{p_T^{\rm stat}(w,P)}.
\]

Large deviations theory expresses this exponent through the rate functions of the sample means \citep{glynn2004}. For one observation $X$ from alternative $i$, let $\psi_i(\theta)=\log\E_{P_i}[e^{\theta X}]$ be its log moment generating function and define the corresponding \emph{large-deviations rate function} (hereafter simply \emph{rate function}) by $I_{P_i}(x)=\sup_{\theta\in\R}\{\theta x-\psi_i(\theta)\}$. Suppose $\psi_i$ is finite in a neighborhood of zero and, for each $i\ne b$, the rate functions $I_{P_b}(\cdot)$ and $I_{P_i}(\cdot)$ are finite and continuous on an open interval containing $[\mu_i,\mu_b]$. Examples include normal, Bernoulli, Poisson, and exponential observations; \citet[Section~2.3]{chenryzhov2023} give the relevant calculations. Let $\bar X_{i,n}$ denote the sample mean of $n$ independent observations from $P_i$. By Cram\'er's theorem, for large $n$ and small $\epsilon>0$,
\begin{equation}
  \Pr_{P_i}\{|\bar X_{i,n}-x|\le\epsilon\}
  \approx \exp\!\left\{-n\inf_{|y-x|\le\epsilon}I_{P_i}(y)\right\}
  \approx \exp\{-n I_{P_i}(x)\},
  \label{eq:cramer-rate}
\end{equation}
so $I_{P_i}(x)$ measures the exponential cost per observation of the sample mean being near $x$. Cram\'er's theorem and independence then give
\begin{equation}
  \Gamma(w,P)=\min_{i\ne b}\ \inf_{\mu_i\le x\le\mu_b}
       \bigl\{w_b I_{P_b}(x)+w_i I_{P_i}(x)\bigr\}.
  \label{eq:cramer-oracle}
\end{equation}

Here, for each alternative $i\ne b$, the inner infimum is the \emph{pairwise exponent}, measuring the smallest exponential cost of making the sample means of $b$ and $i$ meet between $\mu_i$ and $\mu_b$, while the outer minimum selects the least costly competitor to overtake $b$.

The \emph{rate-optimal sample allocation} maximizes the static PICS exponent:
\begin{equation}
  \sbest(P)\in\argmax_{w\in\mathcal S_k}\Gamma(w,P).
  \label{eq:rate-optimal-allocation}
\end{equation}
See \citet{glynn2004} and \citet{chenryzhov2023} for the large-deviations formulation and its optimality conditions in terms of balancing the rate functions, which are summarized in Section~\ref{cls:two-alternative} when they are needed. In a distributional setting where the maximizer is not unique, a procedure would typically be designed to track a particular maximizing allocation $\sbest(P)$. Because this target depends on the unknown distributions, it must be learned from observations.

\subsection{R\&S Procedures: Consistency and Achieving Rate-Optimal Sample Allocations}
\label{sec:requirements}

For an adaptive fixed-budget R\&S procedure $\pi$, the most basic asymptotic requirement is \emph{consistency}: as the sampling budget $T$ grows, the probability of selecting the true best alternative $b(P)$ should converge to one (equivalently, PICS vanishes). Consistency concerns only the final selection $\widehat b_T$.

Motivated by the static large-deviations formulation in Section~\ref{sec:static-allocation}, one often seeks procedures whose observed allocation ratios also converge to the static rate-optimal ratios $\sbest(P)$ in \eqref{eq:rate-optimal-allocation}, or more generally to a predefined target vector $w(P)\in\mathcal S_k$ (which may vary with the configuration $P$ but is nonrandom). We formalize both requirements over the distributional setting $\mathfrak D$ under study, representing the set of $k$-alternative configurations $P=(P_1,\ldots,P_k)$ allowed by the distributional assumptions (with each configuration having finite means and a unique best alternative, as maintained throughout).

\begin{definition}[Consistency and Allocation Tracking]
\label{def:requirements}
Let $\mathfrak D$ be a distributional setting. A procedure $\pi$ is \emph{consistent} on $\mathfrak D$ if, for any fixed $P\in\mathfrak D$,
\[
  \lim_{T\to\infty}\Pr_P\{\widehat b_T\ne b(P)\}=0.
\]
For each $P\in\mathfrak D$, let $w(P)\in\mathcal S_k$ be a predefined vector of allocation ratios. The procedure $\pi$ \emph{tracks} $w$ on $\mathfrak D$ if, for any fixed $P\in\mathfrak D$, every alternative $i$, and every $\epsilon>0$,
\[
  \lim_{T\to\infty}
  \Pr_P\!\left\{
    \left|\frac{N_i(T)}{T}-w_i(P)\right|>\epsilon
  \right\}=0.
\]
\end{definition}

Two features of Definition~\ref{def:requirements} deserve emphasis. First, tracking the problem-specific rate-optimal allocation corresponds to setting $w(P)=\sbest(P)$ for each $P\in\mathfrak D$. We use the term \emph{tracks} broadly to include both procedures explicitly designed to track $\sbest(P)$ and procedures grounded on other principles that are shown to attain these ratios asymptotically. Almost-sure convergence of the allocation ratios implies the convergence in probability required here.

Second, consistency and tracking are universal yet problem-specific requirements across the distributional setting $\mathfrak D$. When establishing such guarantees for an adaptive procedure, one proves them across the entire distributional setting under study (for example, normal distributions or a given exponential family, allowing all parameters that yield a unique best alternative). Accordingly, throughout the paper, $\mathfrak D$ denotes the distributional setting under study. Note also that the limits in Definition~\ref{def:requirements} are pointwise in $P$: the configuration $P\in\mathfrak D$ is held fixed in every limit over $T$, and no uniform convergence over $\mathfrak D$ is required.

\subsection{PICS Exponents of an R\&S Procedure}\label{sec:exponents}

Consistency specifies the limiting value of PCS, but it does not describe the speed of this convergence. Large deviations theory measures that speed through the exponential decay rate of PICS. For an adaptive procedure, this rate need not converge, so we use lower and upper exponents.

\begin{definition}[PICS Exponents]\label{def:exponents}
For a procedure $\pi$ at a configuration $P$, define the lower and upper PICS exponents by
\[
  \underline e_\pi(P)
  =\liminf_{T\to\infty}\frac{1}{T}
     \log\frac{1}{\Pr_P\{\widehat b_T\ne b(P)\}},
  \qquad
  \overline e_\pi(P)
  =\limsup_{T\to\infty}\frac{1}{T}
     \log\frac{1}{\Pr_P\{\widehat b_T\ne b(P)\}}.
\]
We set $\log(1/0)=+\infty$. If the lower and upper exponents agree, their common value is the \emph{PICS exponent}. PICS has \emph{zero exponent} when this common value is zero.
\end{definition}

Here and throughout, all limits are taken for fixed $k$. For a consistent procedure, zero exponent means that $\lim_{T\to\infty}\Pr_P\{\widehat b_T=b(P)\}=1$, but the convergence is not exponentially fast.

One might expect consistency and tracking the problem-specific rate-optimal allocation in Definition~\ref{def:requirements} to yield at least a positive PICS exponent, or perhaps even the optimal static exponent. The next section shows that this expectation fails in a canonical R\&S setting. With normal observations and unknown variances, every consistent procedure that tracks this allocation across the distributional setting has zero exponent at every configuration.

\section{Normal Observations with Unknown Variances}\label{sec:gauss}

We now consider the unknown-variance normal distributional setting $\mathfrak D$, consisting of all configurations $P=(P_1,\ldots,P_k)$ with $P_i=\N(\mu_i(P),\sigma_i^2(P))$, where the mean vector $(\mu_1(P),\ldots,\mu_k(P))$ has a unique largest coordinate and the standard deviations $(\sigma_1(P),\ldots,\sigma_k(P))$ may take arbitrary values in $(0,\infty)^k$. Fix any $P\in\mathfrak D$ and write $b=b(P)$, $\mu_i=\mu_i(P)$, $\sigma_i=\sigma_i(P)$, and $\Delta_i=\mu_b-\mu_i$ for $i\ne b$, following the convention in Section~\ref{sec:problem}. The unique rate-optimal allocation $\sbest(P)$ is characterized by the individual and total balance conditions
\begin{equation}
  \frac{\Delta_i^2}
       {\sigma_b^2/\sbest_b(P)+\sigma_i^2/\sbest_i(P)}
  =
  \frac{\Delta_j^2}
       {\sigma_b^2/\sbest_b(P)+\sigma_j^2/\sbest_j(P)}
  \quad(i,j\ne b),
  \qquad
  \left(\frac{\sbest_b(P)}{\sigma_b}\right)^2
      =\sum_{i\ne b}\left(\frac{\sbest_i(P)}{\sigma_i}\right)^2 .
  \label{eq:balance}
\end{equation}
The first set equalizes the pairwise exponents across the suboptimal alternatives; the second balances the allocation to the best alternative against the allocations to all the others. These conditions follow from \citet[Theorem~1 and Example~1]{glynn2004}; see also \citet{chenryzhov2023}.

Many existing procedures are designed to be consistent and to track the problem-specific rate-optimal allocation. Our first main result shows that any procedure with these two properties has zero PICS exponent in the normal setting. The conclusion holds for any fixed number of alternatives $k\ge2$.

\begin{theorem}\label{thm:zero}
Let $\mathfrak D$ be the unknown-variance normal distributional setting defined above, and let $\sbest(P)$ be the problem-specific rate-optimal allocation in \eqref{eq:balance}. If a procedure $\pi$ is consistent and tracks $\sbest$ on $\mathfrak D$ (Definition~\ref{def:requirements}), then for every configuration $P\in\mathfrak D$ it has zero PICS exponent (Definition~\ref{def:exponents}), that is,
\[
   \lim_{T\to\infty}\frac{1}{T}
   \log\frac{1}{\Pr_P\{\widehat b_T\ne b(P)\}}=0.
\]
For a consistent procedure on $\mathfrak D$, the same conclusion holds if tracking in probability is replaced by convergence of the expected allocation ratios, namely $\lim_{T\to\infty}\E_Q[N_i(T)]/T=\sbest_i(Q)$ for every $Q\in\mathfrak D$ and every $i=1,\ldots,k$.
\end{theorem}

The proof compares a given configuration $P\in\mathfrak D$ with a second configuration $Q\in\mathfrak D$ having a different best alternative ($b(Q)\ne b(P)$, called a \emph{different-best configuration} for $P$). Section~\ref{sec:comparison-bound} first uses a change-of-measure argument to bound the upper PICS exponent at $P$ using the sampling behavior and KL divergences under $Q$. Section~\ref{sec:normal-zero-proof} then uses the normal balance conditions in \eqref{eq:balance} to construct different-best configurations for which this bound shrinks to zero, proving the theorem. What makes this phenomenon particularly striking is that, together with the information structure established in Section~\ref{sec:comparison-bound}, it is the large-deviations balance conditions \eqref{eq:balance} themselves that obstruct exponential PICS convergence.

\subsection{A Change-of-Measure Bound}\label{sec:comparison-bound}

For two distributions $Q_i$ and $P_i$, write $\KL(Q_i\Vert P_i)$ for their Kullback--Leibler (KL) divergence. It is the expected log likelihood ratio $\E_{Q_i}[\log(dQ_i/dP_i)(X)]$ when $Q_i$ is absolutely continuous with respect to $P_i$, and is $+\infty$ otherwise. Absolute continuity means that every event with probability zero under $P_i$ also has probability zero under $Q_i$. The divergence measures the expected information from one observation drawn from $Q_i$ for distinguishing $Q_i$ from $P_i$. The following comparison weights this one-observation information by the frequency with which the procedure samples alternative $i$ under $Q$. It follows from the change-of-measure inequality of \citet[Lemma~1]{kaufmann2016}.

\begin{lemma}\label{lem:comparison}
Suppose the procedure is consistent, $P$ and $Q$ have different unique best alternatives, and $\KL(Q_i\Vert P_i)<\infty$ for every $i$. If $w(Q)\in\mathcal S_k$ satisfies $\lim_{T\to\infty}\E_Q[N_i(T)]/T=w_i(Q)$ for every $i=1,\ldots,k$, then
\begin{equation}
   \limsup_{T\to\infty}\frac{1}{T}
   \log\frac{1}{\Pr_P\{\widehat b_T\ne b(P)\}}
       \ \le\ \sum_{i=1}^k w_i(Q)\,\KL(Q_i\Vert P_i).
   \label{eq:comparison}
\end{equation}
\end{lemma}

Lemma~\ref{lem:comparison} says that selection accuracy at $P$ is constrained by what the procedure must do under any different-best configuration $Q$. Consistency requires the procedure to select $b(Q)$ with probability approaching one under $Q$, but that same selection is incorrect under $P$. The right-hand side of \eqref{eq:comparison} is the asymptotic information rate for distinguishing $Q$ from $P$: alternative $i$ is sampled with limiting ratio $w_i(Q)$ under $Q$, and each of its observations contributes $\KL(Q_i\Vert P_i)$ units of information on average. Thus PICS at $P$ cannot decay faster than the total information collected under $Q$. Importantly, the allocation ratios are evaluated at the different-best configuration $Q$, not at $P$. If this weighted KL information can be made arbitrarily small by varying $Q$, then the PICS exponent at $P$ must be zero. Section~\ref{app:comparison-proof} gives the proof.

We now use the normal balance condition in \eqref{eq:balance} to construct different-best configurations $Q$ for which the weighted KL information $\sum_{i=1}^k\sbest_i(Q)\KL(Q_i\Vert P_i)$ in Lemma~\ref{lem:comparison} is arbitrarily small.

\subsection{Why the Exponent Is Zero}\label{sec:normal-zero-proof}

Lemma~\ref{lem:comparison} reduces the proof to a construction. Fix a configuration $P$. It is enough to find different-best configurations $Q_s$ such that
\begin{equation}
   \lim_{s\downarrow0}\sum_{i=1}^k
      \sbest_i(Q_s)\,\KL((Q_s)_i\Vert P_i)=0.
   \label{eq:normal-vanishing-information}
\end{equation}
We construct $Q_s$ by changing only one suboptimal alternative $j$ of $P$. The change makes alternative $j$ best but gives it standard deviation $s$. The normal balance condition in \eqref{eq:balance} then gives $\sbest_j(Q_s)=O(s)$, while the information in one observation from that alternative satisfies $\KL((Q_s)_j\Vert P_j)=O(\log(1/s))$. Because all other alternatives have the same distributions under $Q_s$ and $P$, their KL divergences are zero. Hence the weighted KL information in \eqref{eq:normal-vanishing-information} is $O(s\log(1/s))$ and converges to zero. We include the full proof below to illustrate the core ideas behind our impossibility mechanism (whereas later, more technical proofs are deferred to the electronic companion).

\proof{Proof of Theorem~\ref{thm:zero}.}
Fix any configuration $P$ and let $b=b(P)$ denote its unique best alternative. Choose an alternative $j\ne b$ and a constant $m>\mu_b$. For every $s>0$, define $Q_s$ by replacing $P_j=\N(\mu_j,\sigma_j^2)$ with $(Q_s)_j=\N(m,s^2)$ and leaving all other alternative distributions unchanged. Thus $j$ is the unique best alternative under $Q_s$, whereas $b$ is the unique best alternative under $P$.

For every $i$, the divergence $\KL((Q_s)_i\Vert P_i)$ is finite: it equals zero for $i\ne j$, while for $i=j$ both distributions are nondegenerate normal distributions. If the procedure tracks $\sbest(Q_s)$ in probability, then $0\le N_i(T)/T\le1$ and convergence in probability to the constant $\sbest_i(Q_s)$ implies convergence in mean. Indeed, for every $\epsilon>0$,
\[
  \E_{Q_s}\!\left[\left|\frac{N_i(T)}{T}-\sbest_i(Q_s)\right|\right]
  \le \epsilon
  +\Pr_{Q_s}\!\left\{
    \left|\frac{N_i(T)}{T}-\sbest_i(Q_s)\right|>\epsilon
  \right\}.
\]
Taking first $\lim_{T\to\infty}$ and then $\epsilon\downarrow0$ gives $\lim_{T\to\infty}\E_{Q_s}[N_i(T)]/T=\sbest_i(Q_s)$ for every $i$. Under the alternative assumption in the theorem, these expected-allocation limits hold directly. Together with consistency and the fact that $P$ and $Q_s$ have different unique best alternatives, these observations verify every hypothesis of Lemma~\ref{lem:comparison}.

Applying the lemma converts the procedure's behavior under $Q_s$ into a bound on the upper PICS exponent at $P$. Since only alternative $j$ has a positive KL divergence, for every $s>0$ this bound is
\begin{equation}
  \limsup_{T\to\infty}\frac{1}{T}
  \log\frac{1}{\Pr_P\{\widehat b_T\ne b(P)\}}
  \le \sbest_j(Q_s)\,\KL((Q_s)_j\Vert P_j).
  \label{eq:gaussian-one-design}
\end{equation}

It remains to show that the right-hand side of \eqref{eq:gaussian-one-design} can be made arbitrarily small. At $Q_s$, the best alternative $j$ has standard deviation $s$. Let $\sigma_{\min,-j}=\min_{i\ne j}\sigma_i>0$. The total balance condition in \eqref{eq:balance} gives
\begin{equation}
 \sbest_j(Q_s)
  =s\left(\sum_{i\ne j}\frac{\sbest_i(Q_s)^2}{\sigma_i^2}\right)^{1/2}
  \le\frac{s}{\sigma_{\min,-j}}\sum_{i\ne j}\sbest_i(Q_s)
  \le\frac{s}{\sigma_{\min,-j}} .
 \label{eq:normal-changed-share}
\end{equation}
Thus, the rate-optimal allocation ratio $\sbest_j(Q_s)$ assigned to alternative $j$, the only changed alternative, is $O(s)$. Lemma~\ref{lem:comparison} uses the KL divergence from the distribution under $Q_s$ to the distribution under $P$. For alternative $j$, the divergence contributed by one observation is therefore
\begin{equation}
   \KL((Q_s)_j\Vert P_j)=\log(\sigma_j/s)
       +\frac{s^2+(m-\mu_j)^2}{2\sigma_j^2}-\frac{1}{2} .
 \label{eq:normal-changed-kl}
\end{equation}
Combining \eqref{eq:normal-changed-share} and \eqref{eq:normal-changed-kl} with $\lim_{s\downarrow0}s\log(1/s)=0$ gives
\begin{equation}
  \lim_{s\downarrow0}
  \sbest_j(Q_s)\,\KL((Q_s)_j\Vert P_j)=0.
 \label{eq:normal-weighted-kl-zero}
\end{equation}
For every $i\ne j$, $(Q_s)_i=P_i$ and hence $\KL((Q_s)_i\Vert P_i)=0$. Equation~\eqref{eq:normal-weighted-kl-zero} therefore proves \eqref{eq:normal-vanishing-information}. To conclude, let $a_T=T^{-1}\log(1/\Pr_P\{\widehat b_T\ne b(P)\})$. The left-hand side of \eqref{eq:gaussian-one-design} does not depend on $s$, and the inequality holds for every $s>0$. Equation~\eqref{eq:normal-weighted-kl-zero} therefore gives
\[
  0\le \liminf_{T\to\infty}a_T
  \le \limsup_{T\to\infty}a_T
  \le \inf_{s>0}\sbest_j(Q_s)\KL((Q_s)_j\Vert P_j)=0.
\]
Hence the limit exists and equals zero.
\Halmos\endproof

The construction exposes the mechanism. Under $Q_s$, the rate-optimal allocation assigns only an $O(s)$ fraction of the budget to the alternative that distinguishes $Q_s$ from $P$, while one observation from that alternative contributes $O(\log(1/s))$ information. Hence the information collected per unit of budget vanishes. We next show the same phenomenon occurs for other predefined allocation ratios and distributional settings.

\section{When Does Tracking Force a Zero Exponent?}\label{sec:information}

The argument in Section~\ref{sec:gauss} extends beyond normal observations. Lemma~\ref{lem:comparison} bounds the upper PICS exponent at $P$ by the weighted KL information $\sum_{i=1}^k w_i(Q)\KL(Q_i\Vert P_i)$ under any different-best configuration $Q$, so PICS cannot decay exponentially whenever this quantity can be made arbitrarily small by varying $Q$. In this section, we establish the exact converse and obtain a complete characterization: whenever the infimum of this weighted KL information over different-best configurations is strictly positive, some consistent procedure tracking $w$ achieves a positive lower PICS exponent at $P$.

We return to the general setup of Section~\ref{sec:problem}; the distributional setting $\mathfrak D$ is no longer restricted to normal distributions. Fix predefined target ratios $w(Q)\in\mathcal S_k$ for each $Q\in\mathfrak D$. These ratios may depend on $Q$ and need not be rate-optimal. Write $\mathfrak T(w)$ for the procedures that are consistent and track $w$ throughout $\mathfrak D$, as in Section~\ref{sec:requirements}, and assume that $\mathfrak T(w)$ is nonempty.

Fix a configuration $P\in\mathfrak D$ at which PICS is evaluated. Let
\[
  \mathcal A_{\rm fin}(P)
  =\bigl\{Q\in\mathfrak D:
      b(Q)\ne b(P),\ 
      \KL(Q_i\Vert P_i)<\infty\text{ for every }i
    \bigr\}
\]
be the set of different-best configurations having finite coordinatewise KL divergence from $P$. Define
\begin{equation}
   B_w(P)=\inf_{Q\in\mathcal A_{\rm fin}(P)}
               \sum_{i=1}^k w_i(Q)\,\KL(Q_i\Vert P_i),
   \qquad \inf\varnothing=+\infty .
   \label{eq:bound}
\end{equation}
Recall from Definition~\ref{def:exponents} that $\underline e_\pi(P)$ and $\overline e_\pi(P)$ denote the lower and upper PICS exponents of procedure $\pi$ at $P$. Because $0\le N_i(T)/T\le1$, tracking in probability also gives $\lim_{T\to\infty}\E_Q[N_i(T)]/T=w_i(Q)$. Lemma~\ref{lem:comparison} therefore gives, for every $\pi\in\mathfrak T(w)$ and $Q\in\mathcal A_{\rm fin}(P)$,
\[
  \overline e_\pi(P)
  \le \sum_{i=1}^k w_i(Q)\,\KL(Q_i\Vert P_i).
\]
Taking the infimum over $Q\in\mathcal A_{\rm fin}(P)$ gives $0\le\underline e_\pi(P)\le\overline e_\pi(P)\le B_w(P)$. Thus $B_w(P)=0$ forces $\underline e_\pi(P)=\overline e_\pi(P)=0$. The following theorem shows that the converse also holds.

\begin{theorem}\label{thm:zero-characterization}
Let $w(Q)\in\mathcal S_k$ ($Q\in\mathfrak D$) be a predefined target allocation mapping such that the class $\mathfrak T(w)$ of procedures that are consistent and track $w$ throughout $\mathfrak D$ (Definition~\ref{def:requirements}) is nonempty. For any configuration $P\in\mathfrak D$, let $\underline e_\pi(P)$ and $\overline e_\pi(P)$ denote the lower and upper PICS exponents of a procedure $\pi$ at $P$ (Definition~\ref{def:exponents}), and let $B_w(P)$ be the weighted-information bound in \eqref{eq:bound}. Then
\begin{equation}
  B_w(P)=0
  \quad\Longleftrightarrow\quad
  \underline e_\pi(P)=\overline e_\pi(P)=0
  \quad\text{for every }\pi\in\mathfrak T(w).
  \label{eq:zero-characterization}
\end{equation}
In particular, whenever $B_w(P)>0$, there exists a procedure $\pi\in\mathfrak T(w)$ with $\underline e_\pi(P)>0$.
\end{theorem}

The theorem therefore reduces the question of whether PICS can decay exponentially at $P$ to evaluating the single quantity $B_w(P)$. If $B_w(P)=0$, every procedure in $\mathfrak T(w)$ has zero PICS exponent at $P$. If $B_w(P)>0$, at least one procedure in $\mathfrak T(w)$ (which may depend on $P$) has a positive lower PICS exponent at $P$. Section~\ref{app:zero-characterization-proof} gives the complete proof.

Related weighted-KL quantities, with the allocation weights evaluated at a different-best configuration, appear in the allocation game of \citet{komiyama2022}. For constant allocation ratios, \citet[Theorem~1]{degenne2023} gives an exact exponent in one-parameter exponential families. In the normal setting with unit variances, \citet[Theorem~3, Eq.~(8)]{komiyama2026} give the infimum in \eqref{eq:bound} and show that their batched procedure attains a constant fraction of this bound when the target ratios are uniformly bounded away from zero. The characterization here does not require uniform positivity: coordinates of $w(Q)$ may approach zero or equal zero.

We now specialize Theorem~\ref{thm:zero-characterization} to the problem-specific rate-optimal allocation mapping $\sbest(\cdot)$ under any distributional setting $\mathfrak D$ satisfying the static large-deviations framework of Section~\ref{sec:static-allocation}.

\begin{corollary}\label{cor:ld-tracking}
Let $\mathfrak D$ be any distributional setting satisfying \eqref{eq:cramer-oracle} (where each configuration has a unique best alternative, as in Section~\ref{sec:problem}), let $\sbest(Q)$ be a rate-optimal allocation satisfying \eqref{eq:rate-optimal-allocation} for each $Q\in\mathfrak D$, and suppose at least one consistent procedure tracks $\sbest(\cdot)$ on $\mathfrak D$ (Definition~\ref{def:requirements}). For every $P\in\mathfrak D$, define the weighted-information bound
\begin{equation}
  B_{\sbest}(P)
  =\inf_{Q\in\mathcal A_{\rm fin}(P)}
    \sum_{i=1}^k \sbest_i(Q)\,\KL(Q_i\Vert P_i).
\label{eq:tracking-weighted-kl}
\end{equation}
All consistent procedures that track $\sbest(\cdot)$ on $\mathfrak D$ have zero PICS exponent at $P$ (Definition~\ref{def:exponents}) if and only if $B_{\sbest}(P)=0$. If $B_{\sbest}(P)>0$, some consistent procedure that tracks $\sbest(\cdot)$ on $\mathfrak D$ has a positive lower PICS exponent at $P$.
\end{corollary}

\proof{Proof.}
Apply Theorem~\ref{thm:zero-characterization} with $w(Q)=\sbest(Q)$ for each $Q\in\mathfrak D$. \Halmos\endproof

To read \eqref{eq:tracking-weighted-kl}, fix a different-best configuration $Q$. Each product $\sbest_i(Q)\KL(Q_i\Vert P_i)$ is the sampling fraction assigned to alternative $i$ at $Q$ times the information cost of changing that alternative from $P_i$ to $Q_i$. Their sum is the total weighted cost of changing $P$ into $Q$, and $B_{\sbest}(P)$ is the smallest such cost among configurations with a different best alternative. Corollary~\ref{cor:ld-tracking} therefore provides a general tool for analyzing whether a positive PICS exponent, or exponential decay of PICS, is possible for a consistent procedure that tracks $\sbest(\cdot)$. Sections~\ref{sec:ld-oracle} and~\ref{sec:applications} apply this tool to common distributional settings.

\section{Rate-Optimal Tracking in One-Parameter Natural Exponential Families}\label{sec:ld-oracle}

In this section, we apply Corollary~\ref{cor:ld-tracking} to common one-parameter natural exponential families. Section~\ref{cls:nef-family} formulates the exponential-family class, parameterizes distributions by their means, and establishes that Corollary~\ref{cor:ld-tracking} applies. Section~\ref{cls:rate-properties} presents the properties and geometric tangent representation of the rate functions, and Section~\ref{cls:classification} states the main result, a complete characterization of zero PICS exponents in terms of directly verifiable boundary tests built from those quantities. Sections~\ref{cls:two-alternative} and~\ref{cls:three-or-more} then develop the underlying balance conditions and insights for $k=2$ and $k\ge3$, respectively. Concrete examples and normal observations with unequal known variances are then studied in Section~\ref{sec:applications}.

\subsection{Model Formulation and Mean Parameterization}\label{cls:nef-family}

Suppose all alternatives belong to a common, nondegenerate one-parameter natural exponential family (NEF). Let $F_\theta$ denote the distribution in this distributional family indexed by the natural parameter $\theta$. With respect to a fixed reference measure, $F_\theta$ has density $f_\theta(x)=h(x)\exp\{\theta x-A(\theta)\}$, where $h$ is a fixed nonnegative function and $A$ is the cumulant function, both determined by the distributional family. The distributional family is known (as in traditional large deviations analysis; see, e.g., \citealt{chenryzhov2023}), and each alternative has its own unknown parameter value. Examples include Bernoulli, Poisson, and exponential distributions, as well as normal distributions with a common known variance. For a normal distributional family with variance $\sigma^2$, taking $h$ to be the zero-mean normal density with that variance gives $\theta=\mu/\sigma^2$ for the distribution with mean $\mu$.

We do not restrict the range of the natural parameter $\theta$; instead, we allow $\theta$ to range over the entire interior of the domain of $A$, $\Theta=(\theta_-,\theta_+)=\operatorname{int}\{\theta\in\R:A(\theta)<\infty\}$, where $-\infty\le\theta_-<\theta_+\le\infty$ denote the lower and upper endpoints of this open interval. On $\Theta$, the cumulant function $A$ is infinitely differentiable (in particular, continuous and twice differentiable), with first and second derivatives $A'(\theta)=\E_{F_\theta}[X]$ and $A''(\theta)=\operatorname{Var}_{F_\theta}(X)>0$ for every $\theta\in\Theta$, where strict positivity follows from the nondegeneracy of the distributional family. Consequently, $A'$ is continuous and strictly increasing on $\Theta$. Let
\[
  \Omega=A'(\Theta)=(\mu_-,\mu_+)
\]
denote the resulting open range of mean values, with endpoints $\mu_-=\lim_{\theta\downarrow\theta_-}A'(\theta)$ and $\mu_+=\lim_{\theta\uparrow\theta_+}A'(\theta)$ satisfying $-\infty\le \mu_-<\mu_+\le\infty$. Because $A'$ maps $\Theta$ bijectively onto $\Omega$, each mean $u\in \Omega$ uniquely identifies a distribution in the distributional family via its inverse natural parameter $\theta_u=(A')^{-1}(u)$. We may therefore parameterize the distributions by their means rather than by $\theta$, writing $F_u=F_{\theta_u}$ for the distribution with mean $u$. In this common-NEF distributional setting $\mathfrak D$, all parameter vectors in $\Theta^k$ (equivalently, all mean vectors in $\Omega^k$) with a unique best alternative are included.

For each $u\in \Omega$, we also write $I_{\mu=u}(\cdot)=I_{F_u}(\cdot)$ for the rate function of its sample mean (see \eqref{eq:cramer-rate} for its meaning). The notation $\mu=u$ in the subscript identifies the true mean of the distribution, whereas the argument of $I_{\mu=u}(\cdot)$ is the value taken by the sample mean. Under this mean parameterization, Cram\'er's sample-mean rate function coincides directly with the KL divergence between distributions in the distributional family.

\begin{lemma}\label{lem:nef-mean-rate}
For every $u,v\in \Omega$,
\begin{equation}
  I_{\mu=u}(v)
  =(\theta_v-\theta_u)v-A(\theta_v)+A(\theta_u)
  =\KL(F_v\Vert F_u).
  \label{eq:nef-mean-rate}
\end{equation}
\end{lemma}

Lemma~\ref{lem:nef-mean-rate} serves as a key bridge between the rate-optimal allocation defined via large-deviations rate functions in \eqref{eq:cramer-oracle}--\eqref{eq:rate-optimal-allocation} and the zero-exponent condition in Corollary~\ref{cor:ld-tracking}, which relies on KL divergences. Specifically, $I_{\mu=u}(v)$ governs how rapidly the sample mean of an alternative with mean $u$ can deviate to $v$ in \eqref{eq:cramer-oracle}, whereas $\KL(F_v\Vert F_u)$ measures the information cost of shifting an alternative's true mean from $u$ to $v$ in \eqref{eq:tracking-weighted-kl}. Because Lemma~\ref{lem:nef-mean-rate} equates these two objects, every ingredient in Corollary~\ref{cor:ld-tracking} (both the rate-optimal weights $\sbest_i(Q)$ and the KL divergences $\KL(Q_i\Vert P_i)$) can be expressed solely in terms of rate functions, enabling a unified analysis built on their properties in Section~\ref{cls:rate-properties}.

In the distributional setting $\mathfrak D$ defined above, Section~\ref{app:oracles} verifies that the static large-deviations formula \eqref{eq:cramer-oracle} holds, the rate-optimal allocation $\sbest(Q)$ is unique and continuous, and at least one consistent procedure tracks $\sbest(\cdot)$ throughout $\mathfrak D$. The hypotheses of Corollary~\ref{cor:ld-tracking} are therefore satisfied. In the subsections that follow, we use \eqref{eq:nef-mean-rate} together with the boundary behavior of the distributional family and the pattern of ties among the means to determine when the weighted-information bound $B_{\sbest}(P)$ in \eqref{eq:tracking-weighted-kl} vanishes.

\subsection{Properties and Geometric Representation of the Rate Functions}\label{cls:rate-properties}

\begin{figure}[H]
\centering
\caption{The blue shaded gap at $x$ in panel (a) is $I_{\mu=u}(x)-\ell_v(x)=I_{\mu=v}(x)$, matching the blue shaded height in panel (b). At $x=u$, the red shaded gap $-\ell_v(u)$ in panel (a) matches the red shaded height $I_{\mu=v}(u)=-\ell_v(u)$ in panel (b). The zero $\zeta_{u,v}$ of the tangent is also marked in panel (a).}
\label{fig:rate-function-geometry}
\includegraphics[width=.90\textwidth]{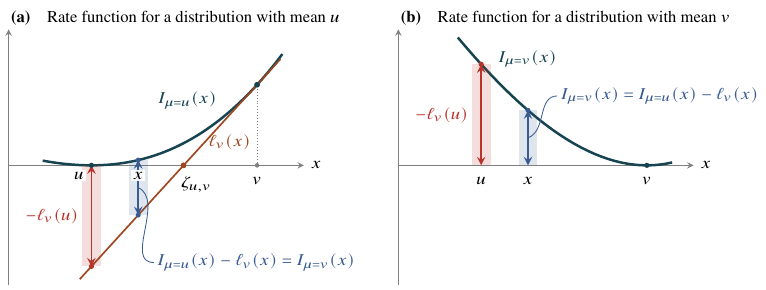}
\end{figure}

Fix a reference mean $u\in \Omega$. An important consequence of Lemma~\ref{lem:nef-mean-rate} is that the single rate function $I_{\mu=u}(\cdot)$ determines both its own shape and, via its tangent lines, the entire rate function $I_{\mu=v}(\cdot)$ for every other mean $v\in \Omega$. Lemma~\ref{lem:nef-rate-properties} collects these properties (proved in Section~\ref{app:oracles}), and Figure~\ref{fig:rate-function-geometry} provides a visual illustration.

\begin{lemma}\label{lem:nef-rate-properties}
Fix $u\in \Omega$. As a function of $x\in \Omega$, the rate function $I_{\mu=u}(x)$ satisfies the following properties.
\begin{enumerate}
\item \textbf{Convexity and derivatives:} $I_{\mu=u}(x)$ is continuous and strictly convex in $x\in \Omega$, nonnegative with $I_{\mu=u}(x)=0$ if and only if $x=u$, strictly decreasing on $(\mu_-,u]$, and strictly increasing on $[u,\mu_+)$. Moreover, it is differentiable on $\Omega$ with derivative $I_{\mu=u}'(x)=\theta_x-\theta_u$ for every $x\in \Omega$ (in particular, $I_{\mu=u}(u)=I_{\mu=u}'(u)=0$).
\item \textbf{Tangent representation:} For any $v\in \Omega$, let $\ell_v(x)=I_{\mu=u}(v)+I_{\mu=u}'(v)(x-v)$ denote the tangent line to $I_{\mu=u}(\cdot)$ at $x=v$. Then, for every $x\in \Omega$,
\begin{equation}
  I_{\mu=v}(x)=I_{\mu=u}(x)-\ell_v(x).
  \label{cls:tangent}
\end{equation}
\item \textbf{Equal-rate crossing point and KL factorization:} For any $v\in \Omega$ with $v\ne u$, the tangent $\ell_v(\cdot)$ crosses the horizontal axis at the unique point
\begin{equation}
  \zeta_{u,v}
  =v-\frac{I_{\mu=u}(v)}{I_{\mu=u}'(v)}
  =\frac{A(\theta_v)-A(\theta_u)}{\theta_v-\theta_u},
  \label{cls:tangent-zero}
\end{equation}
which lies strictly between $u$ and $v$ ($u<\zeta_{u,v}<v$ when $v>u$, and $v<\zeta_{u,v}<u$ when $v<u$) and satisfies $I_{\mu=v}(\zeta_{u,v})=I_{\mu=u}(\zeta_{u,v})$. Moreover, both KL divergences between $F_u$ and $F_v$ factor about $\zeta_{u,v}$:
\begin{equation}
  \KL(F_v\Vert F_u)=I_{\mu=u}(v)=(\theta_v-\theta_u)(v-\zeta_{u,v}),
  \qquad
  \KL(F_u\Vert F_v)=I_{\mu=v}(u)=(\theta_v-\theta_u)(\zeta_{u,v}-u).
  \label{cls:kl-factorization}
\end{equation}
\item \textbf{Boundary comparability across reference means:} For any fixed $u,v\in \Omega$, the ratio $I_{\mu=v}(x)/I_{\mu=u}(x)$ stays bounded above and away from zero as $x\uparrow \mu_+$ and as $x\downarrow \mu_-$:
\begin{equation}
  0<\liminf_{x\uparrow \mu_+}\frac{I_{\mu=v}(x)}{I_{\mu=u}(x)}
  \le\limsup_{x\uparrow \mu_+}\frac{I_{\mu=v}(x)}{I_{\mu=u}(x)}<\infty,
  \qquad
  0<\liminf_{x\downarrow \mu_-}\frac{I_{\mu=v}(x)}{I_{\mu=u}(x)}
  \le\limsup_{x\downarrow \mu_-}\frac{I_{\mu=v}(x)}{I_{\mu=u}(x)}<\infty.
  \label{cls:ref-ratio-bounds}
\end{equation}
\end{enumerate}
\end{lemma}

Part~2 of Lemma~\ref{lem:nef-rate-properties} is the key to our geometric analysis because it reduces an entire family of rate functions to a single curve. Rather than studying $I_{\mu=v}(\cdot)$ separately for each mean $v\in \Omega$, we can fix one reference curve $I_{\mu=u}(\cdot)$ and obtain $I_{\mu=v}(\cdot)$ directly from its tangent line $\ell_v(\cdot)$ at $x=v$. Figure~\ref{fig:rate-function-geometry} illustrates this construction for $v>u$. At every $x\in \Omega$, the blue shaded vertical gap in panel (a) equals $I_{\mu=u}(x)-\ell_v(x)$, which is precisely $I_{\mu=v}(x)$ as plotted in panel (b). Because $I_{\mu=u}(\cdot)$ is strictly convex (Part~1), $\ell_v(\cdot)$ lies strictly below $I_{\mu=u}(\cdot)$ everywhere except at the point of tangency $x=v$. Consequently, the vertical gap in panel (a) is strictly positive for $x\ne v$ and vanishes only at $x=v$, explaining geometrically why $I_{\mu=v}(\cdot)$ shifts its unique zero minimum from $u$ to $v$. Panel (a) also displays both directions of the KL divergence between $F_u$ and $F_v$: $I_{\mu=u}(v)=\KL(F_v\Vert F_u)$ is the height of $I_{\mu=u}(\cdot)$ at $v$, whereas $-\ell_v(u)=I_{\mu=v}(u)=\KL(F_u\Vert F_v)$ is the red shaded vertical gap at $u$, matching the red shaded height at $u$ in panel (b).

Panel (a) also highlights the equal-rate crossing point $\zeta_{u,v}$ from Part~3 of Lemma~\ref{lem:nef-rate-properties}, where the tangent $\ell_v(\cdot)$ crosses the horizontal axis ($\ell_v(\zeta_{u,v})=0$). At $x=\zeta_{u,v}$, the vertical gap in panel (a) equals the curve's own height:
\[
  I_{\mu=u}(\zeta_{u,v})
  =I_{\mu=u}(\zeta_{u,v})-0
  =I_{\mu=u}(\zeta_{u,v})-\ell_v(\zeta_{u,v})
  =I_{\mu=v}(\zeta_{u,v}).
\]
Geometrically, $\zeta_{u,v}$ is therefore the unique point between $u$ and $v$ at which the two rate functions agree; as Section~\ref{cls:two-alternative} will show, this makes $\zeta_{u,v}$ the optimal sample-mean crossing point when allocating observations between distributions with means $u$ and $v$. Because $\ell_v(\cdot)$ has slope $I_{\mu=u}'(v)=\theta_v-\theta_u$ and passes through $(\zeta_{u,v},0)$, the height $\ell_v(v)=\KL(F_v\Vert F_u)$ at $v$ and the red shaded gap $-\ell_v(u)=\KL(F_u\Vert F_v)$ at $u$ yield the two factorizations in \eqref{cls:kl-factorization}, whose sum simplifies to $\KL(F_v\Vert F_u)+\KL(F_u\Vert F_v)=(\theta_v-\theta_u)(v-u)$. When $v<u$, the mirror image holds on the left of $u$, with $v<\zeta_{u,v}<u$ (and \eqref{cls:kl-factorization} remains valid with both factors negative). Figure~\ref{fig:rate-function-geometry} also visualizes Part~4: because $\ell_v(\cdot)$ is a straight line while the strictly convex curve $I_{\mu=u}(\cdot)$ rises more steeply beyond $\max\{u,v\}$ and to the left of $\min\{u,v\}$, the blue vertical gap $I_{\mu=v}(x)=I_{\mu=u}(x)-\ell_v(x)$ in panel~(a) (and height $I_{\mu=v}(x)$ in panel~(b)) stays within constant multiples of the curve height $I_{\mu=u}(x)$ as $x\uparrow \mu_+$ or $x\downarrow \mu_-$. All of these properties depend only on the distributional family; no allocation problem has yet been used.

\subsection{A Complete Characterization of Zero PICS Exponents}\label{cls:classification}

Before deriving the allocation-level mechanics, we state our complete characterization of zero PICS exponents across all common one-parameter natural exponential families. Given any reference mean $u\in \Omega$, recall from Lemma~\ref{lem:nef-rate-properties} the rate function $I_{\mu=u}(\cdot)$, its derivative $I_{\mu=u}'(\cdot)$, and the equal-rate crossing point $\zeta_{u,v}$ between $u$ and $v$ in \eqref{cls:tangent-zero}. Using only these distributional quantities from Section~\ref{cls:rate-properties}, define the three boundary tests
\begin{align}
 S_+
 &=\liminf_{v\uparrow \mu_+}\,(v-\zeta_{u,v})I_{\mu=u}'(\zeta_{u,v}),
 \label{cls:upper-test}\\
 S_-
 &=\liminf_{v\downarrow \mu_-}\,(\zeta_{u,v}-v)\bigl[-I_{\mu=u}'(\zeta_{u,v})\bigr],
 \label{cls:lower-test}\\
 R_-
 &=\liminf_{v\downarrow \mu_-}\,\frac{\zeta_{u,v}-v}{u-v}
 =\liminf_{v\downarrow \mu_-}\,\frac{I_{\mu=u}(v)}{(v-u)I_{\mu=u}'(v)},
 \label{cls:three-test}
\end{align}
where the second expression for $R_-$ follows from $\zeta_{u,v}-v=-I_{\mu=u}(v)/I_{\mu=u}'(v)$ in \eqref{cls:tangent-zero}. Each test relies solely on $I_{\mu=u}$, $I_{\mu=u}'$, and $\zeta_{u,v}$ from Section~\ref{cls:rate-properties}, without invoking any rate-optimal allocation problem, and can therefore be verified directly for a given distributional family.

\begin{theorem}\label{thm:nef-tangent}
Let $\mathfrak D$ be the distributional setting of a common, nondegenerate one-parameter NEF allowing all unique-best configurations in $\Theta^k$ (Section~\ref{cls:nef-family}), and let $S_+$, $S_-$, and $R_-$ be defined by \eqref{cls:upper-test}, \eqref{cls:lower-test}, and \eqref{cls:three-test}, respectively.
For any configuration $P\in\mathfrak D$, all consistent procedures that track $\sbest(\cdot)$ on $\mathfrak D$ (Definition~\ref{def:requirements}) have zero PICS exponent at $P$ (Definition~\ref{def:exponents}) if and only if the corresponding condition below holds.
\begin{enumerate}
\item For $k=2$, $S_-=0$ or $S_+=0$.
\item For $k\ge3$, at least two inferior alternatives at $P$ have the same mean, or $R_-=0$, or $S_+=0$.
\end{enumerate}
Whether $S_-=0$, $S_+=0$, or $R_-=0$ does not depend on the choice of reference mean $u\in\Omega$. When the condition fails, some consistent procedure that tracks $\sbest(\cdot)$ on $\mathfrak D$ has a positive lower PICS exponent at $P$.
\end{theorem}

Section~\ref{app:tangent-classification} gives the complete proof of Theorem~\ref{thm:nef-tangent}, including reference-mean independence and necessity when all means are distinct. In Sections~\ref{cls:two-alternative} and~\ref{cls:three-or-more}, we explain why these boundary conditions and inferior ties govern whether $B_{\sbest}(P)$ vanishes and uncover the geometric and allocation mechanisms behind them; Section~\ref{cls:examples} then evaluates \eqref{cls:upper-test}--\eqref{cls:three-test} on common distributional families.

\subsection{Weighted-Information Setup and the Two-Alternative Case}\label{cls:two-alternative}

We now combine Lemma~\ref{lem:nef-rate-properties} with the static rate-optimal allocation \eqref{eq:rate-optimal-allocation} and Corollary~\ref{cor:ld-tracking}. Consider first any configuration $P=(F_{\mu_1},\ldots,F_{\mu_k})\in\mathfrak D$ with best alternative $b=b(P)$. By the general large-deviations optimality conditions \citep[Theorem~1]{glynn2004} (verified for one-parameter natural exponential families in Sections~\ref{app:oracles} and~\ref{app:tangent-classification}), the unique rate-optimal allocation $w=\sbest(P)$ has strictly positive coordinates and, together with the pairwise optimal crossing points $\{x_j\in(\mu_j,\mu_b):j\ne b\}$ attaining each inner infimum in \eqref{eq:cramer-oracle}, satisfies three balance equations:
\begin{align}
  w_b I_{\mu=\mu_b}'(x_j)+w_j I_{\mu=\mu_j}'(x_j)&=0
  \qquad (j\ne b),
  \label{cls:nef-stationarity}\\
  w_b I_{\mu=\mu_b}(x_j)+w_j I_{\mu=\mu_j}(x_j)&=\Gamma(w,P)
  \qquad (j\ne b),
  \label{cls:nef-individual}\\
  \sum_{j\ne b}\frac{I_{\mu=\mu_b}(x_j)}{I_{\mu=\mu_j}(x_j)}&=1.
  \label{cls:nef-total}
\end{align}
By Part~1 of Lemma~\ref{lem:nef-rate-properties} ($I_{\mu=u}'(x)=\theta_x-\theta_u$), \eqref{cls:nef-stationarity} solves on the natural-parameter scale for $\theta_{x_j}=(w_b\theta_{\mu_b}+w_j\theta_{\mu_j})/(w_b+w_j)$, \eqref{cls:nef-individual} equalizes all pairwise exponents to $\Gamma(w,P)$, and \eqref{cls:nef-total} balances the best alternative against all rivals.

Next, rewrite the weighted-information bound \eqref{eq:tracking-weighted-kl} in Corollary~\ref{cor:ld-tracking} using the rate-function parameterization. Fix $P=(F_{\mu_1},\ldots,F_{\mu_k})\in\mathfrak D$, where $\mu_i=\mu_i(P)$ is the mean of alternative $i$ under $P$. For any configuration $Q=(Q_1,\ldots,Q_k)\in\mathfrak D$, write $\mu_i(Q)$ for the mean of $Q_i$ (so $Q_i=F_{\mu_i(Q)}$). Because every $Q\in\mathfrak D$ has finite coordinatewise KL divergences from $P$, we have $\mathcal A_{\rm fin}(P)=\{Q\in\mathfrak D:b(Q)\ne b(P)\}$, and replacing $\KL(Q_i\Vert P_i)$ by $I_{\mu=\mu_i}(\mu_i(Q))$ via Lemma~\ref{lem:nef-mean-rate} yields
\begin{equation}
  B_{\sbest}(P)
  =\inf_{Q\in\mathfrak D:\,b(Q)\ne b(P)}
   \sum_{i=1}^k \sbest_i(Q)\,I_{\mu=\mu_i}\bigl(\mu_i(Q)\bigr).
  \label{cls:nef-tracking-bound}
\end{equation}
All ingredients in \eqref{cls:kl-factorization}, \eqref{cls:nef-stationarity}--\eqref{cls:nef-total}, and \eqref{cls:nef-tracking-bound} are now expressed solely in terms of the rate functions $I_{\mu=\cdot}(\cdot)$, giving a common framework for both $k=2$ and $k\ge3$.

\subsubsection{Sufficient Condition for $B_{\sbest}(P)=0$ When $k=2$.}
We now apply \eqref{cls:nef-tracking-bound} in the two-alternative case ($k=2$). Fix $P=(F_{\mu_1},F_{\mu_2})\in\mathfrak D$ with $\mu_1>\mu_2$. To construct a different-best configuration $Q\in\mathfrak D$ with $b(Q)\ne b(P)$ in \eqref{cls:nef-tracking-bound}, we can either raise alternative~2 to $v>\mu_1$, setting $Q=(F_{\mu_1},F_v)$, or lower alternative~1 to $v<\mu_2$, setting $Q=(F_v,F_{\mu_2})$. For the unchanged alternative, $\mu_i(Q)=\mu_i$ gives $I_{\mu=\mu_i}(\mu_i(Q))=0$ (Part~1 of Lemma~\ref{lem:nef-rate-properties}), so only a single term remains in \eqref{cls:nef-tracking-bound}:
\begin{equation}
  0\le B_{\sbest}(P)\le \sbest_2(F_{\mu_1},F_v)\,I_{\mu=\mu_2}(v)\quad(v>\mu_1),
  \qquad
  0\le B_{\sbest}(P)\le \sbest_1(F_v,F_{\mu_2})\,I_{\mu=\mu_1}(v)\quad(v<\mu_2).
  \label{cls:two-weighted-information}
\end{equation}
To put both factors in each bound of \eqref{cls:two-weighted-information} on the same pair of means, multiply and divide by $I_{\mu=\mu_1}(v)>0$ when $v>\mu_1$ and by $I_{\mu=\mu_2}(v)>0$ when $v<\mu_2$:
\begin{align}
  0\le B_{\sbest}(P)&\le \sbest_2(F_{\mu_1},F_v)\,I_{\mu=\mu_1}(v)\cdot\frac{I_{\mu=\mu_2}(v)}{I_{\mu=\mu_1}(v)}
  \qquad (v>\mu_1),
  \label{cls:two-aligned-upper}\\
  0\le B_{\sbest}(P)&\le \sbest_1(F_v,F_{\mu_2})\,I_{\mu=\mu_2}(v)\cdot\frac{I_{\mu=\mu_1}(v)}{I_{\mu=\mu_2}(v)}
  \qquad (v<\mu_2).
  \label{cls:two-aligned-lower}
\end{align}
By \eqref{cls:ref-ratio-bounds} in Part~4 of Lemma~\ref{lem:nef-rate-properties}, the ratio factor in \eqref{cls:two-aligned-upper} has finite upper limit as $v\uparrow \mu_+$ and the ratio factor in \eqref{cls:two-aligned-lower} has finite upper limit as $v\downarrow \mu_-$. Consequently, $B_{\sbest}(P)=0$ whenever
\begin{equation}
  \liminf_{v\uparrow \mu_+}\,\sbest_2(F_{\mu_1},F_v)\,I_{\mu=\mu_1}(v)=0
  \qquad\text{or}\qquad
  \liminf_{v\downarrow \mu_-}\,\sbest_1(F_v,F_{\mu_2})\,I_{\mu=\mu_2}(v)=0.
  \label{cls:two-boundary-condition}
\end{equation}

\subsubsection{Rate-Optimal Allocation Ratios and Weighted Information.}
The two limits in \eqref{cls:two-boundary-condition} are symmetric around the fixed unchanged mean. We obtain the upper-boundary allocation ($v\uparrow \mu_+$) geometrically, writing $\sbest_2=\sbest_2(F_{\mu_1},F_v)$ for $v>\mu_1$, and then state the symmetric lower-boundary counterpart ($v\downarrow \mu_-$).

Fix $v>\mu_1$, and assign fraction $w_2\in(0,1)$ to $F_v$ and $1-w_2$ to $F_{\mu_1}$. By \eqref{eq:cramer-oracle} and the tangent representation $I_{\mu=v}(x)=I_{\mu=\mu_1}(x)-\ell_v(x)$ in \eqref{cls:tangent} (with reference mean $\mu_1$), the static exponent is the minimum vertical gap between the curve $I_{\mu=\mu_1}(\cdot)$ and the scaled line $w_2\ell_v(\cdot)$:
\begin{equation}
  \Gamma\bigl((1-w_2,w_2),(F_{\mu_1},F_v)\bigr)
  =\min_{\mu_1\le x\le v}\bigl\{(1-w_2)I_{\mu=\mu_1}(x)+w_2 I_{\mu=v}(x)\bigr\}
  =\min_{\mu_1\le x\le v}\bigl\{I_{\mu=\mu_1}(x)-w_2\ell_v(x)\bigr\}.
  \label{cls:binary-objective}
\end{equation}

\begin{figure}[H]
\centering
\caption{Panel (a) shows that multiplying $\ell_v(\cdot)$ by $w_2$ rotates it about its zero $\zeta_{\mu_1,v}$. In panel (b), the optimal line $\sbest_2\ell_v(\cdot)$ is parallel to the tangent of $I_{\mu=\mu_1}(\cdot)$ at $\zeta_{\mu_1,v}$. Equality of their slopes gives \eqref{cls:binary-optimal-allocation}. For $(F_{\mu_1},F_v)$, the shaded gap at $\zeta_{\mu_1,v}$ is the optimal exponent $\Gamma\bigl((1-\sbest_2,\sbest_2),(F_{\mu_1},F_v)\bigr)$, and the height of the rotated line at $v$ gives the weighted information in \eqref{cls:binary-weighted-geometry}.}
\label{fig:nef-tangent}
\includegraphics[width=.95\textwidth]{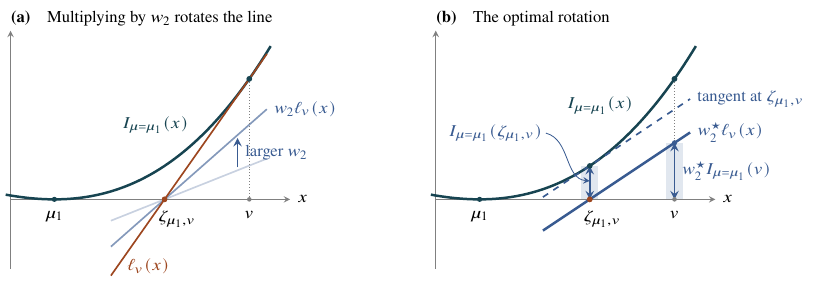}
\end{figure}

Figure~\ref{fig:nef-tangent} illustrates the geometry of maximizing \eqref{cls:binary-objective} over $w_2$. Because $\ell_v(\zeta_{\mu_1,v})=0$ (Part~3 of Lemma~\ref{lem:nef-rate-properties}), every scaled line $w_2\ell_v(\cdot)$ passes through $(\zeta_{\mu_1,v},0)$; varying $w_2$ simply rotates this line about $(\zeta_{\mu_1,v},0)$, as shown in panel~(a). At $x=\zeta_{\mu_1,v}$, the height of $w_2\ell_v(\cdot)$ is zero for every $w_2$, so evaluating the minimum in \eqref{cls:binary-objective} at $x=\zeta_{\mu_1,v}$ bounds every allocation's exponent by
\[
  \Gamma\bigl((1-w_2,w_2),(F_{\mu_1},F_v)\bigr)
  \le I_{\mu=\mu_1}(\zeta_{\mu_1,v}).
\]
To attain this upper bound, rotate $w_2\ell_v(\cdot)$ until it has the same slope as the tangent to $I_{\mu=\mu_1}(\cdot)$ at $\zeta_{\mu_1,v}$, as shown in panel~(b):
\begin{equation}
 \sbest_2 I_{\mu=\mu_1}'(v)=I_{\mu=\mu_1}'(\zeta_{\mu_1,v}),
 \qquad\text{so}\qquad
 \sbest_2=\frac{I_{\mu=\mu_1}'(\zeta_{\mu_1,v})}{I_{\mu=\mu_1}'(v)}.
 \label{cls:binary-optimal-allocation}
\end{equation}
Because $\mu_1<\zeta_{\mu_1,v}<v$ and $I_{\mu=\mu_1}(\cdot)$ is strictly convex (Part~1 of Lemma~\ref{lem:nef-rate-properties}), $0<I_{\mu=\mu_1}'(\zeta_{\mu_1,v})<I_{\mu=\mu_1}'(v)$, so $\sbest_2\in(0,1)$. Shifting $\sbest_2\ell_v(\cdot)$ upward by $I_{\mu=\mu_1}(\zeta_{\mu_1,v})$ gives the tangent line to $I_{\mu=\mu_1}(\cdot)$ at $\zeta_{\mu_1,v}$; by strict convexity, the vertical gap $I_{\mu=\mu_1}(x)-\sbest_2\ell_v(x)$ attains its unique minimum at $x=\zeta_{\mu_1,v}$ with maximal exponent $I_{\mu=\mu_1}(\zeta_{\mu_1,v})$ (see Section~\ref{app:tangent-classification} for the analytic derivation and uniqueness).

Panel~(b) also gives the weighted product $\sbest_2 I_{\mu=\mu_1}(v)$ directly. At $x=v$, $\ell_v(v)=I_{\mu=\mu_1}(v)$, so the height of the optimal rotated line at $v$ is $\sbest_2 I_{\mu=\mu_1}(v)$. Since that line passes through $(\zeta_{\mu_1,v},0)$ with slope $I_{\mu=\mu_1}'(\zeta_{\mu_1,v})$, its height at $v$ equals
\begin{equation}
 \sbest_2 I_{\mu=\mu_1}(v)
 =(v-\zeta_{\mu_1,v})I_{\mu=\mu_1}'(\zeta_{\mu_1,v}).
 \label{cls:binary-weighted-geometry}
\end{equation}

\subsubsection{Boundary Tests $S_+$ and $S_-$.}
Substituting \eqref{cls:binary-weighted-geometry} into the first limit in \eqref{cls:two-boundary-condition} turns the upper-boundary condition into
\[
  \liminf_{v\uparrow \mu_+}\,(v-\zeta_{\mu_1,v})I_{\mu=\mu_1}'(\zeta_{\mu_1,v})=0.
\]
Section~\ref{app:tangent-classification} shows that whether this lower limit vanishes does not depend on the fixed reference mean $\mu_1\in \Omega$, so this condition matches precisely the upper-boundary test $S_+=0$ defined in \eqref{cls:upper-test}. Symmetrically, on the lower side ($v<\mu_2$), $\sbest_1(F_v,F_{\mu_2})\,I_{\mu=\mu_2}(v)=(\zeta_{\mu_2,v}-v)[-I_{\mu=\mu_2}'(\zeta_{\mu_2,v})]$ (see Section~\ref{app:tangent-classification} for the full derivation), and reference-mean independence matches this condition to the lower-boundary test $S_-=0$ in \eqref{cls:lower-test}. Hence, if $S_+=0$ or $S_-=0$, then \eqref{cls:two-boundary-condition} holds and forces $B_{\sbest}(P)=0$, accounting geometrically for the sufficiency direction of Theorem~\ref{thm:nef-tangent} when $k=2$.

\subsection{Insights for Three or More Alternatives}\label{cls:three-or-more}

When $k\ge3$, having three or more alternatives creates two new ways for $B_{\sbest}(P)$ in \eqref{cls:nef-tracking-bound} to vanish beyond the two-alternative case, because two alternatives can compete for first place while starving the remaining ones. We first use individual balance in \eqref{cls:nef-individual} to bound the share of any lower-ranked alternative (proved in Section~\ref{app:oracles}).

\begin{lemma}\label{lem:outsider-share}
Consider any configuration $P=(F_{\mu_1},\ldots,F_{\mu_k})\in\mathfrak D$ with $k\ge3$ and means ordered $\mu_1>\mu_2\ge\cdots\ge\mu_k$ (so $b(P)=1$ is best and alternative~$2$ is a runner-up), and let $w=\sbest(P)$ be its rate-optimal allocation. For every alternative $i\ge3$ with mean $\mu_i<\mu_2$,
\begin{equation}
 w_i I_{\mu=\mu_i}(\mu_2) < I_{\mu=\mu_1}(\mu_2).
 \label{cls:outsider-main}
\end{equation}
\end{lemma}

Applying Lemma~\ref{lem:outsider-share} to different-best configurations in \eqref{cls:nef-tracking-bound} yields the two additional mechanisms specific to $k\ge3$. Throughout the next two constructions, fix $P\in\mathfrak D$ and relabel its alternatives so that $b=1$ is best and $\mu_1>\mu_2\ge\cdots\ge\mu_k$.

\subsubsection{Demoting the Best Alternative Toward $\mu_-$.}
Suppose $\mu_2>\mu_3$ (otherwise $\mu_2=\mu_3$ is already an inferior tie, treated next). Keep alternatives $2,\ldots,k$ fixed at their true means and demote only alternative~$1$ to $v\in(\mu_-,\mu_3)$ to form $Q_v\in\mathfrak D$; under $Q_v$, alternative~$2$ is uniquely best with mean $\mu_2$, alternative~$3$ is a runner-up with mean $\mu_3<\mu_2$, and alternative~$1$ has mean $v<\mu_3$. Writing $w=\sbest(Q_v)$, only alternative~$1$ changes in \eqref{cls:nef-tracking-bound}, so for every $v\in(\mu_-,\mu_3)$,
\[
  0\le B_{\sbest}(P)\le w_1 I_{\mu=\mu_1}(v).
\]
Next, applying Lemma~\ref{lem:outsider-share} to $Q_v$ (with best~$2$ of mean $\mu_2$, runner-up~$3$ of mean $\mu_3$, and alternative~$1$ of mean $v<\mu_3$ in the role of $i$) gives $w_1 I_{\mu=v}(\mu_3) < I_{\mu=\mu_2}(\mu_3)$. Substituting this bound (using $I_{\mu=v}(\mu_3)>0$) and multiplying and dividing by $I_{\mu=\mu_3}(v)>0$ gives
\[
  0\le B_{\sbest}(P)
  \le w_1 I_{\mu=\mu_1}(v)
  < \frac{I_{\mu=\mu_2}(\mu_3)}{I_{\mu=v}(\mu_3)}\,I_{\mu=\mu_1}(v)
  = I_{\mu=\mu_2}(\mu_3)\,\frac{I_{\mu=\mu_3}(v)}{I_{\mu=v}(\mu_3)}\cdot\frac{I_{\mu=\mu_1}(v)}{I_{\mu=\mu_3}(v)}.
\]
Because $\limsup_{v\downarrow \mu_-} I_{\mu=\mu_1}(v)/I_{\mu=\mu_3}(v)<\infty$ by \eqref{cls:ref-ratio-bounds}, $B_{\sbest}(P)=0$ whenever $I_{\mu=\mu_3}(v)/I_{\mu=v}(\mu_3)$ has lower limit zero as $v\downarrow \mu_-$. Using \eqref{cls:kl-factorization} at $u=\mu_3$ cancels $\theta_{\mu_3}-\theta_v$ directly:
\[
 \frac{I_{\mu=\mu_3}(v)}{I_{\mu=v}(\mu_3)}
 =\frac{(\theta_{\mu_3}-\theta_v)(\zeta_{\mu_3,v}-v)}{(\theta_{\mu_3}-\theta_v)(\mu_3-\zeta_{\mu_3,v})}
 =\frac{\zeta_{\mu_3,v}-v}{\mu_3-\zeta_{\mu_3,v}}
 =\frac{(\zeta_{\mu_3,v}-v)/(\mu_3-v)}{1-(\zeta_{\mu_3,v}-v)/(\mu_3-v)}.
\]
Section~\ref{app:tangent-classification} shows that whether $(\zeta_{\mu_3,v}-v)/(\mu_3-v)$ has lower limit zero as $v\downarrow \mu_-$ is independent of the fixed reference mean $\mu_3\in \Omega$, so this ratio vanishes along a sequence $v\downarrow \mu_-$ if and only if $R_-=0$ in \eqref{cls:three-test} (where the second form in \eqref{cls:three-test} follows from \eqref{cls:tangent-zero}).

\subsubsection{Inferior Tie at Any Rank.}
Suppose two inferior alternatives at $P$ share the same mean; relabeling the inferior alternatives if needed (without assuming this shared mean exceeds the remaining inferior means), take them to be alternatives~$2$ and~$3$ with $\mu_2=\mu_3$. Fix $\underline\mu\in(\mu_-,\mu_2)$ and sufficiently small $\epsilon>0$, and assign every other alternative mean $\underline\mu<\mu_2$ to form $Q_\epsilon\in\mathfrak D$:
\[
  \mu_2(Q_\epsilon)=\mu_2+\epsilon,
  \qquad
  \mu_3(Q_\epsilon)=\mu_2,
  \qquad
  \mu_i(Q_\epsilon)=\underline\mu\quad(i\notin\{2,3\}).
\]
Then $Q_\epsilon$ has unique best alternative~$2$ (mean $\mu_2+\epsilon$), runner-up~$3$ (mean $\mu_2=\mu_3$), and every $i\notin\{2,3\}$ at $\underline\mu<\mu_2$. Because alternative~$3$ is unchanged, $I_{\mu=\mu_3}(\mu_3(Q_\epsilon))=0$. Writing $w=\sbest(Q_\epsilon)$ and using $w_2\le 1$, substituting $Q_\epsilon$ into \eqref{cls:nef-tracking-bound} gives
\[
  0\le B_{\sbest}(P)
  \le w_2 I_{\mu=\mu_2}(\mu_2+\epsilon) + \sum_{i\notin\{2,3\}} w_i I_{\mu=\mu_i}(\underline\mu)
  \le I_{\mu=\mu_2}(\mu_2+\epsilon) + \sum_{i\notin\{2,3\}} w_i I_{\mu=\mu_i}(\underline\mu).
\]
By \eqref{cls:kl-factorization} (with $u=\mu_2$ and $v=\mu_2+\epsilon$),
\[
  I_{\mu=\mu_2}(\mu_2+\epsilon)+I_{\mu=\mu_2+\epsilon}(\mu_2)
  =(\theta_{\mu_2+\epsilon}-\theta_{\mu_2})\epsilon
  =O(\epsilon^2),
\]
where the second equality holds because $\theta_x=(A')^{-1}(x)$ is differentiable near $\mu_2$ (so $\theta_{\mu_2+\epsilon}-\theta_{\mu_2}=O(\epsilon)$); since both rate terms are nonnegative, each is $O(\epsilon^2)$. Next, applying Lemma~\ref{lem:outsider-share} to $Q_\epsilon$ (with best~$2$ of mean $\mu_2+\epsilon$, runner-up~$3$ of mean $\mu_2$, and each $i\notin\{2,3\}$ of mean $\underline\mu<\mu_2$) gives $w_i I_{\mu=\underline\mu}(\mu_2) < I_{\mu=\mu_2+\epsilon}(\mu_2)$. Substituting this bound (using $I_{\mu=\underline\mu}(\mu_2)>0$) gives
\[
  0\le B_{\sbest}(P)
  < I_{\mu=\mu_2}(\mu_2+\epsilon) + \sum_{i\notin\{2,3\}}\frac{I_{\mu=\mu_2+\epsilon}(\mu_2)}{I_{\mu=\underline\mu}(\mu_2)}\,I_{\mu=\mu_i}(\underline\mu)
  =O(\epsilon^2),
\]
because each $I_{\mu=\mu_i}(\underline\mu)$ and $I_{\mu=\underline\mu}(\mu_2)>0$ is fixed and independent of $\epsilon$. Sending $\epsilon\downarrow0$ therefore forces $B_{\sbest}(P)=0$ without any mean approaching the boundary of $\Omega$.

Sections~\ref{cls:two-alternative} and~\ref{cls:three-or-more} thus establish the constructive, sufficient mechanisms behind Theorem~\ref{thm:nef-tangent} that force $B_{\sbest}(P)=0$; Section~\ref{app:tangent-classification} supplies the remaining pieces of the proof of Theorem~\ref{thm:nef-tangent}, including reference-mean independence, the $k\ge3$ upper-boundary promotion for $S_+=0$, the implication $S_-=0\implies R_-=0$, and necessity ($B_{\sbest}(P)>0$) when all means are distinct and none of the collapse conditions holds.

\section{Examples and Normal Observations with Known Variances}\label{sec:applications}

We now apply the results of Section~\ref{sec:ld-oracle} to concrete distributional settings. Section~\ref{cls:examples} evaluates the boundary tests of Theorem~\ref{thm:nef-tangent} across four common one-parameter natural exponential families, and Section~\ref{kv:section} moves beyond a common natural exponential family to apply Corollary~\ref{cor:ld-tracking} to normal alternatives with heterogeneous known variances.

\subsection{Examples in Common Natural Exponential Families}\label{cls:examples}

Table~\ref{cls:examples-table} summarizes the boundary quantities $S_-$, $S_+$, and $R_-$ and the resulting exponent regimes at configurations with distinct means for four common distributional families in R\&S (when two inferior means coincide, Theorem~\ref{thm:nef-tangent} forces a zero exponent for every family in the table). Below we evaluate the Bernoulli and exponential cases, which illustrate bounded-domain collapse for all $k\ge2$ and the separation between $k=2$ and $k\ge3$; Section~\ref{app:normal-poisson} evaluates the normal (common known variance) and Poisson cases, completing Table~\ref{cls:examples-table} (and Table~\ref{tab:nef-tracking}).

\begin{table}[tbp]
\caption{Boundary tests and PICS-exponent regimes at configurations with distinct means.}
\label{cls:examples-table}
\centering
\small
\begin{tabular}{@{}lcccll@{}}
\hline
Distributional family & $S_-$ & $S_+$ & $R_-$ & $k=2$ & $k\ge3$ \\
\hline
Normal with common known variance & $\infty$ & $\infty$ & $1/2$ & Positive possible & Positive possible \\
Exponential & $1$ & $\infty$ & $0$ & Positive possible & Zero forced \\
Poisson & $0$ & $\infty$ & $0$ & Zero forced & Zero forced \\
Bernoulli & $0$ & $0$ & $0$ & Zero forced & Zero forced \\
\hline
\end{tabular}
\end{table}

\subsubsection{Bernoulli.}
Take reference mean $u=1/2$ on $\Omega=(0,1)$ and write $I_{1/2}(\cdot)=I_{\mu=1/2}(\cdot)$, so
\[
  I_{1/2}(v)=\log 2+v\log v+(1-v)\log(1-v),
  \qquad
  I_{1/2}'(v)=\log\frac{v}{1-v}
  \qquad(0<v<1).
\]
On the lower side ($0<v<1/2$), letting $v\downarrow0$ gives $I_{1/2}(v)\to\log 2<\infty$ and $-I_{1/2}'(v)=\log[(1-v)/v]\to\infty$. By \eqref{cls:tangent-zero}, $\zeta_{1/2,v}-v=I_{1/2}(v)/[-I_{1/2}'(v)]\to0$, so $\zeta_{1/2,v}=v+(\zeta_{1/2,v}-v)\to0$ and $I_{1/2}(\zeta_{1/2,v})\to\log 2$. Since $\zeta_{1/2,v}\in(v,1/2)$ and $I_{1/2}$ is convex and decreasing on $(0,1/2]$, convexity gives $I_{1/2}(v)\ge I_{1/2}(\zeta_{1/2,v})+I_{1/2}'(\zeta_{1/2,v})(v-\zeta_{1/2,v})$, so
\[
  0\le(\zeta_{1/2,v}-v)\bigl[-I_{1/2}'(\zeta_{1/2,v})\bigr]\le I_{1/2}(v)-I_{1/2}(\zeta_{1/2,v})\longrightarrow 0.
\]
Substituting into \eqref{cls:lower-test} and \eqref{cls:three-test} gives $S_-=0$ and $R_-=\liminf_{v\downarrow0}(\zeta_{1/2,v}-v)/(1/2-v)=0$. By symmetry ($I_{1/2}(1-v)=I_{1/2}(v)$), as $v\uparrow1$ we likewise have $v-\zeta_{1/2,v}\to0$, $\zeta_{1/2,v}\to1$, and $0\le(v-\zeta_{1/2,v})I_{1/2}'(\zeta_{1/2,v})\le I_{1/2}(v)-I_{1/2}(\zeta_{1/2,v})\to0$, giving $S_+=0$ in \eqref{cls:upper-test}. Theorem~\ref{thm:nef-tangent} thus forces a zero PICS exponent for all $k\ge2$.

\subsubsection{Exponential.}
Exponential observations separate $k=2$ from $k\ge3$ even when all means are distinct. Take reference mean $u=1$ and abbreviate $I_1(\cdot)=I_{\mu=1}(\cdot)$, so
\[
  I_1(v)=v-1-\log v,
  \qquad
  I_1'(v)=1-\frac{1}{v}=-\frac{1-v}{v}
  \qquad(v>0).
\]
On the lower side ($0<v<1$), letting $v\downarrow0$ gives $I_1(v)\sim\log(1/v)$ and $-I_1'(v)=(1-v)/v\sim 1/v$, so \eqref{cls:tangent-zero} yields
\[
  \zeta_{1,v}-v=\frac{I_1(v)}{-I_1'(v)}\sim v\log\frac{1}{v}\longrightarrow 0,
  \qquad
  \zeta_{1,v}=v+(\zeta_{1,v}-v)\sim v\log\frac{1}{v}\longrightarrow 0.
\]
Substituting $I_1'(\zeta_{1,v})=-(1-\zeta_{1,v})/\zeta_{1,v}$ into \eqref{cls:lower-test} and \eqref{cls:three-test} gives
\[
  S_-=\liminf_{v\downarrow0}\,(\zeta_{1,v}-v)\frac{1-\zeta_{1,v}}{\zeta_{1,v}}=1,
  \qquad
  R_-=\liminf_{v\downarrow0}\,\frac{\zeta_{1,v}-v}{1-v}=0.
\]
On the upper side ($v>1$), \eqref{cls:tangent-zero} gives $\zeta_{1,v}=v-I_1(v)/I_1'(v)=v\log v/(v-1)\to\infty$ as $v\to\infty$. Thus $v-\zeta_{1,v}\sim v$ and $I_1'(\zeta_{1,v})=1-1/\zeta_{1,v}\to 1$, so \eqref{cls:upper-test} gives $S_+=\infty$. For $k=2$, $S_-=1>0$ and $S_+=\infty$ allow a positive lower exponent at every configuration, whereas for $k\ge3$, $R_-=0$ forces a zero exponent everywhere.

\subsection{Normal Observations with Known Variances}\label{kv:section}

Section~\ref{sec:gauss} showed that when normal variances are unknown, tracking $\sbest(\cdot)$ forces a zero PICS exponent at every configuration, whereas the first example in Table~\ref{cls:examples-table} assumes a common known variance and rules out a zero exponent whenever all means are distinct. To separate variance heterogeneity from variance unknownness, we now allow each alternative to have its own known variance $\sigma_i^2$. Because unequal variances place the alternatives in different one-parameter natural exponential families, Theorem~\ref{thm:nef-tangent} does not apply directly. Fix $\sigma_1,\ldots,\sigma_k>0$ and let $\mathfrak D$ be the known-variance normal distributional setting consisting of all configurations $P_i=\N(\mu_i,\sigma_i^2)$ with a unique best alternative.

\begin{theorem}\label{kv:theorem}
Let $\mathfrak D$ be the normal distributional setting with given known variances $\sigma_1^2,\ldots,\sigma_k^2>0$ (Section~\ref{kv:section}). For any configuration $P\in\mathfrak D$, all consistent procedures that track $\sbest(\cdot)$ on $\mathfrak D$ (Definition~\ref{def:requirements}) have zero PICS exponent at $P$ (Definition~\ref{def:exponents}) if and only if at least two inferior alternatives at $P$ have the same mean (which can occur only when $k\ge3$). When all means are distinct, some consistent procedure that tracks $\sbest(\cdot)$ on $\mathfrak D$ has a positive lower PICS exponent at $P$.
\end{theorem}

With two alternatives, the optimal ratios are $\sigma_i/(\sigma_1+\sigma_2)$ at every configuration. Sampling deterministically in these proportions and selecting the larger sample mean attains the optimal static exponent $(\mu_1-\mu_2)^2/[2(\sigma_1+\sigma_2)^2]$ throughout this setting. More broadly, compared with the unknown-variance setting of Section~\ref{sec:gauss}, knowing the variances removes variance uncertainty and leaves only the alternative means unknown. Because each alternative's precision is known in advance, distinguishing configurations hinges directly on mean separations rather than unknown noise scales: when all means are distinct, no perturbation of the means alone can overturn the best alternative at vanishing weighted information cost, so a zero exponent arises only when $k\ge3$ and at least two inferior means tie. The numerical experiments in Section~\ref{sec:numerical-normal} echo this distinction: on the same underlying normal configurations, the tracking procedures maintain positive finite-budget exponents under known variances when all means are distinct, yet their exponents drop toward zero across all configurations once the variances are unknown and must be estimated.

Table~\ref{kv:summary-table} summarizes the PICS-exponent regimes across the normal variance settings studied in Sections~\ref{sec:gauss}, \ref{cls:classification}, and~\ref{kv:section}: equal and unequal known variances yield the same classification, whereas unknown variances force a zero exponent everywhere. Section~\ref{kv:proof} proves Theorem~\ref{kv:theorem}.

\begin{table}[tbp]
\caption{Summary of PICS-exponent regimes across normal variance settings with a unique best alternative.}
\label{kv:summary-table}
\centering
\small
\begin{tabular}{@{}lccccc@{}}
\hline
 & & & & \multicolumn{2}{c}{$k\ge3$} \\
\cline{5-6}
Variance setting & Result & Common NEF & $k=2$ & \begin{tabular}[b]{@{}c@{}}All means\\[-2pt]distinct\end{tabular} & \begin{tabular}[b]{@{}c@{}}At least two tied\\[-2pt]inferior means\end{tabular} \\
\hline
Common known variance & Theorem~\ref{thm:nef-tangent} & Yes & Positive possible & Positive possible & Zero forced \\
Unequal known variances & Theorem~\ref{kv:theorem} & No & Positive possible & Positive possible & Zero forced \\
Unknown variances & Theorem~\ref{thm:zero} & No & Zero forced & Zero forced & Zero forced \\
\hline
\end{tabular}
\end{table}

\section{The Benefits of Tracking Rate-Optimal Sample Allocations}\label{dt:section}

The negative results in Sections~\ref{sec:gauss}--\ref{sec:applications} show that \emph{pure} tracking of the rate-optimal allocation $\sbest(P)$ often forces $B_{\sbest}(P)=0$ and hence zero PICS exponent. Does this mean that rate-optimal allocations cannot help achieve exponential PICS decay, or improve upon simpler sampling rules, in adaptive procedures? Theorem~\ref{thm:zero-characterization} shows that tracking a target mapping $w(\cdot)$ permits a positive PICS exponent whenever $B_w(P)>0$. To prevent the subexponential collapse of pure tracking, we can combine the rate-optimal allocation with the equal-allocation (EA) weights $w_{\rm EA}=(1/k,\ldots,1/k)$ by setting, for $\lambda\in(0,1)$,
\[
 v_\lambda(P)=(1-\lambda)w_{\rm EA}+\lambda\sbest(P).
\]
This combination is guaranteed to have a positive weighted-information bound: in \eqref{eq:bound}, dropping the nonnegative tracking term $\lambda\sum_{i=1}^k \sbest_i(Q)\KL(Q_i\Vert P_i)\ge 0$ gives $B_{v_\lambda}(P)\ge (1-\lambda)B_{w_{\rm EA}}(P)$, while deterministic equal allocation attains the static exponent $\Gamma(w_{\rm EA},P)>0$ under exponential-tail distributions in \eqref{eq:cramer-oracle} (for instance, $\Gamma(w_{\rm EA},P)=\min_{j\ne b}(\mu_b-\mu_j)^2/[2k(\sigma_b^2+\sigma_j^2)]>0$ for Gaussian alternatives $P_i=\N(\mu_i,\sigma_i^2)$); because deterministic EA tracks $w_{\rm EA}$ consistently, Lemma~\ref{lem:comparison} implies $B_{w_{\rm EA}}(P)\ge \Gamma(w_{\rm EA},P)$ and hence
\[
 B_{v_\lambda}(P)\ge (1-\lambda)\Gamma(w_{\rm EA},P)>0.
\]

Because pure tracking on its own yields a zero exponent at many configurations, one might expect that mixing in $\sbest(P)$ could at best dilute equal allocation to $(1-\lambda)\Gamma(w_{\rm EA},P)$ or leave the exponent no higher than $\Gamma(w_{\rm EA},P)$. We find the opposite: once enforced uniform exploration prevents collapse, tracking $\sbest$ \emph{strictly improves} the PICS exponent beyond equal allocation whenever $k\ge3$. Thus, combined with uniform exploration, rate-optimal tracking delivers genuine improvements in the exponential convergence rate.

To implement this combination through concrete adaptive procedures, we consider OCBA-type plug-in tracking rules \citep{chen2000,glynn2004} that recompute the estimated rate-optimal allocation from the current sample estimates at each step and direct the next observation to the alternative furthest below its target count. First, in Section~\ref{pv:gauss-section}, for independent Gaussian alternatives with known or unknown variances, we reserve a fraction $1-\lambda$ of the budget for an initial EA pilot (in the spirit of \citealt{wuzhou2018}) and allocate the remaining budget by tracking the estimated rate-optimal allocation against cumulative sample counts. Second, in Section~\ref{ea:direct}, we turn to the common one-parameter NEF of Section~\ref{sec:ld-oracle} and analyze two tracking rules that achieve limiting proportions $v_\lambda(P)$: an online mixture rule that tracks the combined target continuously without knowing $T$ in advance, and a two-stage rule that pairs the same EA pilot with post-pilot tracking when $T$ is known.

\subsection{Gaussian Alternatives with Known or Unknown Variances}\label{pv:gauss-section}

We begin with independent Gaussian alternatives whose variances $\sigma_i^2>0$ may be unknown or unequal, extending the normal settings of Sections~\ref{sec:gauss} and~\ref{kv:section} beyond the common one-parameter NEF. Under equal allocation $w_{\rm EA}=(1/k,\ldots,1/k)$, the pairwise exponent for mistaking rival $j\ne b$ for the best alternative $b$ is $G_j=(\mu_b-\mu_j)^2/[2k(\sigma_b^2+\sigma_j^2)]$, the overall PICS exponent is $\Gamma(w_{\rm EA},P)=\min_{j\ne b}G_j$, and $\mathcal A=\{j\ne b:G_j=\Gamma(w_{\rm EA},P)\}$ is the set of active rivals achieving this minimum.

When the total budget $T$ is known in advance, a fraction $1-\lambda$ of the budget can be devoted to uniform exploration upfront via an EA pilot that collects $m=\lfloor(1-\lambda)T/k\rfloor$ observations from each alternative in cyclic order (falling back to equal allocation if $m=0$), leaving $L=T-km$ draws for adaptive tracking. Retaining all pilot observations in subsequent estimates, before each draw $t=km+1,\ldots,T$, we compute the estimated Gaussian rate-optimal allocation $\widehat w(t-1)$ from the current sample means (using the true variances when known, or the sample variances when unknown) and allocate draw $t$ to an alternative whose cumulative count $N_i(t-1)$ is furthest below its total-budget target $t\widehat w_i(t-1)$:
\begin{equation}
 i_t\in\argmax_{1\le i\le k}\bigl\{t\widehat w_i(t-1)-N_i(t-1)\bigr\}
 \qquad(t=km+1,\ldots,T).
 \label{pv:total-clock-rule}
\end{equation}
At budget $T$, the procedure selects an alternative with the largest sample mean across all observations, breaking ties in sampling and selection uniformly at random.

\begin{theorem}\label{pv:gauss}
Let $P$ consist of independent Gaussian alternatives with means $\mu_i$, variances $\sigma_i^2>0$, and a unique best alternative $b$, let $\Gamma(w_{\rm EA},P)=\min_{j\ne b}G_j$ be the equal-allocation exponent with $G_j=(\mu_b-\mu_j)^2/[2k(\sigma_b^2+\sigma_j^2)]$, and let $\mathcal A=\{j\ne b:G_j=\Gamma(w_{\rm EA},P)\}$. Let $\pi_\lambda$ denote rule~\eqref{pv:total-clock-rule} with parameter $\lambda\in(0,1)$ (using true or sample variances). Then there exists a finite constant $C_P>0$ such that, for all sufficiently small $\lambda>0$, its lower and upper PICS exponents (Definition~\ref{def:exponents}) satisfy
\begin{equation}
\begin{aligned}
 \left[1+\lambda\min_{j\in\mathcal A}
     \left(\frac{k\max(\sigma_b^2,\sigma_j^2)}{\sigma_b^2+\sigma_j^2}-1\right)\right]\Gamma(w_{\rm EA},P)-C_P\lambda^2
 &\le \underline e_{\pi_\lambda}(P)
  \le \overline e_{\pi_\lambda}(P) \\
 &\le \left[1+\lambda\min_{j\in\mathcal A}
     \left(\frac{k\max(\sigma_b^2,\sigma_j^2)}{\sigma_b^2+\sigma_j^2}-1\right)\right]\Gamma(w_{\rm EA},P).
\end{aligned}
\label{pv:gslope}
\end{equation}
Moreover, the upper bound in \eqref{pv:gslope} holds for every $\lambda\in(0,1)$ and for any procedure using the same EA pilot that is consistent throughout the known-variance (or unknown-variance) Gaussian distributional setting. Hence $\pi_\lambda$ strictly improves on equal allocation for sufficiently small $\lambda>0$ whenever $k\ge3$, and also for $k=2$ when the variances differ (for $k=2$ with equal variances, equal allocation attains the bound).
\end{theorem}

Theorem~\ref{pv:gauss} closes the loop with Sections~\ref{sec:gauss} and~\ref{kv:section}: even in the unknown-variance Gaussian setting of Theorem~\ref{thm:zero}, where pure tracking forces a zero exponent for every $k\ge2$, pairing an EA pilot with rate-optimal tracking restores exponential PICS decay and strictly outperforms equal allocation whenever $k\ge3$ (and for $k=2$ with unequal variances). Moreover, rule~\eqref{pv:total-clock-rule} achieves the largest first-order slope in $\lambda$ attainable by any consistent procedure after the same EA pilot. For example, with means $(5,0,-2)$ and variances $(4,1,1)$, equal allocation gives $\Gamma(w_{\rm EA},P)=5/6$ (achieved between alternatives~$1$ and~$2$), whereas \eqref{pv:total-clock-rule} raises the exponent to $5/6+7\lambda/6+O(\lambda^2)$. Section~\ref{pv:proofs} proves Theorem~\ref{pv:gauss}. Section~\ref{sec:numerical-ea} illustrates this EA-pilot mechanism numerically for OCBA, BOLD, and plug-in mCEI under unknown variances.

\subsection{One-Parameter Natural Exponential Families}\label{ea:direct}

We now turn to the common one-parameter NEF of Section~\ref{sec:ld-oracle}. In this setting, we analyze two tracking rules whose asymptotic sampling proportions equal $v_\lambda(P)=(1-\lambda)w_{\rm EA}+\lambda\sbest(P)$, differing in whether the total budget $T$ is unknown or known in advance.

\emph{Online mixture tracking (unknown $T$).}
Before observation $t>k$ (after sampling each alternative once in a uniformly random order), compute the estimated rate-optimal allocation $\widehat w(t-1)$ from the current sample means in the common family (setting $\widehat w(t-1)=w_{\rm EA}$ if any sample mean falls outside $\Omega$ and dividing weight equally among tied empirical leaders). With $\widehat v(t-1)=(1-\lambda)w_{\rm EA}+\lambda\widehat w(t-1)$, sample an alternative $i_t$ satisfying
\begin{equation}
 i_t\in\argmax_{1\le i\le k}
        \bigl\{t\widehat v_i(t-1)-N_i(t-1)\bigr\},
 \label{dt:rule}
\end{equation}
so the next observation goes to an alternative whose cumulative count is furthest below its current mixed target. At any budget $T\ge k$, select an alternative with the largest sample mean across all observations, breaking ties in sampling and selection uniformly at random. This procedure operates continuously without knowing $T$ in advance.

\emph{Equal-allocation pilot with post-pilot tracking (known $T$).}
When $T$ is known in advance, we collect the same upfront EA pilot of $m=\lfloor(1-\lambda)T/k\rfloor$ observations per alternative as in Section~\ref{pv:gauss-section} (leaving $L=T-km$ draws for adaptive tracking). Retaining all pilot observations in subsequent estimates, before each post-pilot draw $s=1,\ldots,L$, we recompute the estimated rate-optimal allocation $\widehat w(km+s-1)$ from all sample means collected so far (using the same boundary and tie convention as above), let $M_i(s-1)=N_i(km+s-1)-m$ count only the post-pilot observations of alternative~$i$, and sample
\begin{equation}
 i_s\in\argmax_{1\le i\le k}\bigl\{s\widehat w_i(km+s-1)-M_i(s-1)\bigr\},\qquad s=1,\ldots,L,
 \label{pv:block}
\end{equation}
selecting the largest sample mean across all $T$ observations at budget $T$ and breaking ties uniformly at random. Whereas~\eqref{dt:rule} tracks the mixed target $\widehat v$ continuously, rule~\eqref{pv:block} commits uniform exploration upfront and tracks $\widehat w$ using only the post-pilot counts $M_i$.

\begin{theorem}\label{dt:theorem}
For any configuration $P\in\mathfrak D$ in the common NEF of Section~\ref{cls:nef-family}, there exists $\lambda_0(P)>0$ such that for every $0<\lambda<\lambda_0(P)$, both the online rule~\eqref{dt:rule} and the post-pilot rule~\eqref{pv:block} are consistent and track $v_\lambda$ throughout $\mathfrak D$, and both attain the exact PICS exponent
\begin{equation}
 \lim_{T\to\infty}\frac{1}{T}
    \log\frac{1}{\Pr_P\{\widehat b_T\ne b(P)\}}
 = B_{v_\lambda}(P)
 = \left[1+\left(\frac{k}{2}-1\right)\lambda\right]\Gamma(w_{\rm EA},P).
 \label{dt:exponent}
\end{equation}
This holds for every $k\ge2$ and allows ties among the inferior means. Moreover, for rule~\eqref{pv:block}, consistency and tracking of $v_\lambda$ throughout $\mathfrak D$ hold for every $\lambda\in(0,1)$.
\end{theorem}

Section~\ref{dt:proof} proves Theorem~\ref{dt:theorem} for both the online rule~\eqref{dt:rule} and the post-pilot rule~\eqref{pv:block}, verifying in both cases that finite-sample deviations from the targets preserve the exact exponent for small $\lambda$ (Section~\ref{pv:count-comparison} explains why \eqref{pv:total-clock-rule} uses cumulative counts whereas \eqref{pv:block} uses post-pilot counts). For $k=2$, the exponent in \eqref{dt:exponent} equals $\Gamma(w_{\rm EA},P)$. For $k\ge3$, the improvement is strict even at configurations where pure tracking forces a zero exponent. For instance, with three unit-variance normal alternatives of means $(1,0,0)$, pure tracking gives a zero exponent by Theorem~\ref{thm:nef-tangent}, whereas equal allocation gives $\Gamma(w_{\rm EA},P)=1/12$ and both~\eqref{dt:rule} and~\eqref{pv:block} achieve $(1+\lambda/2)/12$ for all sufficiently small $\lambda>0$.

A limitation of these results is that the small-$\lambda$ threshold depends on the unknown configuration $P$ and therefore cannot be tuned cleanly in advance. Although the exact-exponent guarantee applies for small tracking fractions $\lambda<\lambda_0(P)$, Section~\ref{sec:numerical-ea} illustrates empirically that even reserving half of the budget for the EA pilot ($\lambda=0.5$) yields substantial finite-budget improvements over pure EA.

\section{Illustrative Numerical Experiments}
\label{sec:numerical}

In this section, we conduct numerical experiments to illustrate both the limitations and the benefits of tracking rate-optimal allocations. We first compare the PICS exponents of adaptive procedures with those of the static oracle to examine the gap between static allocation optimality and adaptive selection accuracy in Sections~\ref{sec:numerical-normal} and \ref{sec:numerical-exponential}. We consider settings that differ in the observation distributions, knowledge of the variances, the number of alternatives, and ties among inferior means. We then examine whether tracking rate-optimal allocations after uniform exploration improves selection accuracy in Section~\ref{sec:numerical-ea}.

We evaluate BOLD \citep{chenryzhov2023} and a modified OCBA procedure (hereafter referred to as ``OCBA'' in both the text and figures) because the classical OCBA target can differ from the rate-optimal allocation \citep{chen2000,wuzhou2018}. We modify OCBA by replacing its classical allocation target with the rate-optimal allocation computed from the current parameter estimates. BOLD and OCBA are flexible, adapting to different distributions and knowledge of variances. We also include modified complete expected improvement (mCEI) \citep{chenryzhov2019cei}, which is specific to the normal distribution. For normal observations with unknown variances, we use a plug-in version of mCEI that substitutes sample variances for the known sampling variances in the original rule.

Throughout this section, we report the estimated PICS exponent, obtained by dividing the negative logarithm of PICS by the total sampling budget, that is, $-T^{-1}\log\mathrm{PICS}$. All logarithms are natural. The static oracle knows the true configuration $P$ and uses the rate-optimal allocation $\sbest(P)$, with sample sizes rounded to integers summing to $T$. Its limiting exponent is $\Gamma(\sbest(P),P)$.

In the pure-tracking experiments, each adaptive procedure initially samples each alternative twice. In the EA-pilot experiments, a fixed fraction of the total budget is instead reserved for the initial stage. All initial observations are included in the total budget $T$. After initialization, the procedures allocate one observation at a time, with OCBA assigning each observation to the alternative whose sample size is furthest below its current target. Each simulation-based PICS estimate is based on $200{,}000$ independent replications. We use direct Monte Carlo simulation when incorrect selections occur sufficiently often and importance sampling when they are too rare to estimate reliably by direct simulation. Under importance sampling, each simulated error indicator is weighted by the likelihood ratio.

\subsection{Validation of the PICS Exponents under Normal Sampling}
\label{sec:numerical-normal}

We first examine the convergence behavior of OCBA, BOLD, and mCEI under normal observations. These experiments illustrate the distinction between known and unknown variances summarized in Table~\ref{tab:normal-tracking}, as well as the roles of the number of alternatives and ties among inferior means. By comparing the same configurations with and without variance information, we examine how the need to learn the variances affects the relationship between allocation tracking and selection accuracy.

We consider independent normal alternatives whose variances are all equal to $0.5$. The three mean configurations are $(1,0)$, $(1,0,-0.5)$, and $(1,-0.5,-0.5)$, respectively, so alternative~1 is uniquely best in every configuration. The total sampling budget $T$ ranges from $50$ to $2{,}400$. The corresponding static oracle exponents $\Gamma(\sbest(P),P)$ are approximately $0.25000$, $0.21883$, and $0.38604$, respectively.

\subsubsection{Known Variances.}

\begin{figure}[t]
  \centering
  \caption{Estimated PICS exponents under normal observations with known variances. Solid curves with diamond, square, and filled-circle markers show OCBA, BOLD, and mCEI, respectively; black dashed curves with open circles show the static oracle, and black dotted lines mark its limiting exponent. With two alternatives, the three procedures' curves coincide with the static oracle near its limiting exponent. With three distinct means, they drop below the oracle and remain at positive levels over the reported budgets. With tied inferior means, they decrease toward zero. These results are consistent with the distinction between distinct and tied inferior means in the first row of Table~\ref{tab:normal-tracking}.}
  \label{fig:numerical-normal-known}
  \includegraphics[width=\textwidth]{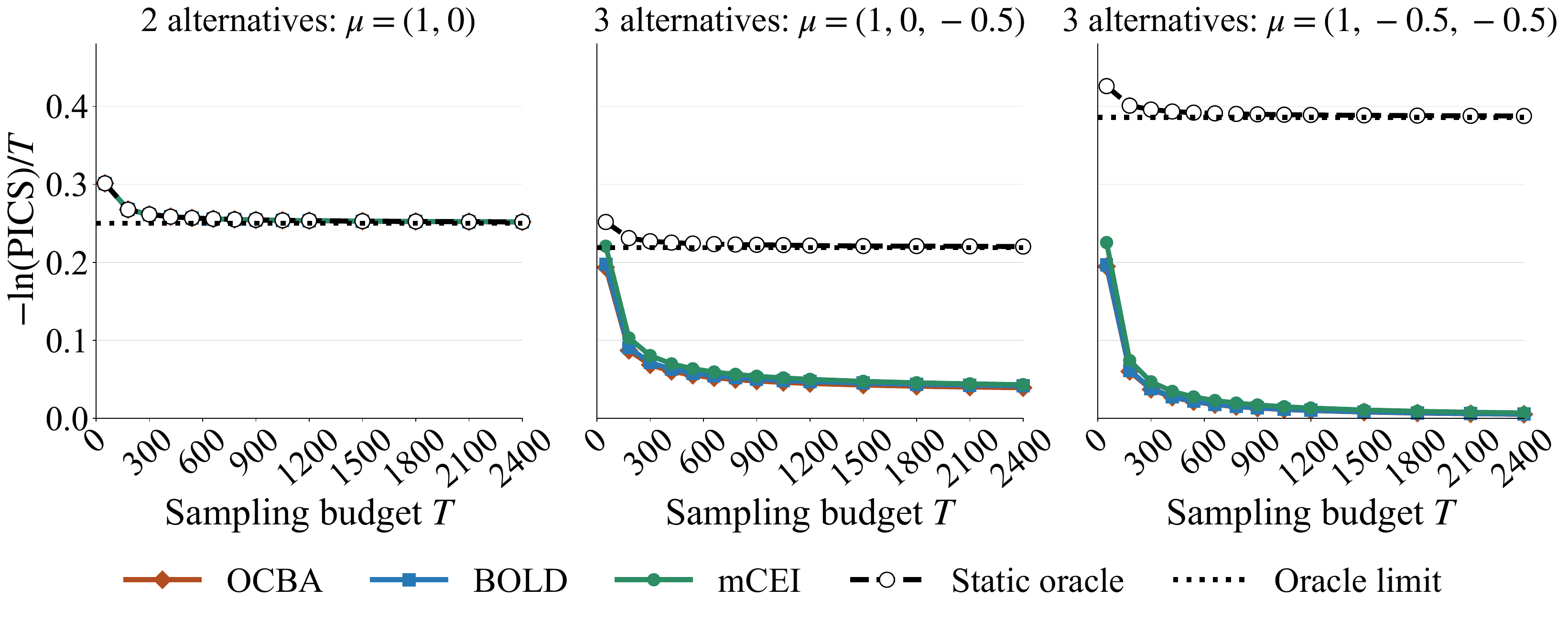}
\end{figure}

Figure~\ref{fig:numerical-normal-known} reports the results with known variances. In the two-alternative configuration, the three procedures allocate equally at the reported budgets, and their PICS is therefore evaluated using the normal tail probability. Since EA is rate-optimal at this configuration, the three procedures' curves coincide with the oracle curve up to simulation error in the oracle estimates, approaching the limiting exponent of $0.25$. With three distinct means, the PICS exponent estimates initially decrease and then change more slowly, remaining at positive levels below the oracle over the reported range. When the inferior means tie, however, the adaptive estimates continue to decrease and reach values close to zero, while the oracle remains near its positive limit. These observations illustrate the distinction established in the first row of Table~\ref{tab:normal-tracking}: under normal sampling with known variances and distinct means, a consistent procedure that tracks the rate-optimal allocation achieves exponential PICS decay, whereas tied inferior means force a zero exponent for every such procedure.

\subsubsection{Unknown Variances.}

\begin{figure}[t]
  \centering
  \caption{Estimated PICS exponents under normal observations with unknown variances. Colored solid curves show OCBA, BOLD, and mCEI, respectively. Black dashed curves show the static oracle, and black dotted lines mark its limiting exponent. The PICS exponents of the three adaptive procedures decrease to values close to zero in all three configurations, corresponding to the second row of Table~\ref{tab:normal-tracking}.}
  \label{fig:numerical-normal-unknown}
  \includegraphics[width=\textwidth]{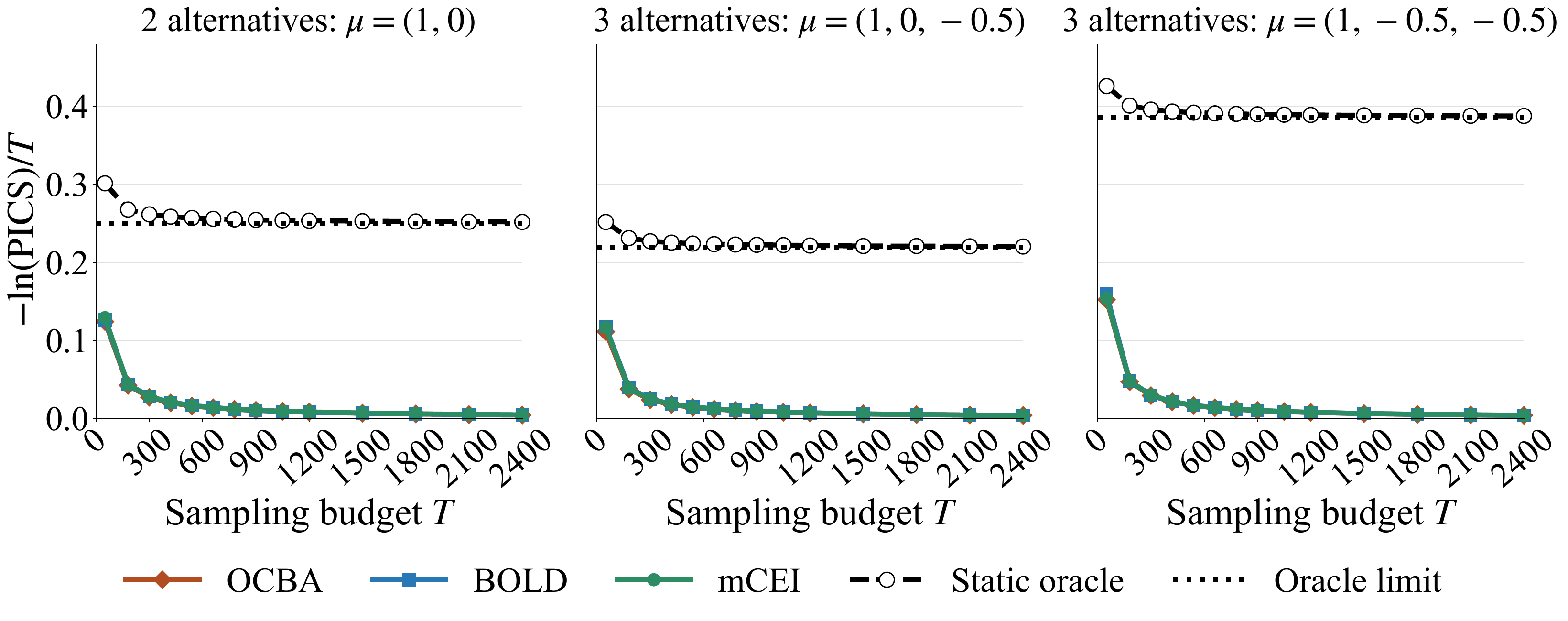}
\end{figure}

Figure~\ref{fig:numerical-normal-unknown} shows the results when the variances are unknown. The estimated exponents of all three procedures decrease substantially as $T$ increases and become close to zero in every configuration. This decline occurs even with two alternatives and with three distinct means, where the known-variance curves remain near positive levels over the reported range. The oracle curves are unchanged because the true configurations are unchanged. These observations provide an illustration of the second row of Table~\ref{tab:normal-tracking}, which establishes that a procedure that is consistent and tracks the problem-specific rate-optimal allocation throughout the unknown-variance normal distributional setting has a zero PICS exponent.

\subsection{Validation of the PICS Exponents under Exponential Sampling}
\label{sec:numerical-exponential}

We next examine the PICS exponents under exponential sampling to illustrate the corresponding theoretical results summarized in Table~\ref{tab:nef-tracking}. We evaluate distribution-specific versions of BOLD and OCBA. BOLD uses the optimality conditions for exponential observations, while OCBA tracks the rate-optimal allocation computed from the current sample means under the exponential distributional family.

We consider independent exponential alternatives with three mean configurations: $(1,0.5)$, $(1,0.5,0.25)$, and $(1,0.25,0.25)$. The corresponding static oracle exponents $\Gamma(\sbest(P),P)$ are approximately $0.05966$, $0.05708$, and $0.17387$, respectively.

\begin{figure}[t]
  \centering
  \caption{Estimated PICS exponents under exponential observations. Solid curves with diamond and square markers show OCBA and BOLD, respectively; black dashed curves with open circles show the static oracle, and black dotted lines mark its limiting exponent. With two alternatives, the OCBA and BOLD curves closely follow and nearly coincide with the static oracle near its positive limit across the reported budgets. With three alternatives, both procedures' curves decrease rapidly toward zero, both with distinct means and with tied inferior means. These patterns align with the exponential row of Table~\ref{tab:nef-tracking}.}
  \label{fig:numerical-exponential}
  \includegraphics[width=\textwidth]{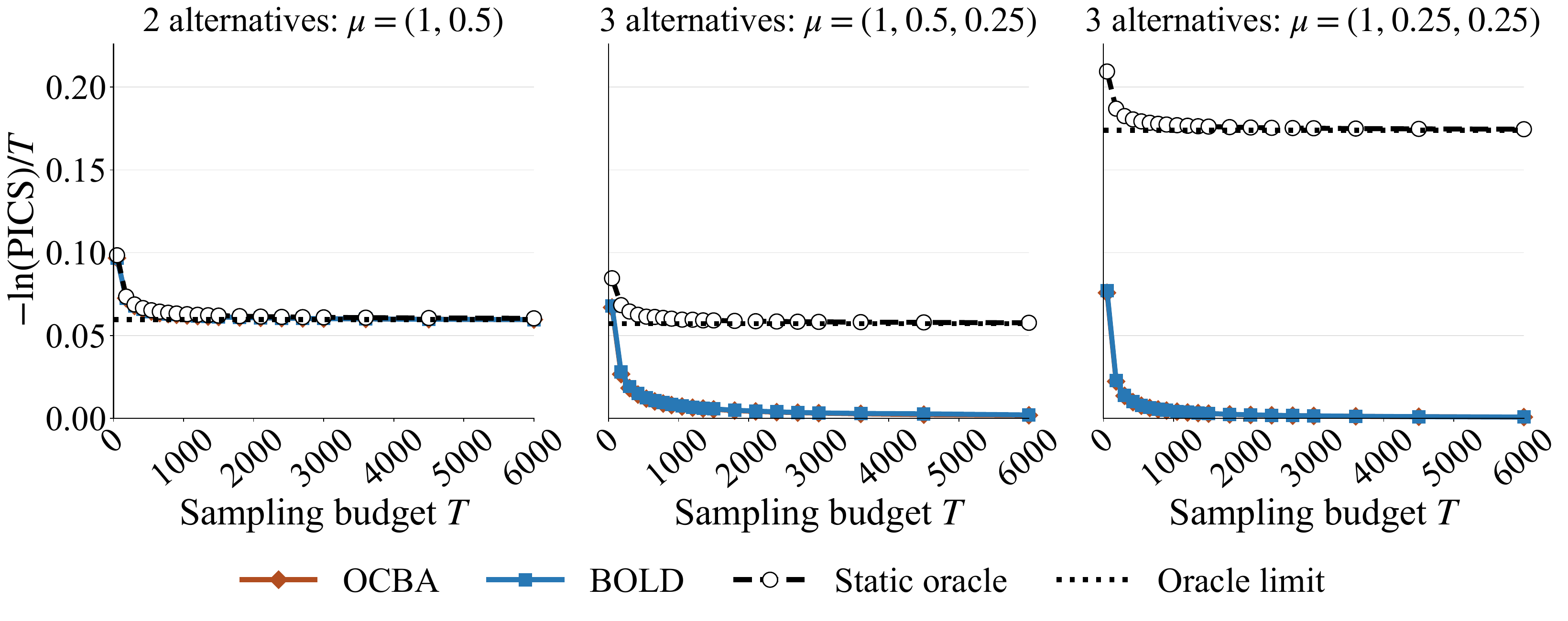}
\end{figure}

Figure~\ref{fig:numerical-exponential} shows that, with two exponential alternatives, both procedures maintain estimated exponents around $0.06$, close to the oracle. With three alternatives, their estimates decrease to values close to zero, regardless of whether the inferior means are distinct or tied, while the oracle remains near its positive limit. Unlike the known-variance normal case, distinct means therefore do not prevent this decline. These results are consistent with the exponential results of Table~\ref{tab:nef-tracking} and illustrate the limitation of rate-optimal sample allocation under exponential sampling when $k \geq 3$.

\subsection{Illustration of the Benefits of Tracking Rate-Optimal Sample Allocations}
\label{sec:numerical-ea}

We now examine the value of rate-optimal allocation information. Motivated by Section~\ref{dt:section}, we combine an EA pilot with OCBA, BOLD, or mCEI and compare the resulting procedures with pure EA.

We consider three independent normal alternatives with means $(1,0,-0.5)$ and variances all equal to $0.5$. Both the means and variances are unknown to the adaptive procedures. The total sampling budget $T$ ranges from $50$ to $2{,}400$. For each budget, approximately $50\%$ is allocated equally among the alternatives before adaptive sampling begins. Specifically, the pilot collects $\lfloor T/6\rfloor$ observations from each alternative. The remaining budget is allocated sequentially using OCBA, BOLD, or mCEI, giving the three hybrid procedures EA-OCBA, EA-BOLD, and EA-mCEI.

We compare these procedures with pure EA and the static oracle. At this configuration, equal allocation $w_{\rm EA}=(1/3,1/3,1/3)$ has exponent $\Gamma(w_{\rm EA},P)=1/6\approx0.16667$. The oracle allocation $\sbest(P)$ assigns proportions of approximately $0.44666$, $0.42901$, and $0.12433$ to alternatives~1, 2, and~3, respectively, with exponent $\Gamma(\sbest(P),P)\approx0.21883$.

\begin{figure}[htbp]
  \centering
  \caption{Estimated PICS exponents for EA-OCBA, EA-BOLD, and EA-mCEI under normal observations with unknown variances. Each subfigure compares one EA-initialized procedure with pure EA and the static oracle. Colored solid curves show EA-OCBA, EA-BOLD, and EA-mCEI, respectively. Gray dash-dotted curves show pure EA, black dashed curves show the oracle, and black dotted lines mark its limiting exponent. The hybrid procedures attain larger finite-budget exponents than pure EA over the reported range, illustrating the additional benefit of tracking rate-optimal allocations after uniform exploration.}
  \label{fig:numerical-ea}
  \includegraphics[width=\textwidth]{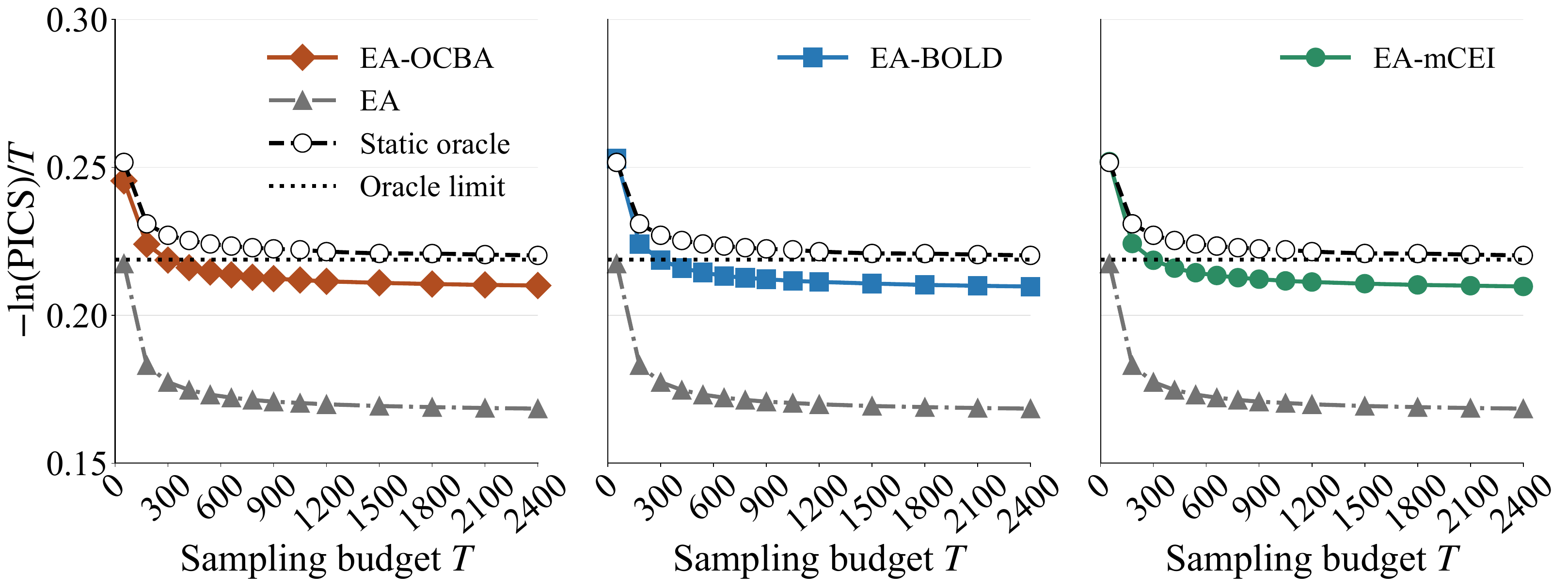}
\end{figure}

Figure~\ref{fig:numerical-ea} shows that all three EA-initialized procedures attain larger finite-budget exponents than pure EA over the reported budgets. At $T=2{,}400$, their estimated exponents are approximately $0.21$, compared with approximately $0.17$ for EA and $0.22$ for the oracle. The hybrid curves change only modestly at larger budgets and remain above the EA curve. This behavior contrasts with the corresponding pure-tracking results in the middle subfigure of Figure~\ref{fig:numerical-normal-unknown}, where the estimated exponents decrease to values close to zero. These findings illustrate the benefit emphasized in Section~\ref{dt:section}: although pure tracking alone has a zero exponent at this configuration, using rate-optimal allocation information after uniform exploration can significantly improve selection accuracy over both pure EA and pure-tracking procedures.

\section{Conclusion and Discussion}\label{sec:conclusion}

This paper examines what tracking the rate-optimal static allocation implies for the selection accuracy of an adaptive fixed-budget R\&S procedure, uncovering both its fundamental limits and its constructive value. On the negative side, for independent normal alternatives with unknown means and variances, every consistent procedure that tracks the problem-specific rate-optimal allocation has zero PICS exponent at any configuration with a unique best alternative. More generally, we give a necessary and sufficient weighted-KL condition for tracking any predefined allocation ratios to force a zero exponent, and translate this condition into a complete classification for common one-parameter natural exponential families, showing that exponential, Poisson, and Bernoulli alternatives frequently suffer subexponential collapse under pure tracking. On the positive side, these limits do not render rate-optimal allocations uninformative: once systematic uniform exploration prevents this collapse, tracking preserves exponential convergence of PICS and strictly improves on equal allocation alone.

These findings call for a nuanced interpretation of allocations derived from large-deviations analysis. The rate-optimal allocation solves an important static planning problem and remains a valuable guide for adaptive sampling, and procedures motivated by such allocations have demonstrated strong finite-budget performance in numerical studies \citep{chen2000,shin2018,chenryzhov2023}. Our results do not contradict that evidence. Rather, they show that asymptotic convergence of allocation ratios alone does not capture finite-budget behavior, and that understanding the finite-budget dynamics of adaptive sampling rules remains an important direction for future work.

Another direction is to determine how broadly these conclusions extend beyond the families studied here. A result encompassing every setting covered by the static large-deviations analysis may be too broad, because support geometry and the finiteness of KL divergence can differ substantially across families. A more tractable next step is to study particular families of interest beyond the natural exponential family. The quantity $B_{\sbest}(P)$ in Corollary~\ref{cor:ld-tracking} provides a direct way to organize this investigation.

\bibliographystyle{informs2014}
\bibliography{rate-optimal-algorithms}

\ECSwitch
\ECHead{Electronic Companion}
\renewcommand{\theHsection}{ec.\arabic{section}}
\renewcommand{\theHsubsection}{ec.\arabic{section}.\arabic{subsection}}
\renewcommand{\theHlemma}{ec.\arabic{lemma}}
\renewcommand{\theHtheorem}{ec.\arabic{theorem}}
\renewcommand{\theHcorollary}{ec.\arabic{corollary}}
\renewcommand{\theHproposition}{ec.\arabic{proposition}}
\section{Proof of Lemma~\ref{lem:comparison}}\label{app:comparison-proof}

\proof{Proof of Lemma~\ref{lem:comparison}.}
See also \citet[Lemma~7 and Appendix~D.1]{barrier2023} for the exponent form of the change-of-measure inequality. Let $Q^T$ and $P^T$ denote the joint distributions, under $Q$ and $P$, respectively, of the procedure's random seed, sampling decisions, observations, and final selection. The random seed has the same distribution in both experiments. Given that seed and the history through time $t-1$, the procedure chooses alternative $\iota_t$ by the same rule under either configuration. Conditional on $\iota_t=i$ and the history, the next observation has distribution $Q_i$ under $Q$ and $P_i$ under $P$. The chain rule for KL divergence adds $\KL(Q_i\Vert P_i)$ whenever alternative $i$ is sampled. The final selection adds no divergence because it is determined by the same rule given the seed and the observed history. Taking expectations under $Q$ gives
\begin{equation}
   \KL(Q^T\Vert P^T)
     =\sum_{t=1}^T \E_Q[\KL(Q_{\iota_t}\Vert P_{\iota_t})]
     =\sum_{i=1}^k \E_Q[N_i(T)]\,\KL(Q_i\Vert P_i).
   \label{eq:transcript-kl}
\end{equation}

Consider the event $E_T=\{\widehat b_T=b(Q)\}$ and write $q_T=\Pr_Q\{E_T\}$ and $r_T=\Pr_P\{E_T\}$. Consistency at $Q$ gives $\lim_{T\to\infty}q_T=1$. Since $b(Q)\ne b(P)$, the selection on $E_T$ is incorrect under $P$, so $r_T\le \Pr_P\{\widehat b_T\ne b(P)\}$. Under $Q$ and $P$, the indicator of $E_T$ has Bernoulli distributions with success probabilities $q_T$ and $r_T$, respectively. Its KL divergence is $\kl(q_T,r_T)=q_T\log(q_T/r_T)+(1-q_T)\log((1-q_T)/(1-r_T))$ (with $0\log 0=0$ and $0\log(0/0)=0$). The data-processing inequality gives $\KL(Q^T\Vert P^T)\ge\kl(q_T,r_T)$ because retaining only whether $E_T$ occurs cannot increase KL divergence. For all sufficiently large $T$ we have $q_T>0$, and finiteness in \eqref{eq:transcript-kl} then implies $r_T>0$ (otherwise $\kl(q_T,0)=+\infty$). If $q_T=r_T=1$, then $\Pr_P\{\widehat b_T\ne b(P)\}=1$ and the left-hand side of \eqref{eq:comparison} is zero; otherwise finiteness of $\kl(q_T,r_T)$ also implies $r_T<1$, and expanding the binary divergence gives
\[
 \kl(q_T,r_T)
   =q_T\log\frac{1}{r_T}+(1-q_T)\log\frac{1}{1-r_T}-h_{\rm bin}(q_T)
   \ge q_T\log\frac{1}{\Pr_P\{\widehat b_T\ne b(P)\}}-\log 2 ,
\]
where $h_{\rm bin}(q)=-q\log q-(1-q)\log(1-q)\le\log 2$ is the entropy of the binary outcome, with its endpoint values defined by continuity. Combining this inequality with \eqref{eq:transcript-kl} gives
\[
 \frac{1}{T}\log\frac{1}{\Pr_P\{\widehat b_T\ne b(P)\}}
 \le
 \frac{\sum_{i=1}^k \E_Q[N_i(T)]\KL(Q_i\Vert P_i)+\log 2}
      {Tq_T}.
\]
Because $\lim_{T\to\infty}q_T=1$ and $\lim_{T\to\infty}\E_Q[N_i(T)]/T=w_i(Q)$ for every $i$, taking the upper limit in this inequality yields \eqref{eq:comparison}.
\Halmos\endproof

\section{Proof of Theorem~\ref{thm:zero-characterization}}
\label{app:zero-characterization-proof}

\subsection{The Zero-Exponent Direction}
\label{sec:information-zero-direction}

\proof{Proof that $B_w(P)=0$ forces zero exponent.}
Fix a configuration $P$ satisfying $B_w(P)=0$ and a procedure $\pi\in\mathfrak T(w)$. Because $\inf\varnothing=+\infty$ in \eqref{eq:bound}, $B_w(P)=0$ implies $\mathcal A_{\rm fin}(P)\ne\varnothing$. Fix a different-best configuration $Q\in\mathcal A_{\rm fin}(P)$. For every alternative $i$, tracking gives
\[
  0\le \frac{N_i(T)}{T}\le1,
  \qquad
  \lim_{T\to\infty}\frac{N_i(T)}{T}=w_i(Q)
  \quad\text{in probability}.
\]
Boundedness therefore implies $\lim_{T\to\infty}\E_Q[N_i(T)]/T=w_i(Q)$. By Lemma~\ref{lem:comparison},
\[
  \overline e_\pi(P)
  \le \sum_{i=1}^k w_i(Q)\KL(Q_i\Vert P_i).
\]
Taking the infimum over $Q\in\mathcal A_{\rm fin}(P)$ gives
\[
  0\le \underline e_\pi(P)\le \overline e_\pi(P)\le B_w(P)=0.
\]
Therefore $\lim_{T\to\infty}\frac{1}{T}\log\frac{1}{\Pr_P\{\widehat b_T\ne b(P)\}}=0$ for every $\pi\in\mathfrak T(w)$. \Halmos\endproof

\subsection{Constructing a Positive Lower PICS Exponent}
\label{sec:information-positive-direction}

\proof{Proof.}
Fix a configuration $P$ satisfying $B_w(P)>0$, and choose a finite constant $g$ with $0<g<B_w(P)$. Start with any procedure in $\mathfrak T(w)$, called the base procedure, and write $\widetilde b_T$ for its provisional selection. We first suppose that every different-best configuration belongs to $\mathcal A_{\rm fin}(P)$. To obtain a consistent modification in $\mathfrak T(w)$ with lower PICS exponent at least $g$ at $P$, it is enough to construct a test statistic $L_T$ satisfying
\begin{equation}
  \Pr_P\{L_T\ge e^{gT}\}\le e^{-gT},
  \qquad
  \lim_{T\to\infty}\Pr_Q\{L_T\ge e^{gT}\}=1
  \quad\text{for every different-best configuration }Q.
  \label{eq:test-requirements}
\end{equation}
The first property in \eqref{eq:test-requirements} controls PICS at $P$, while the second preserves consistency under any fixed different-best configuration. Section~\ref{sec:infinite} extends the construction when some different-best configuration has an infinite coordinatewise KL divergence. The connection between distinguishability and KL divergence also arises in classical i.i.d.\ hypothesis testing \citep[Theorem~2.1]{bjelakovic2005}.

\emph{A Mixture Statistic.} Write $\iota_t$ for the alternative sampled at stage $t$ and $X_t$ for the observation collected at that stage. For each alternative $i$, choose a bounded measurable function $f_i$ and define
\begin{equation}
  h_i(x)=f_i(x)-\log\E_{P_i}[e^{f_i(X)}],
  \qquad
  L_n^f=\exp\!\left\{\sum_{t=1}^n h_{\iota_t}(X_t)\right\},
  \qquad L_0^f=1.
  \label{eq:component-martingale}
\end{equation}
Let $\mathcal F_n$ be the history through stage $n$, including the procedure's private randomization. The choice $\iota_n$ is $\mathcal F_{n-1}$-measurable, and $X_n$ has conditional distribution $P_{\iota_n}$ under $P$. Because $\E_{P_i}[e^{h_i(X)}]=1$ for every $i$,
\[
  \E_P\!\left[e^{h_{\iota_n}(X_n)}\mid\mathcal F_{n-1}\right]=1.
\]
Thus $L_n^f$ is a nonnegative mean-one martingale even under adaptive sampling.

A particular choice of $f=(f_1,\ldots,f_k)$ may detect one different-best configuration better than another. We combine a countable family of choices so that the procedure can use the same statistic for every $Q$. Let $\mathcal C$ be the countable algebra generated by intervals with rational endpoints, which generates the usual $\sigma$-algebra on $\R$. Use all rational-valued simple functions whose level sets belong to $\mathcal C$. The resulting vectors $f$ form a countable family. Assign weights and define
\begin{equation}
  \rho_f>0,
  \qquad \sum_f\rho_f=1,
  \qquad L_n=\sum_f\rho_f L_n^f.
  \label{eq:mixture-statistic}
\end{equation}
Conditional monotone convergence permits the countable sum to pass through conditional expectation, so
\[
  \E_P[L_n\mid\mathcal F_{n-1}]
  =\sum_f\rho_f\E_P[L_n^f\mid\mathcal F_{n-1}]
  =\sum_f\rho_f L_{n-1}^f=L_{n-1}.
\]
Also $\E_P[L_n]=\sum_f\rho_f=1$. The countable nonnegative sum in \eqref{eq:mixture-statistic} is measurable, and its finite expectation implies $L_n<\infty$ almost surely under $P$. Under another configuration it may equal $+\infty$, in which case it simply crosses every finite threshold. Thus $L_n$ is a nonnegative mean-one martingale under $P$. The martingale maximal inequality gives
\begin{equation}
  \Pr_P\!\left\{\max_{n\le T}L_n\ge e^{gT}\right\}\le e^{-gT}.
  \label{eq:mixture-maximal}
\end{equation}

\emph{Growth of the Mixture Statistic Under a Different-Best Configuration.} Fix $Q$ and consider any sequence of experiments whose allocation ratios satisfy $\lim_{n\to\infty}N_i(n)/n=v_i$ in probability for every $i$. This general form will be used with $v=w(Q)$ for the base procedure and with cyclic equal allocation $v=w_{\rm EA}$ in Section~\ref{sec:infinite}. Let $X_{i,r}$ denote the $r$th observation in the i.i.d.\ stream from alternative $i$, and define
\[
  \overline h_{i,q}
  =\frac{1}{q}\sum_{r=1}^{q}h_i(X_{i,r})\quad(q\ge1),
  \qquad \overline h_{i,0}=0.
\]
For a fixed $f$, Equation~\eqref{eq:component-martingale} gives
\[
  \frac{1}{n}\log L_n^f
  =\sum_{i=1}^k \frac{N_i(n)}{n}
    \overline h_{i,N_i(n)}.
\]
If $v_i>0$, then for every fixed $q_0$, $\lim_{n\to\infty}\Pr_Q\{N_i(n)<q_0\}=0$. For every $\epsilon>0$, the adaptive sample mean satisfies
\[
  \Pr_Q\!\left\{
    \left|\overline h_{i,N_i(n)}-\E_{Q_i}[h_i]\right|>\epsilon\right\}
  \le \Pr_Q\{N_i(n)<q_0\}
  +\Pr_Q\!\left\{
    \sup_{q\ge q_0}\left|\overline h_{i,q}
      -\E_{Q_i}[h_i]\right|>\epsilon\right\}.
\]
For fixed $q_0$, the first probability converges to zero as $n$ tends to infinity. The strong law of large numbers along the entire i.i.d.\ stream from alternative $i$ makes the second probability converge to zero as $q_0$ tends to infinity. Taking these limits in that order proves convergence at the adaptive count $N_i(n)$. If $v_i=0$, then because $h_i$ is bounded, its absolute contribution is at most $\|h_i\|_\infty N_i(n)/n$, which converges to zero in probability. Therefore
\begin{equation}
  \lim_{n\to\infty}\frac{1}{n}\log L_n^f
    =\sum_{i=1}^k v_i\E_{Q_i}[h_i]
  \qquad\text{in probability.}
  \label{eq:fixed-f-growth}
\end{equation}

We now determine how large the limiting mean log-likelihood ratio can be. By the Donsker--Varadhan variational formula for KL divergence \citep{gray2011},
\begin{equation}
   \sup_{f_i}\{\E_{Q_i}[f_i]-\log\E_{P_i}[e^{f_i}]\}
       =\KL(Q_i\Vert P_i),
   \label{eq:dv-variational}
\end{equation}
where the supremum ranges over all bounded measurable functions and may equal $+\infty$. Indeed, if $Q_i\ll P_i$, truncating $\log(dQ_i/dP_i)$ to $[-M,M]$ yields scores converging to $\KL(Q_i\Vert P_i)$ by dominated convergence in $L^1(P_i)$ and $L^1(Q_i)$; if $Q_i\not\ll P_i$, choosing a measurable set $E$ with $P_i(E)=0<Q_i(E)$ and taking $f_i=M\mathbf1_E$ gives score $M Q_i(E)\to\infty$. Notice that the objective in \eqref{eq:dv-variational} is exactly $\E_{Q_i}[h_i]$.

The countable family used to construct $L_n$ attains the same supremum. By the monotone class theorem, measurable sets can be approximated by sets in $\mathcal C$ under the finite measure $P_i+Q_i$. Approximating any bounded measurable function $f_i$ first by simple functions and then their values by rationals yields uniformly bounded approximants $f_{i,m}$ in $L^1(P_i+Q_i)$ from our countable coordinate family. Since the exponential function is Lipschitz on every bounded interval,
\[
  \lim_{m\to\infty}\E_{Q_i}[f_{i,m}]=\E_{Q_i}[f_i],
  \qquad
  \lim_{m\to\infty}\E_{P_i}[e^{f_{i,m}}]=\E_{P_i}[e^{f_i}].
\]
Because $\E_{P_i}[e^{f_i}]>0$, continuity of the logarithm also gives $\lim_{m\to\infty}\log\E_{P_i}[e^{f_{i,m}}] =\log\E_{P_i}[e^{f_i}]$. Hence the countable family has the same variational supremum \eqref{eq:dv-variational} as all bounded measurable functions.

Take any finite $c<\sum_{i:v_i>0}v_i\KL(Q_i\Vert P_i)$. Because our countable family consists of all product tuples $f=(f_1,\ldots,f_k)$ whose coordinates are rational-valued simple functions on $\mathcal C$, we can choose each coordinate $f_i$ separately to obtain a tuple $f$ in the family satisfying $\sum_{i=1}^k v_i\E_{Q_i}[h_i]>c$ (if some alternative with $v_i>0$ has $\KL(Q_i\Vert P_i)=+\infty$, we can make $v_i\E_{Q_i}[h_i]>c$ on that coordinate alone and take $f_j\equiv 0$, hence $h_j\equiv 0$, on every other coordinate). Although this choice of $f$ may depend on $Q$ in the analysis, it is included in the sum \eqref{eq:mixture-statistic} defining $L_n$, so
\[
  \frac{1}{n}\log L_n
  \ge \frac{1}{n}\log L_n^f+\frac{1}{n}\log\rho_f,
  \qquad
  \lim_{n\to\infty}\frac{1}{n}\log\rho_f=0.
\]
Combining this lower bound with \eqref{eq:fixed-f-growth} gives
\begin{equation}
   \lim_{n\to\infty}\Pr_Q\!\left\{\frac{1}{n}\log L_n>c\right\}=1 .
   \label{eq:test}
\end{equation}

\emph{The Final Selection.} Suppose every different-best configuration belongs to $\mathcal A_{\rm fin}(P)$. Run the base procedure for all $T$ observations and define
\begin{equation}
  \widehat b_T=
  \begin{cases}
    \widetilde b_T, & L_T\ge e^{gT},\\
    b(P),           & L_T<e^{gT}.
  \end{cases}
  \label{eq:finite-kl-selection-rule}
\end{equation}
This modification leaves the allocation ratios unchanged. Under \eqref{eq:finite-kl-selection-rule},
\[
  \{\widehat b_T\ne b(P)\}\subseteq\{L_T\ge e^{gT}\},
\]
and therefore, at $P$,
\[
  \Pr_P\{\widehat b_T\ne b(P)\}
  \le \Pr_P\{L_T\ge e^{gT}\}
  \le e^{-gT}
\]
by \eqref{eq:mixture-maximal}. If $b(Q)=b(P)$, then $\{\widehat b_T\ne b(Q)\}\subseteq\{\widetilde b_T\ne b(Q)\}$. If $b(Q)\ne b(P)$, then
\[
  \sum_{i=1}^k w_i(Q)\KL(Q_i\Vert P_i)\ge B_w(P)>g,
  \qquad
  \lim_{T\to\infty}\Pr_Q\{L_T<e^{gT}\}=0,
\]
where the second statement follows from \eqref{eq:test} with $v=w(Q)$ and $c=g$. In this case,
\[
  \{\widehat b_T\ne b(Q)\}
  \subseteq
  \{\widetilde b_T\ne b(Q)\}\cup\{L_T<e^{gT}\}.
\]
Both probabilities on the right converge to zero, so the modified procedure remains consistent.

This completes the construction when every different-best configuration belongs to $\mathcal A_{\rm fin}(P)$. The next subsection modifies it to handle different-best configurations with an infinite coordinatewise KL divergence. \endproof

\subsection{Different-Best Configurations with Infinite KL Divergence}
\label{sec:infinite}

\proof{Completion of the proof of Theorem~\ref{thm:zero-characterization}.}

Continue to fix $P$, choose a finite $g$ with $0<g<B_w(P)$, and choose a base procedure in $\mathfrak T(w)$. Equation~\eqref{eq:test} detects a different-best configuration $Q$ through the weighted KL information on coordinates with $w_i(Q)>0$. If every infinite KL divergence occurs on a coordinate with $w_i(Q)=0$, the base procedure need not sample that coordinate often enough. We therefore add a cyclic test that samples every alternative, and then use a final independent run of the base procedure to recover the target allocation ratios.

For this construction, view the observations from each alternative as a pre-generated infinite i.i.d.\ sequence, with the $k$ sequences mutually independent, and give each phase an independent private random seed. A nonanticipative phase reveals only the next unused observation from a selected sequence. Consequently, conditional on the history accumulated at the end of any phase, the unused tails remain independent and have their original laws.

\emph{The sampling and selection rule.} Choose $0<\alpha<1$ such that $g/\alpha<B_w(P)$. If $B_w(P)<\infty$, any $\alpha\in(g/B_w(P),1)$ works; if $B_w(P)=+\infty$, any $\alpha\in(0,1)$ works. For small budgets $T<1/\alpha$, simply run the base procedure directly on all $T$ observations; for the remainder of the proof we take $T\ge 1/\alpha$ so that $n=\lfloor\alpha T\rfloor\ge1$. Run a base procedure in $\mathfrak T(w)$ with budget $n$. Let $\mathcal F_T^{(1)}$ be the history generated by this run, including its private randomization, and let $\widetilde b_n$ and $L_n$ be its provisional selection and the mixture statistic in \eqref{eq:mixture-statistic}. Both are $\mathcal F_T^{(1)}$-measurable. Define the cyclic-trigger event
\[
  \mathcal E_T=\{\widetilde b_n\ne b(P),\ L_n<e^{gT}\}.
\]
If $\widetilde b_n=b(P)$, select $b(P)$. If $\widetilde b_n\ne b(P)$ and $L_n\ge e^{gT}$, select $\widetilde b_n$. Only on $\mathcal E_T$ do we execute the cyclic test.

The cyclic test uses unused observations from the $k$ streams and fresh private randomization, all independent of $\mathcal F_T^{(1)}$. Conditional on $\mathcal F_T^{(1)}$, these observations therefore have the same law as fresh independent streams under the current configuration. On the full probability space, whether or not $\mathcal E_T$ occurs, use these fresh streams to define a copy $L_r^{\mathrm{cyc}}$ of \eqref{eq:mixture-statistic}, starting from $L_0^{\mathrm{cyc}}=1$ and sampling alternatives in the order $1,\ldots,k,1,\ldots$. Define
\[
  \tau_T=\inf\{r\ge1:L_r^{\mathrm{cyc}}\ge e^{gT}\},
  \qquad \inf\varnothing=+\infty.
\]
The procedure executes this cyclic process only on $\mathcal E_T$. On that event, select $\widetilde b_n$ if $\tau_T\le T-n$ and select $b(P)$ otherwise. Let
\[
  M_T=
  \begin{cases}
    0, & \mathcal E_T^c,\\
    \min\{\tau_T,T-n\}, & \mathcal E_T,
  \end{cases}
\]
where $\min\{+\infty,T-n\}=T-n$.

After the selection has been determined, let $\mathcal F_T^{\rm pre}$ denote the complete history accumulated so far and set $H_T=T-n-M_T$. The random budget $H_T$ is $\mathcal F_T^{\rm pre}$-measurable. Spend these remaining observations on a new run of the base procedure, using unused observation tails and private randomization independent of $\mathcal F_T^{\rm pre}$. Conditional on $H_T=q$, this final run therefore has exactly the law of an independent base procedure with deterministic budget $q$. Because $n+M_T+H_T=T$, the three phases use exactly the available budget. Each sampling decision uses only the observations already revealed in its phase and fresh private randomization. The selection determined before the final run is retained and reported after all $T$ observations have been collected; the final run is used only to restore the target allocation ratios. Thus the construction is a valid nonanticipative fixed-budget procedure.

\emph{PICS at $P$.} An incorrect selection under $P$ requires either $L_n\ge e^{gT}$ or a threshold crossing by the cyclic test. More precisely,
\[
  \{\widehat b_T\ne b(P)\}
  \subseteq
  \{L_n\ge e^{gT}\}
  \cup\bigl(\mathcal E_T\cap\{\tau_T\le T-n\}\bigr).
\]
Markov's inequality applied to the mean-one statistic $L_n$ bounds the first probability by $e^{-gT}$. Because $\mathcal E_T\in\mathcal F_T^{(1)}$ and conditional on $\mathcal F_T^{(1)}$ the cyclic statistic is a nonnegative mean-one martingale, Doob's maximal inequality yields $\Pr_P\{\tau_T\le T-n\mid\mathcal F_T^{(1)}\}\le e^{-gT}$, and integrating over $\mathcal E_T$ gives $\Pr_P\{\tau_T\le T-n,\,\mathcal E_T\}\le e^{-gT}$. Hence
\begin{equation}
  \Pr_P\{\widehat b_T\ne b(P)\}
  \le \Pr_P\{L_n\ge e^{gT}\}
     +\Pr_P\{\tau_T\le T-n,\ \mathcal E_T\}
  \le 2e^{-gT}.
  \label{eq:infinite-pics-bound}
\end{equation}
Thus, the constructed procedure has a lower PICS exponent of at least $g$ at $P$.

\emph{Consistency.} First suppose $b(Q)=b(P)$. Because $\widetilde b_n=b(Q)=b(P)$ guarantees both $\widehat b_T=b(P)=b(Q)$ and $\mathcal E_T^c$, we have
\begin{equation}
  \{\widehat b_T\ne b(Q)\}\cup \mathcal E_T
  \subseteq \{\widetilde b_n\ne b(Q)\}.
  \label{eq:same-best-inclusion}
\end{equation}
Because $\lim_{T\to\infty}n=\infty$, consistency of the base procedure makes the probability on the right converge to zero.

Next let $Q\in\mathcal A_{\rm fin}(P)$. Choose $g/\alpha<c<B_w(P)$. Since the first run tracks $w(Q)$ and
\[
  \sum_{i=1}^k w_i(Q)\KL(Q_i\Vert P_i)\ge B_w(P)>c,
\]
Equation~\eqref{eq:test}, applied at budget $n$, gives $\lim_{T\to\infty}\Pr_Q\{\log L_n>nc\}=1$. Moreover, $\lim_{T\to\infty}n/T=\alpha$ and $\alpha c>g$, so $nc>gT$ for all sufficiently large $T$. Therefore
\begin{equation}
  \{\widehat b_T\ne b(Q)\}\cup \mathcal E_T
  \subseteq
  \{\widetilde b_n\ne b(Q)\}\cup\{L_n<e^{gT}\},
  \qquad
  \lim_{T\to\infty}\Pr_Q\{\mathcal E_T\}=0.
  \label{eq:finite-competing-events}
\end{equation}
Both events on the right have probabilities converging to zero, which proves consistency under any fixed finite-KL different-best configuration.

It remains to consider a different-best configuration $Q$ with infinite KL divergence on at least one coordinate. Fix $0<\delta<1-\alpha$ and set $m=\lfloor\delta T\rfloor$. Under cyclic sampling, every limiting allocation ratio is $1/k$. The corresponding weighted KL information is therefore infinite. Choose a finite $c>g/\delta$. Equation~\eqref{eq:test} applied to the fresh cyclic streams gives
\[
  \lim_{T\to\infty}\Pr_Q\{\log L_m^{\mathrm{cyc}}>mc\}=1.
\]
For all sufficiently large $T$, $mc>gT$ and $m\le T-n$. A value $L_m^{\mathrm{cyc}}>e^{gT}$ implies that the cyclic statistic crossed the threshold by time $m$, so
\begin{equation}
  \Pr_Q\{\tau_T>m\}
  \le \Pr_Q\{\log L_m^{\mathrm{cyc}}\le gT\}
  \le \Pr_Q\{\log L_m^{\mathrm{cyc}}\le mc\},
  \qquad
  \lim_{T\to\infty}\Pr_Q\{\tau_T>m\}=0.
  \label{eq:cyclic-crossing}
\end{equation}
Because the cyclic test uses fresh observation tails independent of $\mathcal F_T^{(1)}$,
\[
  \Pr_Q\{\tau_T>m\mid\mathcal F_T^{(1)}\}
  =\Pr_Q\{\tau_T>m\}
  \qquad\text{almost surely}.
\]
Thus the conclusion in \eqref{eq:cyclic-crossing} is unaffected by the first-run history.

When $\widetilde b_n=b(Q)$ (which differs from $b(P)$), the rule selects $b(Q)$ directly if $L_n\ge e^{gT}$, and also selects $b(Q)$ after the cyclic test if $L_n<e^{gT}$ and $\tau_T\le T-n$. Since $m\le T-n$ for all large $T$,
\begin{equation}
  \{\widehat b_T\ne b(Q)\}
  \subseteq
  \{\widetilde b_n\ne b(Q)\}\cup\{\tau_T>T-n\}
  \subseteq
  \{\widetilde b_n\ne b(Q)\}\cup\{\tau_T>m\}.
  \label{eq:infinite-consistency-events}
\end{equation}
Consistency of the base procedure and \eqref{eq:cyclic-crossing} make the two probabilities on the right converge to zero. Thus the modified procedure is consistent under any fixed different-best configuration, including one outside $\mathcal A_{\rm fin}(P)$.

The cyclic test also uses a negligible fraction of the budget. For $m=\lfloor\delta T\rfloor$,
\begin{equation}
  \Pr_Q\{M_T>m\}
  \le \Pr_Q\{\tau_T>m\},
  \qquad
  \lim_{T\to\infty}\Pr_Q\{M_T>m\}=0.
  \label{eq:cyclic-budget}
\end{equation}
Given any $\epsilon>0$, choose $0<\delta<\min\{\epsilon,1-\alpha\}$. For all sufficiently large $T$, $m<\epsilon T$, so \eqref{eq:cyclic-budget} implies $\lim_{T\to\infty}\Pr_Q\{M_T/T>\epsilon\}=0$. Under any configuration $Q$ with $b(Q)=b(P)$ or $Q\in\mathcal A_{\rm fin}(P)$, $M_T=0$ on $\mathcal E_T^c$ and $\lim_{T\to\infty}\Pr_Q\{\mathcal E_T\}=0$ (by \eqref{eq:finite-competing-events} in the latter case and \eqref{eq:same-best-inclusion} in the former). Consequently,
\begin{equation}
  \lim_{T\to\infty}\frac{M_T}{T}=0
  \qquad\text{in probability under every }Q\in\mathfrak D.
  \label{eq:cyclic-budget-negligible}
\end{equation}

\emph{Tracking.} Equation~\eqref{eq:cyclic-budget-negligible} and $\lim_{T\to\infty}n/T=\alpha$ give
\[
  \lim_{T\to\infty}\frac{H_T}{T}=1-\alpha>0
  \qquad\text{in probability}.
\]
Fix an alternative $i$ and $\epsilon>0$. Let $N_i^{\rm base}(q)$ be the number of observations assigned to alternative $i$ by a base-procedure run with deterministic budget $q$, and define
\[
  \eta_{i,q}(\epsilon)
  =\Pr_Q\!\left\{
      \left|\frac{N_i^{\rm base}(q)}{q}-w_i(Q)\right|>\epsilon
    \right\}.
\]
Tracking by the base procedure gives $\lim_{q\to\infty}\eta_{i,q}(\epsilon)=0$. Let $C_{i,T}^{(3)}$ be the number of observations of alternative $i$ in the final run, setting $C_{i,T}^{(3)}/H_T=0$ when $H_T=0$. Because $H_T$ is $\mathcal F_T^{\rm pre}$-measurable and the final run is independent of $\mathcal F_T^{\rm pre}$, conditioning on $\mathcal F_T^{\rm pre}$ yields, for every integer $K\ge1$,
\begin{equation}
  \Pr_Q\!\left\{
    \left|\frac{C_{i,T}^{(3)}}{H_T}-w_i(Q)\right|>\epsilon
  \right\}
  \le \Pr_Q\{H_T<K\}+\sup_{q\ge K}\eta_{i,q}(\epsilon).
  \label{eq:random-budget-tracking}
\end{equation}
For each fixed $K$, the first term tends to zero as $T$ tends to infinity; the second tends to zero as $K$ tends to infinity. Hence, in probability,
\[
  \lim_{T\to\infty}\frac{C_{i,T}^{(3)}}{H_T}=w_i(Q),
  \qquad
  \lim_{T\to\infty}\frac{C_{i,T}^{(3)}}{T}
  =\lim_{T\to\infty}
    \left\{\frac{H_T}{T}\frac{C_{i,T}^{(3)}}{H_T}\right\}
  =(1-\alpha)w_i(Q).
\]

Let $C_{i,T}^{(1)}$ and $C_{i,T}^{(2)}$ be the counts from the first base run and the cyclic test. Tracking in the first run and $\lim_{T\to\infty}n/T=\alpha$ give $\lim_{T\to\infty}C_{i,T}^{(1)}/T=\alpha w_i(Q)$ in probability, while $0\le C_{i,T}^{(2)}\le M_T$ and \eqref{eq:cyclic-budget-negligible} give $\lim_{T\to\infty}C_{i,T}^{(2)}/T=0$ in probability. Therefore
\[
  \lim_{T\to\infty}\frac{N_i(T)}{T}
  =\lim_{T\to\infty}
   \left\{
     \frac{C_{i,T}^{(1)}}{T}
     +\frac{C_{i,T}^{(2)}}{T}
     +\frac{C_{i,T}^{(3)}}{T}
   \right\}
  =w_i(Q)
  \qquad\text{in probability}.
\]
The constructed procedure is therefore consistent and belongs to $\mathfrak T(w)$. By \eqref{eq:infinite-pics-bound}, its lower PICS exponent at $P$ is at least $g$. Since $g$ was an arbitrary finite constant in $(0,B_w(P))$ (which is nonempty whenever $B_w(P)>0$, including $B_w(P)=+\infty$), this constructs a procedure in $\mathfrak T(w)$ with $\underline e_\pi(P)>0$ whenever $B_w(P)>0$, establishing both the final assertion of Theorem~\ref{thm:zero-characterization} and the reverse implication in \eqref{eq:zero-characterization}. \Halmos\endproof

\section{Rate-Optimal Allocations in Natural Exponential Families}
\label{app:oracles}

We first verify Lemma~\ref{lem:nef-mean-rate}, the static large-deviations formula, and the existence of a consistent procedure that tracks the problem-specific rate-optimal allocation in the natural exponential family setting of Section~\ref{sec:ld-oracle}. For $F_\theta$ in that distributional family, the log moment generating function at $s$ is $A(\theta+s)-A(\theta)$ whenever $\theta+s\in\Theta$, which is finite on a neighborhood of zero because $\Theta$ is open. Setting $u=A'(\theta)$ (so $\theta=\theta_u$), \eqref{eq:cramer-rate} yields
\[
  I_{\mu=u}(v)=\sup_{s:\,\theta+s\in\Theta}\bigl\{sv-[A(\theta+s)-A(\theta)]\bigr\}
  \qquad(v\in\Omega).
\]
Because $A$ is strictly convex on $\Theta$ and $A'(\theta_v)=v$, the objective is strictly concave in $s$ with derivative $v-A'(\theta+s)$ vanishing uniquely at $s=\theta_v-\theta_u$. Evaluating at that maximizer gives the first equality in \eqref{eq:nef-mean-rate}. Taking the log-ratio of the densities $f_{\theta_v}$ and $f_\theta$ and using $\E_{F_v}[X]=v$ gives $\KL(F_v\Vert F_u)=(\theta_v-\theta)v-A(\theta_v)+A(\theta)$, proving Lemma~\ref{lem:nef-mean-rate} \citep[see also][Lemma~11]{degenne2023}. The rate function $I_{\mu=u}(x)$ is finite and jointly continuous in the interior mean and parameter intervals; its derivative in $x$ is $\theta_x-\theta_u$, so it is strictly convex and has its unique minimum at the mean $u$. These properties give the hypotheses of \eqref{eq:cramer-oracle}.

\proof{Proof of Lemma~\ref{lem:nef-rate-properties}.}
For Part~1, differentiating \eqref{eq:nef-mean-rate} (written as $I_{\mu=u}(x)=(\theta_x-\theta_u)x-A(\theta_x)+A(\theta_u)$) with respect to $x$ and using $A'(\theta_x)=x$ and $d\theta_x/dx=1/A''(\theta_x)>0$ cancels the $\theta_x'$ terms and yields $I_{\mu=u}'(x)=\theta_x-\theta_u$ and $I_{\mu=u}''(x)=1/A''(\theta_x)>0$ on $\Omega$. Hence $I_{\mu=u}(x)$ is continuous, differentiable, and strictly convex in $x\in \Omega$, with $I_{\mu=u}'(u)=0$, $I_{\mu=u}'(x)<0$ on $(\mu_-,u)$, and $I_{\mu=u}'(x)>0$ on $(u,\mu_+)$. By \eqref{eq:nef-mean-rate}, $I_{\mu=u}(x)=\KL(F_x\Vert F_u)\ge0$, with equality if and only if $x=u$.

For Part~2, subtracting \eqref{eq:nef-mean-rate} at $v$ from \eqref{eq:nef-mean-rate} at $u$ gives
\[
\begin{aligned}
  I_{\mu=u}(x)-I_{\mu=v}(x)
  &= (\theta_v-\theta_u)x-A(\theta_v)+A(\theta_u)\\
  &= I_{\mu=u}(v)+(\theta_v-\theta_u)(x-v)
   = I_{\mu=u}(v)+I_{\mu=u}'(v)(x-v)
   =\ell_v(x),
\end{aligned}
\]
where the second equality uses \eqref{eq:nef-mean-rate} and the third uses $I_{\mu=u}'(v)=\theta_v-\theta_u$. Rearranging yields \eqref{cls:tangent}.

For Part~3, when $v\ne u$ we have $I_{\mu=u}'(v)=\theta_v-\theta_u\ne0$, so $\ell_v(x)=0$ has the unique solution $\zeta_{u,v}=v-I_{\mu=u}(v)/I_{\mu=u}'(v)$ in \eqref{cls:tangent-zero}. Strict convexity from Part~1 keeps the tangent strictly below $I_{\mu=u}(\cdot)$ at $u$, so $\ell_v(u)<I_{\mu=u}(u)=0$, while $\ell_v(v)=I_{\mu=u}(v)>0$. Thus $\zeta_{u,v}$ lies strictly between $u$ and $v$ in both cases ($v>u$ and $v<u$), and \eqref{cls:tangent} at $x=\zeta_{u,v}$ gives $I_{\mu=v}(\zeta_{u,v})=I_{\mu=u}(\zeta_{u,v})$. Furthermore, since $\ell_v(\cdot)$ has slope $\theta_v-\theta_u$ and vanishes at $\zeta_{u,v}$, comparing $\ell_v(x)=(\theta_v-\theta_u)(x-\zeta_{u,v})$ with $\ell_v(x)=(\theta_v-\theta_u)x-A(\theta_v)+A(\theta_u)$ from Part~2 yields $\zeta_{u,v}=[A(\theta_v)-A(\theta_u)]/(\theta_v-\theta_u)$. Evaluating $\ell_v(v)=I_{\mu=u}(v)=\KL(F_v\Vert F_u)$ and $-\ell_v(u)=I_{\mu=v}(u)=\KL(F_u\Vert F_v)$ then gives \eqref{cls:kl-factorization}.

For Part~4, fix $u,v\in \Omega$ and any interior point $x_0\in(\max\{u,v\},\mu_+)$. By Part~2, $I_{\mu=u}(x)-I_{\mu=v}(x)=\ell_v(x)$ is affine in $x$, so $|\ell_v(x)|=O(1+|x|)$. For every $x\in[x_0,\mu_+)$, convexity from Part~1 gives $I_{\mu=u}(x)\ge I_{\mu=u}(x_0)+(\theta_{x_0}-\theta_u)(x-x_0)$ with $I_{\mu=u}(x_0)>0$ and $\theta_{x_0}-\theta_u>0$, and the same linear lower bound holds for $I_{\mu=v}(x)$. Hence $I_{\mu=u}(x)$ and $I_{\mu=v}(x)$ are each bounded below by a positive multiple of $1+|x|$ on $[x_0,\mu_+)$, so $I_{\mu=v}(x)/I_{\mu=u}(x)$ is bounded between two positive constants on $[x_0,\mu_+)$, establishing the first chain in \eqref{cls:ref-ratio-bounds}. The second chain follows symmetrically using any fixed interior point below $\min\{u,v\}$. \Halmos\endproof

\proof{Proof of Lemma~\ref{lem:outsider-share}.}
Write $\Gamma^*=\Gamma(\sbest(P),P)$ for the optimal static exponent of this configuration. For any $i\ge3$ with $\mu_i<\mu_2$, individual balance in \eqref{cls:nef-individual} at its pairwise crossing point $x_i\in(\mu_i,\mu_1)$ gives $\Gamma^*=w_1 I_{\mu=\mu_1}(x_i)+w_i I_{\mu=\mu_i}(x_i)$. Meanwhile, evaluating the runner-up comparison in \eqref{eq:cramer-oracle} at $x=\mu_2$ (where $I_{\mu=\mu_2}(\mu_2)=0$) gives $\Gamma^*\le w_1 I_{\mu=\mu_1}(\mu_2)$. Combining these two relations yields
\[
  w_1 I_{\mu=\mu_1}(x_i)+w_i I_{\mu=\mu_i}(x_i)
  =\Gamma^*
  \le w_1 I_{\mu=\mu_1}(\mu_2).
\]
Because $w_i I_{\mu=\mu_i}(x_i)>0$, dropping this term gives $w_1 I_{\mu=\mu_1}(x_i)<w_1 I_{\mu=\mu_1}(\mu_2)$, and canceling $w_1>0$ yields $I_{\mu=\mu_1}(x_i)<I_{\mu=\mu_1}(\mu_2)$. Since $I_{\mu=\mu_1}(\cdot)$ is strictly decreasing on $(\mu_-,\mu_1]$ (Part~1 of Lemma~\ref{lem:nef-rate-properties}), this forces $x_i>\mu_2$. Because $I_{\mu=\mu_i}(\cdot)$ is strictly increasing on $[\mu_i,\mu_+)$, we then obtain $I_{\mu=\mu_i}(\mu_2)<I_{\mu=\mu_i}(x_i)$. Using $w_1\le 1$ therefore gives
\[
  w_i I_{\mu=\mu_i}(\mu_2)
  < w_i I_{\mu=\mu_i}(x_i)
  < \Gamma^*
  \le w_1 I_{\mu=\mu_1}(\mu_2)
  \le I_{\mu=\mu_1}(\mu_2),
\]
which proves \eqref{cls:outsider-main}. \Halmos\endproof

We next establish uniqueness and continuity of the rate-optimal allocation $\sbest(P)$ and construct a consistent tracking procedure on $\mathfrak D$. For $i\ne b$, write
\begin{equation}
  \gamma_i(w,P)
  =\inf_{\mu_i\le x\le\mu_b}
    \{w_b I_{P_b}(x)+w_i I_{P_i}(x)\},
  \label{eq:pairwise-exponent}
\end{equation}
so that $\Gamma(w,P)=\min_{i\ne b}\gamma_i(w,P)$.

\emph{Interior allocations and equal pairwise exponents.} Consider any configuration with unique best alternative $b$. The continuity of the rate functions and compactness of each meeting interval in \eqref{eq:pairwise-exponent} make $\Gamma(w,P)$ continuous on the closed simplex, so a maximizing allocation exists. Equal allocation $w_{\rm EA}$ gives a positive pairwise exponent for every competitor: the two rate functions have distinct unique zeros at $\mu_b$ and $\mu_i$, so their positively weighted sum is strictly positive on the compact interval $[\mu_i,\mu_b]$, giving $\Gamma(w_{\rm EA},P)>0$.

By contrast, every boundary allocation has zero exponent. If $w_b=0$, evaluating any pairwise objective at the competitor's mean gives zero; if $w_i=0$ for some $i\ne b$, evaluating that pairwise objective at the best alternative's mean gives zero. Thus every optimal allocation lies in the interior of the simplex.

We next show that every pairwise exponent at an optimal allocation equals the optimal static exponent $\Gamma^*=\max_{w\in\mathcal S_k}\Gamma(w,P)>0$. Suppose instead that $\gamma_i(w,P)>\Gamma^*$ for some optimal $w$ and some $i\ne b$. For sufficiently small $\epsilon>0$, define
\[
  \widetilde w_i=w_i-\epsilon,
  \qquad
  \widetilde w_j=c_\epsilon w_j\quad(j\ne i),
  \qquad
  c_\epsilon=1+\frac{\epsilon}{1-w_i}>1.
\]
The coordinates of $\widetilde w$ sum to one. By continuity, $\gamma_i(\widetilde w,P)>\Gamma^*$ for all sufficiently small $\epsilon$. For every $j\ne b,i$, homogeneity of \eqref{eq:pairwise-exponent} gives
\[
  \gamma_j(\widetilde w,P)
  =c_\epsilon\gamma_j(w,P)
  \ge c_\epsilon \Gamma^*>\Gamma^*.
\]
Thus $\Gamma(\widetilde w,P)>\Gamma^*$, contradicting optimality. Hence all pairwise exponents equal $\Gamma^*$ at every optimal allocation.

\emph{Uniqueness.} For positive $w_b$ and $w_i$, the pairwise objective $w_b I_{P_b}(x)+w_i I_{P_i}(x)$ is strictly convex between the two means. Its derivative is negative at $\mu_i$ and positive at $\mu_b$, so it has a unique interior minimizer. At that point,
\[
  \frac{w_i}{w_b}=-\frac{I_{P_b}'(x)}{I_{P_i}'(x)}.
\]
Thus, two different values of $w_i/w_b$ give different minimizing points.

Suppose two distinct allocations $w$ and $w'$ were optimal, and put $\overline w=(w+w')/2$. Because $\sum_{j\ne b} (w_j/w_b)=1/w_b-1$, equal ratios across all $j\ne b$ would force $w_b=w_b'$ and hence $w=w'$; thus the ratios $w_i/w_b$ and $w_i'/w_b'$ must differ for some $i\ne b$. The corresponding pairwise objectives have no common minimizer, so the infimum at $\overline w$ is strictly larger than the average of the two infima. Concavity gives the weak inequality for every other pair:
\[
  \gamma_i(\overline w,P)>
    \frac{\gamma_i(w,P)+\gamma_i(w',P)}{2},
  \qquad
  \gamma_j(\overline w,P)\ge
    \frac{\gamma_j(w,P)+\gamma_j(w',P)}{2}
  \quad (j\ne i).
\]
All pairwise exponents at $w$ and $w'$ equal the optimal static exponent $\Gamma^*$. Consequently, $\gamma_i(\overline w,P)>\Gamma^*$ and $\gamma_j(\overline w,P)\ge \Gamma^*$ for every other competitor $j$. If all these inequalities are strict, then $\Gamma(\overline w,P)>\Gamma^*$, contradicting the optimality of $\Gamma^*$. Otherwise $\Gamma(\overline w,P)=\Gamma^*$, so $\overline w$ is optimal, yet its $i$th pairwise exponent strictly exceeds $\Gamma^*$, contradicting the equal-pairwise-exponent property just established. The first case covers, in particular, $k=2$, when there is no other competitor.

\emph{Continuity and tracking.} Near a configuration with a unique best alternative, the index of that alternative remains the same. In each comparison, write $x=(1-t)\mu_i+t\mu_b$ so that the minimization is over the interval $0\le t\le1$. The objective is jointly continuous in $t$, the means, and the allocation ratios. Compactness of the interval gives continuity of the pairwise exponent, and taking the minimum over the finitely many pairwise exponents preserves continuity. For any sequence of mean vectors satisfying $\lim_{n\to\infty}\mu^{(n)}=\mu$, compactness of the simplex implies that every subsequential limit of their optimal allocations maximizes the objective at $\mu$. Uniqueness therefore gives continuity of the optimal allocation.

Choose integers $m_T$ satisfying
\[
  \lim_{T\to\infty}m_T=\infty,
  \qquad
  \lim_{T\to\infty}\frac{m_T}{T}=0,
  \qquad
  km_T\le T,
\]
and take $m_T$ pilot observations from each alternative. If $\widehat\mu_i(T)$ is the pilot mean, the strong law gives $\lim_{T\to\infty}\widehat\mu_i(T)=\mu_i(P)$ almost surely. The mean-parameter space $A'(\Theta)$ is open, and every true mean lies in this space. Hence, almost surely, all pilot means eventually lie in $A'(\Theta)$. Moreover, the unique-best gap $\Delta=\mu_b-\max_{i\ne b}\mu_i$ is positive. Eventually $\max_i|\widehat\mu_i(T)-\mu_i|<\Delta/3$, so $b$ is also the unique best alternative for the pilot mean vector. The pilot means therefore eventually determine a valid configuration $\widehat P_T\in\mathfrak D$. Continuity of the optimal allocation then gives
\[
  \lim_{T\to\infty}\widehat\mu_i(T)=\mu_i(P)\quad\text{for every }i
  \quad\Longrightarrow\quad
  \lim_{T\to\infty}\sbest(\widehat P_T)=\sbest(P)
  \qquad\text{almost surely}.
\]
Allocate the remaining $T-km_T$ observations in the allocation ratios $\sbest(\widehat P_T)$, up to integer rounding, and use the pilot means to make the final selection. Use equal allocation whenever any pilot mean lies outside $A'(\Theta)$ or the largest pilot mean is tied. At any fixed configuration with interior means and a unique best alternative, these exceptional events eventually cease almost surely. For each alternative $i$, the total allocation then satisfies
\begin{equation}
  \frac{N_i(T)}{T}
  =\frac{m_T}{T}
   +\frac{T-km_T}{T}\,\sbest_i(\widehat P_T)
   +O\!\left(\frac{1}{T}\right).
  \label{eq:pilot-tracking}
\end{equation}
Taking $\lim_{T\to\infty}$ in \eqref{eq:pilot-tracking} gives $\lim_{T\to\infty}N_i(T)/T=\sbest_i(P)$ almost surely. The final selection also equals $b(P)$ eventually almost surely because it is based on the pilot means. The procedure is therefore consistent and tracks the problem-specific rate-optimal allocation in this setting.

\section{An Allocation Bound for Three or More Alternatives}\label{app:allocation-bound}

\begin{lemma}\label{lem:allocation-bound}
Let $k\ge3$ and choose an alternative $a$. Fix the distributions $(Q_i:i\ne a)$ of the other alternatives, with unique best alternative $h$, and write $\mu_i$ for the mean of $Q_i$. Let $v$ be an optimal static allocation for this reduced problem and let $G\in(0,\infty)$ be its static exponent. Choose any $j\notin\{a,h\}$ and let $y=\mu_j$. Form $Q$ by assigning distribution $R$ to alternative $a$, with mean below $y$ and $0<I_R(y)<\infty$. Suppose both problems satisfy the rate-function finiteness and continuity assumptions of Section~\ref{sec:static-allocation}, and use the continuous extension of \eqref{eq:cramer-oracle} to the allocation simplex. Then every optimal static allocation at $Q$ satisfies
\begin{equation}
   \sbest_a(Q)\le \frac{G}{G+I_R(y)}.
   \label{eq:restored-allocation}
\end{equation}
\end{lemma}

\proof{Proof.}
For any full allocation $w$ with $w_a<1$, the ratios $w_i/(1-w_a)$, $i\ne a$, form a feasible allocation in the reduced problem. Homogeneity of the pairwise exponents in \eqref{eq:cramer-oracle} therefore gives
\begin{equation}
  \Gamma(w,Q)\le (1-w_a)G.
  \label{eq:reduced-problem-upper-bound}
\end{equation}
The same bound holds when $w_a=1$, because every comparison in the reduced problem then receives zero observations.

Set $u=G/(G+I_R(y))$, and define the full allocation $\widetilde w$ by $\widetilde w_a=u$ and $\widetilde w_i=(1-u)v_i$ for $i\ne a$. Because the mean of $R$ is strictly less than $y<\mu_h$, alternative $h$ remains the unique best alternative at $Q$. The smallest exponent among the comparisons excluding $a$ is $(1-u)G$. For the new comparison between $h$ and $a$, the reduced comparison of $h$ and $j$ gives
\[
  G\le \inf_{y\le x\le\mu_h}
       \{v_h I_{Q_h}(x)+v_j I_{Q_j}(x)\}
  \le v_h I_{Q_h}(y)+v_j I_{Q_j}(y)
  =v_h I_{Q_h}(y),
\]
where the second inequality evaluates the pairwise objective at the mean $y$ of alternative $j$, and the equality uses $I_{Q_j}(y)=0$. At any meeting value $x$ of the sample means of $h$ and $a$,
\[
  (1-u)v_h I_{Q_h}(x)+uI_R(x)
  \ge
  \begin{cases}
    (1-u)v_h I_{Q_h}(y)\ge(1-u)G, & x\le y,\\
    uI_R(y)=(1-u)G, & x\ge y.
  \end{cases}
\]
Here the first line uses that $I_{Q_h}(\cdot)$ is nonincreasing to the left of its mean, and the second uses that $I_R(\cdot)$ is nondecreasing to the right of its mean. Thus the new comparison also has exponent at least $(1-u)G$, and hence
\[
  (1-u)G
  =\Gamma(\widetilde w,Q)
  \le\Gamma(\sbest(Q),Q)
  \le(1-\sbest_a(Q))G,
\]
where the last inequality is \eqref{eq:reduced-problem-upper-bound}. Dividing by $G>0$ gives $\sbest_a(Q)\le u$, which is \eqref{eq:restored-allocation}. \Halmos\endproof

For the normal setting of Section~\ref{sec:gauss}, take $R=\N(m,s^2)$ with $m<y$. Substituting $I_R(y)=(y-m)^2/(2s^2)$ into \eqref{eq:restored-allocation} gives the explicit bound $\sbest_a(Q)\le 2Gs^2/(2Gs^2+(y-m)^2)$.

\section{Proof of Theorem~\ref{thm:nef-tangent}}\label{app:tangent-classification}

\proof{Proof.}
Write $P_i=F_{\eta_i}$, $p_i=A'(\eta_i)$, and $b=b(P)$. For any different-best configuration $Q$, write $Q_i=F_{t_i}$, $q_i=A'(t_i)$, and $h=b(Q)\ne b$. With $A^*$ denoting the convex conjugate of $A$, set
$J_t(x)=I_{\mu=A'(t)}(x)=A^*(x)-tx+A(t)$ and
$\mathcal D(t,a)=\KL(F_t\Vert F_a)=J_a(A'(t))$. All such KL divergences are finite, and Section~\ref{app:oracles} verifies the tracking hypothesis of Corollary~\ref{cor:ld-tracking}. Hence a zero PICS exponent is forced at $P$ if and only if
\begin{equation}
 \label{cls:comp}
 B_{\sbest}(P)=
 \inf_{Q:\,b(Q)\ne b}\ \sum_{i=1}^k \sbest_i(Q)\,\mathcal D(t_i,\eta_i)=0,
\end{equation}
where the infimum ranges over configurations $Q\in\mathfrak D$ (each having a unique best alternative). We establish Theorem~\ref{thm:nef-tangent} in three parts: tangent and balance identities together with reference-parameter bounds, sufficiency of the boundary or inferior-tie conditions, and necessity when all true means are distinct.

\emph{Tangent identities and rate-optimal binary fractions.}
For any two distinct parameters $t,a\in\Theta$, let $\sbest_{\rm bin}(t,a)$ denote the rate-optimal static fraction assigned to $F_t$ in the two-alternative configuration $(F_t,F_a)$ (regardless of which mean is larger). At the maximizing fraction $w=\sbest_{\rm bin}(t,a)$, the two pairwise terms agree at the common meeting mean $x=x(t,a)$ where $\frac{d}{dx}[w J_t(x)+(1-w)J_a(x)]=0$. Because $J_t(x)-J_a(x)=(a-t)x+A(t)-A(a)$ vanishes at $x(t,a)$ and $J_t'(x)=\theta_x-t$, solving $J_t(x)=J_a(x)$ and $w(\theta_x-t)+(1-w)(\theta_x-a)=0$ gives
\begin{equation}
 \label{cls:secant}
 x(t,a)=\frac{A(t)-A(a)}{t-a},\qquad
 \sbest_{\rm bin}(t,a)=\frac{\theta_{x(t,a)}-a}{t-a}.
\end{equation}
For any reference mean $u\in \Omega$, use $\sigma=+1$ for the upper side and $\sigma=-1$ for the lower side, set $L_+=\mu_+-u$ and $L_-=u-\mu_-$, and define
\[
 H_\sigma(d)=I_{\mu=u}(u+\sigma d),\qquad
 g_\sigma(d)=\frac{H_\sigma(d)}{H_\sigma'(d)},\qquad
 z_\sigma(d)=d-g_\sigma(d)
 \qquad (0<d<L_\sigma).
\]
Take $a=\theta_u$ and $A'(t)=u+\sigma d$ (writing $v=A'(t)$), and put $g=g_\sigma(d)$ and $z=z_\sigma(d)$. By Lemma~\ref{lem:nef-mean-rate}, $H_\sigma(d)=(t-a)(v-x)$ with $x=x(t,a)$ and $H_\sigma'(d)=\sigma(t-a)$, so $g=\sigma(v-x)$, $z=\sigma(x-u)$, and therefore $x=\zeta_{u,v}=u+\sigma z$, giving $g_+=\,v-\zeta_{u,v}$ and $g_-=\,\zeta_{u,v}-v$. Thus $S_+=\liminf_{d\uparrow L_+}g_+(d)H_+'(z_+(d))$, $S_-=\liminf_{d\uparrow L_-}g_-(d)H_-'(z_-(d))$, and $R_-=\liminf_{d\uparrow L_-}g_-(d)/d$. Using $H_\sigma'(z)=\sigma(\theta_x-a)$ and $\mathcal D(a,t)=J_t(u)=(t-a)(x-u)$ in \eqref{cls:secant} gives the pointwise identities
\begin{equation}
 \label{cls:identities}
 gH_\sigma'(z)=\sbest_{\rm bin}(t,a)\mathcal D(t,a),\qquad
 \frac{g}{d}=
 \frac{\mathcal D(t,a)}{\mathcal D(t,a)+\mathcal D(a,t)} .
\end{equation}
In particular, \eqref{cls:identities} shows that $R_-=0$ precisely when $\mathcal D(t,a)/\mathcal D(a,t)$ has lower limit zero as $t\to\theta_-$. Moreover, $S_-=0$ implies $R_-=0$: differentiating $z_-(d)=d-H_-(d)/H_-'(d)$ gives $z_-'(d)=H_-(d)H_-''(d)/[H_-'(d)]^2>0$ on $(0,L_-)$, so $z_-(d)$ and hence $H_-'(z_-(d))$ increase strictly on $(0,L_-)$ and stay bounded away from zero as $d\uparrow L_-$. Therefore $S_-=\liminf_{d\uparrow L_-}g_-(d)H_-'(z_-(d))=0$ forces $\liminf_{d\uparrow L_-}g_-(d)=0$, which implies $R_-=\liminf_{d\uparrow L_-}g_-(d)/d=0$ (since $d\to L_->0$ as $d\uparrow L_-$).

For general $k$, put $w=\sbest(Q)$ and let $\Gamma^*=\Gamma(w,Q)>0$ be the optimal static exponent at $Q$. By Section~\ref{app:oracles}, all coordinates $w_j$ are positive and each pairwise minimum $\min_x\{w_hJ_{t_h}(x)+w_jJ_{t_j}(x)\}$ equals $\Gamma^*$. If $x_j$ is the unique minimizer for rival $j\ne h$, stationarity in $x$ gives the first identity in \eqref{cls:balance}; to obtain the second, let $f_j(\xi)=\min_x\{J_{t_h}(x)+\xi J_{t_j}(x)\}$ (with unique minimizer $x_j(\xi)$), which increases from $0$ to $J_{t_h}(q_j)$ with derivative $f_j'(\xi)=J_{t_j}(x_j(\xi))>0$. At common level $s=\Gamma^*/w_h$, the ratios $\xi_j(s)=w_j/w_h=f_j^{-1}(s)$ yield objective $s/(1+\sum_{j\ne h}\xi_j(s))$, whose interior maximizer satisfies $\sum_{j\ne h}(s\xi_j'(s)-\xi_j(s))=1$. Since $\xi_j'(s)=1/J_{t_j}(x_j)$ and $s=J_{t_h}(x_j)+\xi_j J_{t_j}(x_j)$, we have $s\xi_j'(s)-\xi_j(s)=J_{t_h}(x_j)/J_{t_j}(x_j)$, giving
\begin{equation}
 \label{cls:balance}
 \theta_{x_j}=\frac{w_ht_h+w_jt_j}{w_h+w_j},\qquad
 \sum_{j\ne h}\frac{J_{t_h}(x_j)}{J_{t_j}(x_j)}=1 .
\end{equation}
Because $J_{t_h}(x)/J_{t_j}(x)$ decreases strictly between $q_j$ and $q_h$ and equals $1$ at $x(t_h,t_j)$, each summand in \eqref{cls:balance} being at most $1$ implies $x_j\ge x(t_h,t_j)$, and since $\theta_x$ is strictly increasing in $x$ with $t_h>t_j$, \eqref{cls:secant} and \eqref{cls:balance} yield $w_h/(w_h+w_j)=(\theta_{x_j}-t_j)/(t_h-t_j)\ge\sbest_{\rm bin}(t_h,t_j)$.

We also record two reference-parameter comparison bounds used throughout the remainder of the proof. Fix any compact interior interval $K\subset\Theta$ and a point $a\in\Theta$. First, for $x$ approaching either endpoint of $\Omega$ ($\mu_+$ or $\mu_-$), the rate functions $J_e(x)$ and $J_a(x)$ stay comparable uniformly over $e\in K$:
\begin{equation}
 \label{cls:ref-forward}
 0 < c_K \le \frac{J_e(x)}{J_a(x)} \le C_K < \infty
 \qquad(e\in K,\ x\in\Omega\text{ outside a compact neighborhood of }A'(K)\cup\{A'(a)\}).
\end{equation}
Indeed, $J_e(x)-J_a(x)=(a-e)x+A(e)-A(a)$ grows at most linearly in $|x|$. Near a finite endpoint of $\Omega$, both $J_e(x)$ and $J_a(x)$ remain bounded away from zero uniformly for $e\in K$; if an endpoint is infinite, choosing a fixed supporting parameter $\tau_0\in\Theta$ strictly beyond $K\cup\{a\}$ toward that endpoint gives $J_e(x)=J_{\tau_0}(x)+(\tau_0-e)x-A(\tau_0)+A(e)\ge (\tau_0-e)x-O(1)\ge c|x|-O(1)$ uniformly for $e\in K$ (since $J_{\tau_0}(x)\ge0$), and likewise for $J_a(x)$. Setting $x=A'(t)$ in \eqref{cls:ref-forward} yields uniform comparability of $\mathcal D(t,e)$ and $\mathcal D(t,a)$ near either endpoint of $\Theta$. Second, on the lower side near $\theta_-$, the reverse divergences are uniformly comparable:
\begin{equation}
 \label{cls:ref-reverse}
 0 < \tilde c_K \le \frac{\mathcal D(e,t)}{\mathcal D(a,t)} \le \tilde C_K < \infty
 \qquad(e\in K,\ t\text{ near }\theta_-).
\end{equation}
When $\theta_-$ is finite, both $\mathcal D(e,t)=J_t(A'(e))$ and $\mathcal D(a,t)=J_t(A'(a))$ are bounded away from zero and bounded above for $e\in K$ and $t$ near $\theta_-$. When $\theta_-=-\infty$, choosing an interior mean $\underline m$ strictly below $\{A'(e):e\in K\}\cup\{A'(a)\}$ gives $J_t(u)-J_t(\underline m)=A^*(u)-A^*(\underline m)-t(u-\underline m)\ge c|t|-O(1)$ for some $c>0$ and all $u\in A'(K)\cup\{A'(a)\}$, while $|J_t(u)-J_t(v)|=O(1+|t|)$, proving \eqref{cls:ref-reverse}. In particular, applying \eqref{cls:identities} with \eqref{cls:ref-reverse} shows that whether $R_-=0$ holds is independent of the reference mean $u\in\Omega$.

\emph{Sufficiency.}
Let $r$ be the runner-up mean at $Q$, and consider any $i$ with $q_i<r$. Evaluating the runner-up comparison at $r$ gives $\Gamma^*\le w_h J_{t_h}(r)$, whereas the comparison with $i$ at $x_i$ gives $\Gamma^*=w_h J_{t_h}(x_i)+w_i J_{t_i}(x_i)>w_h J_{t_h}(x_i)$. Hence $J_{t_h}(x_i)<J_{t_h}(r)$ forces $x_i>r$, and therefore $J_{t_i}(r)<J_{t_i}(x_i)$, giving
\begin{equation}
 \label{cls:outsider}
 w_i J_{t_i}(r) < w_i J_{t_i}(x_i) < \Gamma^* \le w_h J_{t_h}(r) \le J_{t_h}(r).
\end{equation}
(One can alternatively obtain \eqref{cls:outsider} from Lemma~\ref{lem:allocation-bound} by removing alternative $i$.)

If two inferior alternatives at $P$ share a mean $\tilde\mu$, form $Q$ by placing them at $\tilde\mu+\varepsilon$ and $\tilde\mu$ (for small $\varepsilon>0$) and placing every other alternative at a fixed interior mean $m_0\in\Omega$ with $m_0<\tilde\mu$. Then the best alternative at $Q$ is the first formerly tied inferior alternative, so $b(Q)\ne b(P)$, and the runner-up mean at $Q$ is $\tilde\mu$. Because $J_{\theta_{\tilde\mu+\varepsilon}}(\tilde\mu)=O(\varepsilon^2)$ while $J_{\theta_{m_0}}(\tilde\mu)>0$ is fixed, \eqref{cls:outsider} shows that every alternative with mean $m_0$ has weight $w_i=O(\varepsilon^2)$ while its divergence $\mathcal D(\theta_{m_0},\eta_i)$ is fixed and finite; meanwhile the top two alternatives have vanishing KL costs, proving \eqref{cls:comp}.

If all means are distinct and $k\ge3$, form $Q$ by moving only the true best alternative $b$ to parameter $t\to\theta_-$, letting $h$ and $j$ denote the best and second-best among the unchanged alternatives. Eventually $A'(t)<p_j$, so \eqref{cls:outsider} bounds the total weighted cost in \eqref{cls:comp} by
\[
 w_b\mathcal D(t,\eta_b)\le \frac{J_{\eta_h}(p_j)\,\mathcal D(t,\eta_b)}{J_t(p_j)} = \frac{J_{\eta_h}(p_j)\,\mathcal D(t,\eta_b)}{\mathcal D(\eta_j,t)}.
\]
By \eqref{cls:ref-forward}, \eqref{cls:ref-reverse}, and \eqref{cls:identities}, this upper bound converges to zero along a subsequence whenever $R_-=0$.

Suppose next that $S_+=0$ for some $a\in\Theta$, so there is a sequence $t\to\theta_+$ with $\sbest_{\rm bin}(t,a)\mathcal D(t,a)\to0$. Necessarily $\theta_+=\infty$, because at a finite upper endpoint the secant $x(t,a)$ in \eqref{cls:secant} stays bounded above $x(t_0,a)>A'(a)$ for fixed $t_0>a$ and all $t\ge t_0$, keeping $\sbest_{\rm bin}(t,a)$ bounded away from zero. Set $x=x(t,a)$ and $\tau=\theta_x$. Since $x$ is the average of $A'$ over $[a,t]$, we have $x\to\mu_+$ and $\tau\to\infty$.

Choose any inferior alternative $h\ne b$. By \eqref{cls:ref-forward}, there exists $M_0<\infty$ such that $J_a(x)\le M_0 J_{\eta_j}(x)$ for all $j\ne h$ and all $x$ sufficiently near $\mu_+$; pick a fixed $0<c<\min\{1,1/((k-1)M_0)\}$, so that $c J_a(x)<J_{\eta_j}(x)/(k-1)$ holds eventually for every $j\ne h$. Form $Q$ by changing only $h$ to parameter $t'=\tau+c(t-\tau)$. Since $t'=(1-c)\tau+ct$ with $J_\tau(x)=0$ and $\theta\mapsto J_\theta(x)$ is convex, we obtain $J_{t'}(x)\le c J_t(x)=c J_a(x)$. By \eqref{cls:balance}, at least one rival $j\ne h$ satisfies $J_{t'}(x_j)/J_{\eta_j}(x_j)\ge 1/(k-1)$. Because $p_j<x<A'(t')$ eventually, if we had $x_j>x$, monotonicity would yield $J_{t'}(x_j)/J_{\eta_j}(x_j)<J_{t'}(x)/J_{\eta_j}(x)\le c J_a(x)/J_{\eta_j}(x)<1/(k-1)$, a contradiction; hence $x_j\le x$, i.e., $\theta_{x_j}\le \tau$, and \eqref{cls:balance} gives
\begin{equation}
 \label{cls:upper-share}
 w_h\le \frac{w_h}{w_h+w_j}=\frac{\theta_{x_j}-\eta_j}{t'-\eta_j}\le \frac{\tau-\eta_j}{t'-\eta_j}.
\end{equation}
Since $\tau\to\infty$ and $t'\ge ct+(1-c)\tau$, the bound in \eqref{cls:upper-share} is $O((\tau-a)/(t-a))=O(\sbest_{\rm bin}(t,a))$ uniformly over the finitely many rivals $j\ne h$. Moreover, $\partial_t\mathcal D(t,\eta_h)=(t-\eta_h)A''(t)>0$ for $t>\eta_h$, so $\eta_h<t'\le t$ eventually gives $\mathcal D(t',\eta_h)\le\mathcal D(t,\eta_h)=O(\mathcal D(t,a))$ by \eqref{cls:ref-forward}. Because only $h$ was changed, $w_h\mathcal D(t',\eta_h)=O(\sbest_{\rm bin}(t,a)\mathcal D(t,a))\to 0$, proving \eqref{cls:comp} for all $k\ge2$.

To see reference-mean independence, applying \eqref{cls:upper-share} to a single opponent $F_e$ ($k=2$) shows $\sbest_{\rm bin}(t',e)\mathcal D(t',e)\to0$ for every fixed $e\in\Theta$, so whether $S_+=0$ does not depend on $a\in\Theta$. Finally, for $k=2$, reflecting both observations ($X\mapsto -X$) reverses the true order while preserving binary rate-optimal fractions and KL divergences, proving reference-mean independence of $S_-$ and sufficiency of $S_-=0$ when $k=2$.

\emph{Necessity.}
It remains to consider distinct true means. Suppose $B_{\sbest}(P)=0$ and take a sequence of different-best configurations $Q\in\mathfrak D$ with vanishing weighted cost in \eqref{cls:comp}. After passing to a subsequence, we may fix the best alternative $h=b(Q)\ne b$, the runner-up $r$, and an alternative $i$ satisfying $w_i\ge 1/k$. Since $w_i\mathcal D(t_i,\eta_i)\to0$, strict convexity of $J_{\eta_i}(\cdot)$ with unique zero at $p_i$ gives $q_i\to p_i$ (hence $t_i\to\eta_i$). Because $q_h\ge q_i$, after passing to a further subsequence, $q_h$ converges either to an endpoint of $\Omega$ or to an interior mean $\bar q\in\Omega$.

If $q_h$ escapes to $\mu_+$, necessarily $i\ne h$, and we write $t=t_h\to\theta_+$ and $a_n=t_i\to\eta_i$. By \eqref{cls:balance} and $x_i\ge x(t,a_n)$, we have $\sbest_{\rm bin}(t,a_n)\mathcal D(t,\eta_h)\le kw_h\mathcal D(t,\eta_h)\to0$. By continuity of $x(t_0,\cdot)$ for fixed $t_0>\eta_i$, $x(t,a_n)$ stays bounded above $A'(a_n)$ uniformly for $a_n$ near $\eta_i$, so necessarily $\theta_+=\infty$. Choosing a fixed $a_0<\eta_i$ in $\Theta$, monotonicity of the integral average $x(t,\cdot)$ in \eqref{cls:secant} gives $x(t,a_0)\le x(t,a_n)$ eventually. Because both secants converge to $\mu_+$, $\theta_{x(t,a_n)}\to\infty$ and $t\to\infty$ while $a_n\to\eta_i$ is bounded, so $\theta_{x(t,a_0)}\le\theta_{x(t,a_n)}$ and \eqref{cls:secant} yield $\sbest_{\rm bin}(t,a_0)=O(\sbest_{\rm bin}(t,a_n))$. Applying \eqref{cls:ref-forward} then gives $\sbest_{\rm bin}(t,a_0)\mathcal D(t,a_0)\to0$, hence $S_+=0$. (Because $q_h\ge q_i\to p_i\in\Omega$, $q_h$ cannot escape to $\mu_-$.)

Suppose $k=2$ and $q_h\to\bar q\in\Omega$. If the rival $r$ escapes to $\mu_-$, reflecting both alternatives ($X\mapsto -X$) reverses the mean order while preserving binary rate-optimal fractions and KL divergences, so the upper-escape argument just given implies $S_-=0$. Otherwise, both $q_h$ and $q_r$ converge to interior limits in $\Omega$. By \eqref{cls:secant}, both binary fractions $w_h=\sbest_{\rm bin}(t_h,t_r)$ and $w_r=1-w_h$ stay bounded away from zero (including at a common interior limit $\bar t$, where Taylor expansion of $A$ around $\bar t$ gives $\sbest_{\rm bin}(t_h,t_r)\to1/2$). Vanishing weighted cost in \eqref{cls:comp} then forces $(q_h,q_r)\to(p_h,p_r)$, contradicting $p_h<p_b$ (since $q_h>q_r=q_b$). Thus $S_+=0$ or $S_-=0$ whenever $k=2$.

Finally, let $k\ge3$ with $q_h\to\bar q\in\Omega$ (an interior limit). We show that $\Gamma^*=\Gamma(w,Q)$ stays bounded away from zero along this subsequence. For $\nu_1>\nu_2$ in $\Omega$, set $J(\nu_1,\nu_2)=\min_x\{J_{\theta_{\nu_1}}(x)+J_{\theta_{\nu_2}}(x)\}$. Its unique minimizer $x\in(\nu_2,\nu_1)$ satisfies $\partial_{\nu_2}J(\nu_1,\nu_2)=(\nu_2-x)/A''(\theta_{\nu_2})<0$. Under equal allocation, the hardest pairwise comparison at $Q$ is therefore with the runner-up $r$, giving $\Gamma^*\ge J(q_h,q_r)/k$. If $\Gamma^*\to0$ along a further subsequence, monotonicity and continuity of $J$ force $q_r\to\bar q$. On a compact neighborhood of $\bar q$, let $0<c_0<C_0<\infty$ bound $(A^*)''$ below and above. Comparing equal allocation and evaluating the $h,r$ comparison at $q_h$ or $q_r$ gives the quadratic bounds
\[
 \frac{c_0(q_h-q_r)^2}{4k}\le \Gamma^*\le
 \frac{C_0(q_h-q_r)^2}{2}\min(w_h,w_r).
\]
Because $h=b(Q)$ is unique, $q_h-q_r>0$; canceling $(q_h-q_r)^2$ therefore gives $\min(w_h,w_r)\ge c_0/(2kC_0)>0$, so both $w_h$ and $w_r$ stay bounded away from zero. Vanishing KL costs in \eqref{cls:comp} then imply $p_h=p_r=\bar q$, contradicting distinct true means. Hence $\Gamma^*$ stays bounded away from zero.

If $h\ne i$, evaluating the comparison between $h$ and $i$ at $q_i$ gives $\Gamma^*\le w_h J_{t_h}(q_i)$, where $J_{t_h}(q_i)$ stays bounded (since $q_h\to\bar q$ and $q_i\to p_i$); if $h=i$, $w_h\ge 1/k$ directly. Thus $w_h$ stays bounded away from zero, so $w_h\mathcal D(t_h,\eta_h)\to0$ forces $q_h\to p_h$. Next, because $q_b<q_h\to p_h<p_b$, $q_b$ stays bounded away from $p_b$, keeping $\mathcal D(t_b,\eta_b)=J_{\eta_b}(q_b)$ bounded away from zero; hence $w_b\mathcal D(t_b,\eta_b)\to0$ forces $w_b\to0$. Since $\Gamma^*\le w_b J_{t_b}(q_h)$ and $\Gamma^*$ is bounded away from zero, we must have $J_{t_b}(q_h)\to\infty$, which forces $q_b\to\mu_-$ (any interior subsequential limit keeps $J_{t_b}(q_h)$ bounded). Using \eqref{cls:ref-reverse} on the compact set containing $q_h\to p_h$ and \eqref{cls:identities} gives
\[
 \frac{\mathcal D(t_b,\eta_b)}{J_{t_b}(q_h)}
 = \frac{\mathcal D(t_b,\eta_b)}{\mathcal D(\theta_{q_h},t_b)}
 \le \frac{w_b\mathcal D(t_b,\eta_b)}{\Gamma^*}\longrightarrow0,
\]
whence $R_-=0$. Thus every sequence of different-best configurations with vanishing weighted cost forces $S_+=0$, $S_-=0$ (for $k=2$), or $R_-=0$ (for $k\ge3$, noting $S_-=0\implies R_-=0$). The positive lower exponent in the complementary cases follows from Corollary~\ref{cor:ld-tracking}. \Halmos\endproof

\section{Boundary-Test Calculations for Normal and Poisson Observations}\label{app:normal-poisson}

Here we evaluate the boundary tests $S_-$, $S_+$, and $R_-$ in \eqref{cls:upper-test}, \eqref{cls:lower-test}, and \eqref{cls:three-test} for normal observations with a common known variance and for Poisson observations, completing Table~\ref{cls:examples-table} (and Table~\ref{tab:nef-tracking}).

\emph{Normal observations with a common known variance ($\Omega=\R$).}
Fix a common known variance $\sigma^2>0$ and any reference mean $u\in\R$. The normal rate function and its derivative are
\[
  I_{\mu=u}(v)=\frac{(v-u)^2}{2\sigma^2},
  \qquad
  I_{\mu=u}'(v)=\frac{v-u}{\sigma^2}.
\]
Substituting these expressions into \eqref{cls:tangent-zero} gives $\zeta_{u,v}=v-I_{\mu=u}(v)/I_{\mu=u}'(v)=(u+v)/2$. Consequently $v-\zeta_{u,v}=(v-u)/2$ and $I_{\mu=u}'(\zeta_{u,v})=(v-u)/(2\sigma^2)$, so
\[
  (v-\zeta_{u,v})I_{\mu=u}'(\zeta_{u,v})=\frac{(v-u)^2}{4\sigma^2}\longrightarrow\infty
  \quad(v\to\pm\infty),
  \qquad
  \frac{\zeta_{u,v}-v}{u-v}=\frac{1}{2}.
\]
Thus $S_-=S_+=\infty$ and $R_-=1/2$. By Theorem~\ref{thm:nef-tangent}, a positive lower PICS exponent is possible at every configuration when $k=2$ and whenever all means are distinct when $k\ge3$.

\emph{Poisson observations ($\Omega=(0,\infty)$).}
Take reference mean $u=1$ and write $I_1(\cdot)=I_{\mu=1}(\cdot)$, so
\[
  I_1(v)=v\log v-v+1,
  \qquad
  I_1'(v)=\log v
  \qquad(v>0).
\]
On the lower side ($0<v<1$), letting $v\downarrow0$ gives $I_1(v)\to1$ and $-I_1'(v)=-\log v\to\infty$. By \eqref{cls:tangent-zero}, $\zeta_{1,v}-v=I_1(v)/[-I_1'(v)]\to0$, so $\zeta_{1,v}=v+(\zeta_{1,v}-v)\to0$ and $I_1(\zeta_{1,v})\to1$. Since $\zeta_{1,v}\in(v,1)$ and $I_1$ is convex and decreasing on $(0,1]$, convexity gives $I_1(v)\ge I_1(\zeta_{1,v})+I_1'(\zeta_{1,v})(v-\zeta_{1,v})$, so
\[
  0\le(\zeta_{1,v}-v)\bigl[-I_1'(\zeta_{1,v})\bigr]\le I_1(v)-I_1(\zeta_{1,v})\longrightarrow 0.
\]
Hence $S_-=0$ in \eqref{cls:lower-test}, and $(\zeta_{1,v}-v)/(1-v)\to0$ gives $R_-=0$ in \eqref{cls:three-test}. On the upper side ($v>1$), \eqref{cls:tangent-zero} gives $\zeta_{1,v}=v-I_1(v)/I_1'(v)=(v-1)/\log v\to\infty$ as $v\to\infty$. Thus $v-\zeta_{1,v}\sim v$ and $I_1'(\zeta_{1,v})=\log\zeta_{1,v}\to\infty$, so \eqref{cls:upper-test} yields $S_+=\infty$. By Theorem~\ref{thm:nef-tangent}, a zero PICS exponent is forced for all $k\ge2$.

\section{Proof of Theorem~\ref{kv:theorem}}\label{kv:proof}

\proof{Proof.}
Fix $P$ with means $\mu_i$, and set $\delta(P)=\min_{i<j}|\mu_i-\mu_j|$, $\sigma_{\min}=\min_i\sigma_i$, and $\sigma_{\max}=\max_i\sigma_i$. We first establish the two-sided bound on the weighted-information bound $B_{\sbest}(P)$ in \eqref{eq:tracking-weighted-kl}:
\begin{equation}
 \frac{\delta(P)^2}{16k\sigma_{\max}^2}
 \ \le\ B_{\sbest}(P)\ \le\
 \frac{\delta(P)^2}{8\sigma_{\min}^2}.
 \label{kv:gap-bound}
\end{equation}
Notice that, because the best alternative at $P$ is unique, $\delta(P)=0$ if and only if at least two inferior alternatives share the same mean.

Fix a different-best configuration $Q$ with means $q_i$. Put $w=\sbest(Q)$ and $g=\Gamma(w,Q)$. With known variances, the information cost of $Q$ in \eqref{eq:bound} is
\[
  C(Q,P)=\sum_i \frac{w_i(q_i-\mu_i)^2}{2\sigma_i^2}.
\]
Write $h=b(Q)$ and let $r$ be an alternative with the second-largest mean at $Q$. Put $d=q_h-q_r$. Under $Q$, the joint rate function of the sample means at proportions $w$ is $\sum_i w_i(x_i-q_i)^2/(2\sigma_i^2)$. Its infimum over $\{x:\max_{i\ne h}x_i\ge x_h\}$ is $g$. The vector of means at $P$ belongs to this set, so $C(Q,P)\ge g$.

The Gaussian pairwise exponent for $h$ and $i$ at proportions $w$ is $(q_h-q_i)^2/[2(\sigma_h^2/w_h+\sigma_i^2/w_i)]$. Equal allocation gives a lower bound on $g$, and the pair $h,r$ at the optimal proportions gives an upper bound. Together these yield
\[
  \frac{d^2}{4k\sigma_{\max}^2}\le g\le\frac{d^2}{2V},
  \qquad V=\frac{\sigma_h^2}{w_h}+\frac{\sigma_r^2}{w_r}.
\]
By Cauchy--Schwarz,
\[
 |(\mu_h-\mu_r)-d|
 = \left|\frac{\sigma_h}{\sqrt{w_h}}\cdot\frac{\sqrt{w_h}(q_h-\mu_h)}{\sigma_h}
        -\frac{\sigma_r}{\sqrt{w_r}}\cdot\frac{\sqrt{w_r}(q_r-\mu_r)}{\sigma_r}\right|
 \le \sqrt{2C(Q,P)V}
 \le d\sqrt{C(Q,P)/g}.
\]
Since $C(Q,P)\ge g$, the triangle inequality gives
$\delta(P)\le|\mu_h-\mu_r|\le d+|(\mu_h-\mu_r)-d|\le 2d\sqrt{C(Q,P)/g}
\le 4\sigma_{\max}\sqrt{kC(Q,P)}$.
Taking the infimum over $Q$ proves the lower bound in \eqref{kv:gap-bound}.

For the upper bound, choose a pair with gap $\delta(P)$ and label its members $h,r$ so that $h\ne b(P)$. Put
\[
  z=\frac{\sigma_r\mu_h+\sigma_h\mu_r}{\sigma_h+\sigma_r}.
\]
For $\epsilon>0$, form a different-best configuration $Q_\epsilon$ with means $q_h=z+\epsilon$, $q_r=z$, and $q_i=z-1$ for every $i\notin\{h,r\}$. As $\epsilon\downarrow0$, the divergences on $h$ and $r$ both tend to $\delta(P)^2/[2(\sigma_h+\sigma_r)^2]$. We show that the optimal allocation places asymptotically all observations on these two alternatives.

Write $w=\sbest(Q_\epsilon)$ and $g=\Gamma(w,Q_\epsilon)$. The $h,r$ comparison gives $g\le\epsilon^2/[2(\sigma_h+\sigma_r)^2]$, since $w_h+w_r\le1$. For every $i\notin\{h,r\}$, the total balance condition $w_h^2/\sigma_h^2=\sum_{\ell\ne h}w_\ell^2/\sigma_\ell^2\ge w_i^2/\sigma_i^2$ in \eqref{eq:balance} gives $w_i/w_h\le \sigma_i/\sigma_h$. By individual balance, the $h,i$ pairwise exponent also equals $g$, whence
\[
  w_i
  = \frac{2g}{(q_h-q_i)^2}
     \left(\sigma_i^2+\frac{w_i}{w_h}\sigma_h^2\right)
  \le \frac{2g\,\sigma_i(\sigma_i+\sigma_h)}{(1+\epsilon)^2}
  =O(\epsilon^2).
\]
The divergences on these alternatives are fixed and finite. Thus $C(Q_\epsilon,P)$ tends to $\delta(P)^2/[2(\sigma_h+\sigma_r)^2]$, which is at most the upper bound in \eqref{kv:gap-bound}. The same limit holds for $k=2$, when the two chosen alternatives receive the whole budget.

Finally, a consistent tracker exists on this distributional setting $\mathfrak D$. Take $m_T$ pilot observations per alternative with $m_T\to\infty$ and $m_T/T\to0$, select the largest pilot mean, and allocate the remaining observations in the optimal proportions at the pilot means, up to rounding. The Gaussian objective in \eqref{eq:cramer-oracle} is continuous in the means and the allocation, and its maximizer is unique at each unique-best configuration. The optimal proportions therefore converge in probability to $\sbest(P)$, and the pilot selection is consistent. The equivalence and the existence of a consistent tracker now follow from Corollary~\ref{cor:ld-tracking} and uniqueness of the best alternative. \Halmos\endproof

\section{Proof of Theorem~\ref{pv:gauss}}\label{pv:proofs}

\subsection{Likelihood Comparison and Concentration Preliminaries}\label{pv:prelims}

We first give a likelihood comparison used in the proofs of Theorems~\ref{pv:gauss} and~\ref{dt:theorem}. For each alternative let $I_i(\cdot)=I_{P_i}(\cdot)$, write $b=b(P)$, and let $J_j=\min_x\{I_b(x)+I_j(x)\}$ for $j\ne b$.

\begin{lemma}[Comparison at fixed terminal counts]\label{lem:fixed-count-crossing}
Suppose the alternatives belong to the common family of Section~\ref{sec:ld-oracle}, or are Gaussian with positive, possibly unequal variances. Fix a rival $j$ and a terminal count vector $n$ with $n_b,n_j>0$ and $\sum_i n_i=T$. Let $z_n$ minimize $n_b I_b(x)+n_j I_j(x)$ and write $K_j(n)$ for the minimum. Under $R_{j,n}$, tilt $P_b,P_j$ to mean $z_n$, retaining the Gaussian variances when applicable, and leave the other laws unchanged. For any nonanticipating sampler, on $N(T)=n$ the likelihood ratio of the observed record satisfies
\begin{equation}
 \frac{d\Pr_P}{d\Pr_{R_{j,n}}}
   =\exp\{-K_j(n)-\beta_n T d\},\qquad
 d=\bar X_j(T)-\bar X_b(T),\qquad
 \beta_n=\frac{n_j}{T} I_j'(z_n)>0 .
 \label{dt:crossing}
\end{equation}
In particular, $K_j(n)\ge\min(n_b,n_j)J_j$. For any history event $E\subseteq\{N(T)=n,\ d\ge0\}$ we have $\Pr_P(E)\le e^{-K_j(n)}\Pr_{R_{j,n}}(E)$. More generally, for any history event $E\subseteq\{N(T)=n\}$ and measurable $Z$ such that $K_j(n)+\beta_nTd\ge\gamma T-Z$ on $E$,
\begin{equation}
 \Pr_P(E)\le e^{-\gamma T}\E_{R_{j,n}}[e^Z\mathbf1_E].
 \label{dt:transfer}
\end{equation}
\end{lemma}

\proof{Proof.}
Strict convexity places $z_n$ strictly between $\mu_j$ and $\mu_b$, with $n_b I_b'(z_n)+n_j I_j'(z_n)=0$ (and $I_j'(z_n)>0$ since $z_n>\mu_j$). For $i=b,j$, the log density of $P_i$ relative to its tilt at one observation $y$ is $-I_i(z_n)-I_i'(z_n)(y-z_n)$. The sampling and randomization rules are the same under both laws, and the remaining alternatives have identical laws under $P$ and $R_{j,n}$. Summing over the draws actually made on $\{N(T)=n\}$ gives log likelihood ratio
\[
  -\bigl[n_b I_b(z_n)+n_j I_j(z_n)\bigr]
  -n_b I_b'(z_n)\bigl(\bar X_b(T)-z_n\bigr)
  -n_j I_j'(z_n)\bigl(\bar X_j(T)-z_n\bigr)
  =-K_j(n)-n_j I_j'(z_n)\bigl(\bar X_j(T)-\bar X_b(T)\bigr),
\]
where the last equality substitutes $n_b I_b'(z_n)=-n_j I_j'(z_n)$. This yields \eqref{dt:crossing}. The bound on $K_j(n)$ follows from nonnegativity of the rates. Integrating over $E$ under the full law $R_{j,n}$ gives both probability bounds. \Halmos\endproof

In the common family $P_i=F_{\eta_i}$, the tilted parameter is $t_n=(n_b\eta_b+n_j\eta_j)/(n_b+n_j)$, so $z_n=A'(t_n)$ and $\beta_n=(\eta_b-\eta_j)n_bn_j/[T(n_b+n_j)]$. For Gaussian variances $v_i$,
\begin{equation}
 z_n=\frac{n_b\mu_b/v_b+n_j\mu_j/v_j}{n_b/v_b+n_j/v_j},
 \qquad K_j(n)=\frac{(\mu_b-\mu_j)^2}{2(v_b/n_b+v_j/n_j)} .
 \label{dt:gauss-cost}
\end{equation}
An incorrect empirical-best selection requires a crossing $d\ge0$. There are at most $(k-1)(T+1)^k$ choices of $j,n$, so polynomial factors in their probability bounds do not affect the exponent.

We will also use concentration under these same crossing laws. Generate the independent streams in advance and write $\bar X_{i,r}$ for the mean of the first $r$ observations of stream $i$. For fixed $P$, the tilted parameters lie in compact interior intervals. Given $\kappa>0$ and a fixed $x_0>0$, for some $C,h>0$ independent of $j,n,T$,
\begin{equation}
 \Pr_{R_{j,n}}\!\left\{\max_{i,\,\kappa T\le r\le T}
 |\bar X_{i,r}-\mu_i(R_{j,n})|\ge x\right\}
 \le CT e^{-h T x^2},\qquad 0<x\le x_0 .
 \label{dt:prefix}
\end{equation}
Indeed, the centered log moment generating functions on these intervals are uniformly bounded by a quadratic near zero. A Chernoff bound with tilt proportional to $x$, followed by a union over deterministic prefixes, proves \eqref{dt:prefix}. For Gaussian observations, the same fixed-neighborhood exponential bound holds for the sample variances by their chi-squared distributions.

\subsection{Completing the Proof of Theorem~\ref{pv:gauss}}

\proof{Proof of Theorem~\ref{pv:gauss}.}
Put $v_i=\sigma_i^2$ and $\Delta_j=\mu_b-\mu_j$. We first establish the universal upper bound in \eqref{pv:gslope} and then derive the matching first-order lower bound for $\pi_\lambda$.

\emph{Upper bound.} The least costly Gaussian crossing under EA has means $b,j$ equal to
\[
 z_j=\frac{v_j\mu_b+v_b\mu_j}{v_b+v_j},\qquad
 d_i^j=\frac{(\mu_i-z_j)^2}{2v_i}
      = kG_j\frac{v_i}{v_b+v_j},\quad i\in\{b,j\}.
\]
For $j\in\mathcal A$, every outsider mean is strictly less than $z_j$. Indeed, if $\mu_\ell\ge z_j$ for some $\ell\notin\{b,j\}$, then $\Delta_\ell=\mu_b-\mu_\ell\le\mu_b-z_j=v_b\Delta_j/(v_b+v_j)$, so $G_\ell=\Delta_\ell^2/[2k(v_b+v_\ell)]\le G_j v_b^2/[(v_b+v_j)(v_b+v_\ell)]<G_j$, contradicting $j\in\mathcal A$.

Perturb the pair to means $z_j-\delta,z_j+\delta$, respectively, using the same variances and keeping the outsiders unchanged, and call the resulting configuration $Q_\delta$. For small $\delta>0$, $j$ wins. As in the proof of Lemma~\ref{lem:comparison}, the record KL divergence sums the per-draw divergences weighted by the expected counts under $Q_\delta$; since the $m$ pilot counts per alternative are fixed and at most $L=T-km$ further draws go to $\{b,j\}$ (while all outsiders are unchanged),
\[
  \KL(Q_\delta^T\Vert P^T)
  \le m\bigl[\KL(Q_{\delta,b}\Vert P_b)+\KL(Q_{\delta,j}\Vert P_j)\bigr]
     +L\max\bigl(\KL(Q_{\delta,b}\Vert P_b),\KL(Q_{\delta,j}\Vert P_j)\bigr).
\]
Since $m/T\to(1-\lambda)/k$, $L/T\to\lambda$, and $\KL(Q_{\delta,i}\Vert P_i)\to d_i^j$ as $\delta\downarrow0$ with $d_b^j+d_j^j=kG_j$, consistency and binary data processing followed by $\delta\downarrow0$ give, for every procedure $\pi$ using the same EA pilot that is consistent throughout the known-variance (or unknown-variance) Gaussian distributional setting (including $\pi=\pi_\lambda$),
\begin{equation}
 \overline e_\pi(P)\le(1-\lambda)G_j+\lambda\max(d_b^j,d_j^j)
 \qquad (j\in\mathcal A).
 \label{pv:gauss-upper}
\end{equation}
At $k=2$ with equal variances, both divergences equal $\Gamma(w_{\rm EA},P)$.

\emph{Inactive rivals and non-exit neighborhood.} For the lower bound under $\pi_\lambda$, Lemma~\ref{lem:fixed-count-crossing} gives $K_j(n)\ge mJ_j=kmG_j$. Since $km/T=1-\lambda+O(1/T)$, this already covers inactive rivals ($G_j>\Gamma(w_{\rm EA},P)$) for sufficiently small $\lambda$.

For $j\in\mathcal A$, \eqref{dt:gauss-cost} gives $z_n=z_j+O(\lambda)+O(1/T)$ uniformly in $n$. Fix a sufficiently small radius $\varepsilon>0$ defining a neighborhood in which $|\bar X_i-z_j|\le\varepsilon$ for $i\in\{b,j\}$, $|\bar X_\ell-\mu_\ell|\le\varepsilon$ for $\ell\notin\{b,j\}$, and $|\widehat v_i-v_i|\le\varepsilon$ for all $i$ (taking $\widehat v_i=v_i$ when variances are known). Under $R_{j,n}$, every post-pilot step uses at least $m$ observations per alternative, so for small $\lambda$, \eqref{dt:prefix} and the chi-squared sample-variance bound give probability $O(e^{-hT})$ for exiting this neighborhood at any post-pilot step, for some $h>0$ independent of $n$ and small $\lambda$. By Lemma~\ref{lem:fixed-count-crossing}, any crossing with an exit therefore has $P$-probability at most $e^{-kmG_j}O(e^{-hT})=O(e^{-[(1-\lambda)G_j+h]T})$.

\emph{Second-stage routing off the exit event.} Exact empirical ties and nonpositive sample variances have probability zero by a union over deterministic Gaussian prefixes. Off this null set, whenever all post-pilot estimates stay inside the $\varepsilon$-neighborhood, the top two empirical means are $b$ and $j$ with gap $|\bar X_b(t-1)-\bar X_j(t-1)|\le 2\varepsilon$, while every outsider $\ell\notin\{b,j\}$ stays strictly separated from $z_j$ (since $\mu_\ell<z_j$). Comparing the top two empirical means bounds the estimated optimal exponent by $\widehat\Gamma=O(\varepsilon^2)$, so individual and total balance~\eqref{eq:balance} force each outsider weight to satisfy $\widehat w_\ell(t-1)=O(\varepsilon^2)$ (as in Section~\ref{kv:proof}). Total balance between the top two empirical means therefore reduces to
\[
 \frac{\widehat w_b(t-1)^2}{\widehat v_b(t-1)}-\frac{\widehat w_j(t-1)^2}{\widehat v_j(t-1)}=O(\varepsilon^4),
 \qquad \widehat w_b(t-1)+\widehat w_j(t-1)=1-O(\varepsilon^2),
\]
which gives $\widehat w_i(t-1)=\sqrt{v_i}/(\sqrt{v_b}+\sqrt{v_j})+O(\varepsilon)$ for $i\in\{b,j\}$.

Choosing $\varepsilon$ small enough yields a fixed margin $\eta>0$ as follows. Since $N_i(t-1)\le m+L$ and $N_\ell(t-1)\ge m$, any two current counts satisfy $N_i(t-1)-N_\ell(t-1)\le L$. When $v_b\ne v_j$, the higher-variance member $H$ of $\{b,j\}$ satisfies $\widehat w_H(t-1)-\max_{i\ne H}\widehat w_i(t-1)\ge\eta$ throughout the neighborhood; while $t\ge km$, the score of $H$ in \eqref{pv:total-clock-rule} exceeds every other score by at least $t\eta-L\ge km\eta-L>0$ for small $\lambda$, so every second-stage draw goes to $H$. When $v_b=v_j$, $\min(\widehat w_b(t-1),\widehat w_j(t-1))-\max_{\ell\notin\{b,j\}}\widehat w_\ell(t-1)\ge\eta$, so both pair scores strictly beat every outsider score and all second-stage draws remain within $\{b,j\}$ (for $k=2$ there are no outsiders).

\emph{Completing the lower bound.} For a fixed count vector $n$ and rival $j\in\mathcal A$, let $u_i=(n_i-m)/L$ be the fraction of second-stage draws on $i$, so $n_i/T=(1-\lambda)/k+\lambda u_i+O(1/T)=1/k+\lambda(u_i-1/k)+O(1/T)$. A Taylor expansion of $K_j(n)/T$ about $n_i/T=1/k$ gives, uniformly over all feasible $n$,
\[
 K_j(n)/T=G_j+\lambda(u_bd_b^j+u_jd_j^j-G_j)+O(\lambda^2)+O(1/T),
\]
since the partial derivatives of $K_j(n)/T$ at equal allocation are $d_b^j$ and $d_j^j$ and $d_b^j+d_j^j=kG_j$. Whenever the non-exit crossing event in $\{N(T)=n,d\ge 0\}$ has positive probability, we must have $u_b d_b^j+u_j d_j^j=\max(d_b^j,d_j^j)$ (either $u_H=1$ when $v_b\ne v_j$, or $u_b+u_j=1$ with $d_b^j=d_j^j=\max(d_b^j,d_j^j)$ when $v_b=v_j$), so Lemma~\ref{lem:fixed-count-crossing} bounds its probability by $e^{-K_j(n)}$.

Set
\[
 S=\min_{j\in\mathcal A}\max(d_b^j,d_j^j)-\Gamma(w_{\rm EA},P)
  =\Gamma(w_{\rm EA},P)\min_{j\in\mathcal A}\left(\frac{k\max(v_b,v_j)}{v_b+v_j}-1\right).
\]
For small $\lambda$, the exit exponent $(1-\lambda)G_j+h$ strictly exceeds $G_j+\lambda S$. Summing over the polynomial number of count vectors $n$ and all rivals $j$ gives $\underline e_{\pi_\lambda}(P)\ge \Gamma(w_{\rm EA},P)+\lambda S-C_P\lambda^2$ for some finite $C_P>0$ and all sufficiently small $\lambda>0$. Together with the upper bound $\overline e_{\pi_\lambda}(P)\le \Gamma(w_{\rm EA},P)+\lambda S$ established in \eqref{pv:gauss-upper} for every $\lambda\in(0,1)$ and every consistent continuation after the EA pilot, this proves \eqref{pv:gslope}. The strict positivity of $S$ in the stated cases follows from $k\max(v_b,v_j)/(v_b+v_j)-1\ge k/2-1$, which is strictly positive whenever $k\ge3$, and also for $k=2$ when $v_b\ne v_j$. \Halmos\endproof

\section{Proof of Theorem~\ref{dt:theorem} and Rule Comparison}\label{dt:proof}

Using Lemma~\ref{lem:fixed-count-crossing} and the concentration bound \eqref{dt:prefix} from Section~\ref{pv:prelims}, we prove Theorem~\ref{dt:theorem} first for the online rule~\eqref{dt:rule} and then for the post-pilot rule~\eqref{pv:block}.

\proof{Proof of Theorem~\ref{dt:theorem} (online rule~\eqref{dt:rule}).}
Put $a=(1-\lambda)/k$ and $c=a+\lambda/2$. Consistent with Section~\ref{ea:direct}, when computing $\widehat w$, if all sample means lie in $\Omega=A'(\Theta)$ and several tie for largest, we divide the weight equally among those tied leaders, and if any sample mean lies outside $\Omega$, we set $\widehat w=w_{\rm EA}$. All constants in this proof may depend on $P,k$ and can be chosen uniformly for all sufficiently small positive $\lambda$.

\emph{Tracking.} Before draw $t$, the deficits $t\widehat v_i(t-1)-N_i(t-1)$ in \eqref{dt:rule} sum to $t-(t-1)=1$, so the selected alternative $i_t$ must have strictly positive pre-draw deficit, i.e., $N_{i_t}(t-1)<t\widehat v_{i_t}(t-1)$. Adding that draw gives $N_{i_t}(t)<t\widehat v_{i_t}(t-1)+1$. Applying this bound at each alternative's last draw and using $\sum_i N_i(t)=t$ gives
\begin{equation}
 [1-(k-1)(a+\lambda)]t-O(k)\le N_i(t)\le(a+\lambda)t+O(1).
 \label{dt:count-bounds}
\end{equation}
For every $0<\lambda\le 1/[2(k-1)^2]$ (which does not depend on the configuration $Q\in\mathfrak D$), the lower coefficient in \eqref{dt:count-bounds} satisfies $[1-(k-1)^2\lambda]/k\ge1/(2k)$. Since \eqref{dt:count-bounds} then guarantees $N_i(t)\to\infty$ for every alternative $i$, at every unique-best $Q\in\mathfrak D$, the strong law of large numbers and continuity of $\sbest$ give consistency and $\widehat v(t)\to v_\lambda(Q)$ almost surely. Each alternative is drawn infinitely often, so the same last-draw bound gives $\limsup_t N_i(t)/t\le v_{\lambda,i}(Q)$. Because the fractions sum to one, $N_i(t)/t\to v_{\lambda,i}(Q)$ almost surely, proving tracking throughout $\mathfrak D$.

\emph{The oracle near a decisive tie.} Set $t_j=(\eta_b+\eta_j)/2$, $z_j=A'(t_j)$, and $J=\min_{j\ne b}J_j=k\Gamma(w_{\rm EA},P)$, where $w_{\rm EA}=(1/k,\ldots,1/k)$. Because $I_b'(z_j)+I_j'(z_j)=2\theta_{z_j}-\eta_b-\eta_j=0$, we have $J_j=I_b(z_j)+I_j(z_j)=A(\eta_b)+A(\eta_j)-2A(t_j)$, whose derivative in $\eta_j$ is $\mu_j-z_j<0$. Thus the minimizing rivals are the runners-up. For any such $j$, set the means of $b$ and $j$ equal to $z_j$ and keep the others at their true values to form $x^j$. Only $b$ and $j$ tie for first at $x^j$. Let $p_b^j=p_j^j=1/2$ and $p_i^j=0$ otherwise. On a bounded rectangle $U$ about $x^j$ whose closure lies in $\Omega^k$, the estimated rate-optimal allocation satisfies
\begin{equation}
 |\widehat w_i(x)-p_i^j|\le C_0|x_b-x_j| \qquad(x\in U),
 \label{dt:local-weights}
\end{equation}
for a finite constant $C_0>0$.

To verify \eqref{dt:local-weights}, write $g=|x_b-x_j|>0$ and denote the two empirical leaders by $h,r$ in decreasing order. Because all outsider gaps are bounded away from zero on $U$, \eqref{cls:outsider} implies $w_\ell=O(g^2)$ for every $\ell\notin\{h,r\}$. Moreover, each comparison mean $x_\ell$ lies in $(x_r,x_h)$ (as in Lemma~\ref{lem:outsider-share}), so $J_{t_h}(x_\ell)=O(g^2)$ while $J_{t_\ell}(x_\ell)$ stays bounded away from zero; hence its term in \eqref{cls:balance} is also $O(g^2)$.

For the $h,r$ comparison, let $s$ be the minimizing parameter, so stationarity gives $(\theta_{x_h}-s)/(s-\theta_{x_r})=w_r/w_h$. For any $\alpha,s\in\Theta$, Taylor expansion gives $J_\alpha(A'(s))=A(\alpha)-A(s)-A'(s)(\alpha-s)=\frac{1}{2}A''(\tilde s)(\alpha-s)^2$ for an intermediate point $\tilde s$. Because $A''$ is bounded away from zero and $A'''$ is bounded locally on $U$ while $|\theta_{x_h}-\theta_{x_r}|=O(g)$, the two curvature values have ratio $A''(\tilde s_h)/A''(\tilde s_r)=1+O(g)$, and \eqref{cls:balance} yields
\[
 \frac{A''(\tilde s_h)}{A''(\tilde s_r)}\left(\frac{w_r}{w_h}\right)^2
 =(1+O(g))\left(\frac{w_r}{w_h}\right)^2
 =1-O(g^2).
\]
Thus $w_r/w_h=1+O(g)$. Since $w_h+w_r=1-O(g^2)$, we obtain $w_h,w_r=1/2+O(g)$, proving \eqref{dt:local-weights} (the tie convention covers $g=0$).

At the configuration obtained by giving $b,j$ means $z_j-\epsilon,z_j+\epsilon$ while keeping all other alternatives unchanged, all means lie in $\Omega$ and alternative~$j$ becomes the unique best alternative for small $\epsilon>0$ (since $j$ is a runner-up, every other mean is at most $\mu_j<z_j$). Their mixed shares approach $c$ as $\epsilon\downarrow0$ by \eqref{dt:local-weights}, and their KL costs approach $I_b(z_j),I_j(z_j)$. Hence \eqref{eq:bound} bounds the weighted-information bound by $B_{v_\lambda}(P)\le cJ$.

\emph{The lower bound.} By \eqref{dt:count-bounds}, every feasible $n_i/T=1/k+O(\lambda)+O(1/T)$, so $\beta_n$ in \eqref{dt:crossing} is bounded away from zero for large $T$. If $J_j>J$, Lemma~\ref{lem:fixed-count-crossing} gives $K_j(n)\ge\min(n_b,n_j)J_j>cJT$ with a strict margin for small $\lambda$.

For a runner-up, call a history good if the empirical means before every draw after $T/2$ lie in $U$. Since $t_n=t_j+O(\lambda)+O(1/T)$, the mean vector under $R_{j,n}$ lies a fixed distance inside $U$, and every late prefix has length at least a fixed multiple of $T$ by \eqref{dt:count-bounds}. Thus \eqref{dt:prefix} bounds the $R_{j,n}$-probability of an exit by $Ce^{-hT}$. On any crossing event $E\subseteq\{N(T)=n,\,d\ge0\}$ with an exit, Lemma~\ref{lem:fixed-count-crossing} gives $\Pr_P(E)\le e^{-K_j(n)}\Pr_{R_{j,n}}(E)\le Ce^{-K_j(n)-hT}$, where $K_j(n)+hT\ge[1/k-O(\lambda)]JT+hT-O(1)>cJT$ for small $\lambda$.

It remains to bound the terminal counts on a good crossing. Put $v_i^j=a+\lambda p_i^j$ and let $Y_t$ denote the observation on draw $t$. If $i$ was last drawn by $T/2$, then $n_i\le(a+\lambda)T/2+O(1)\le aT+O(1)\le Tv_i^j+O(1)$. At a later last draw $s>T/2$, the quota bound and \eqref{dt:local-weights} give $n_i\le s v_i^j+1+C_0\lambda s|\bar X_b(s-1)-\bar X_j(s-1)|$. For $\ell\in\{b,j\}$, the pathwise identity
\begin{equation}
 \bar X_\ell(T)-\bar X_\ell(s-1)=\frac{1}{n_\ell}\sum_{t=s}^T
  \mathbf1\{i_t=\ell\}\big(Y_t-\bar X_\ell(s-1)\big)
 \label{dt:pathwise-mean}
\end{equation}
bounds this mean change by $C T^{-1}\sum_{t=s}^T(1+|Y_t|)$, since $\bar X_\ell(s-1)\in U$ and $n_\ell\ge T/(2k)-O(k)$. Since $|\bar X_b(s-1)-\bar X_j(s-1)|\le d+\sum_{\ell\in\{b,j\}}|\bar X_\ell(T)-\bar X_\ell(s-1)|$ (recall $d=\bar X_j(T)-\bar X_b(T)\ge0$) and $v_i^j\ge a$ (so $s v_i^j\le T v_i^j-a(T-s)$), substituting $l=T-s+1$ gives, for a finite constant $C_1>0$,
\[
 n_i\le T v_i^j+O(1)+C_0\lambda T d+H_T,\qquad
 H_T=\max_{1\le l\le T}\left[C_1\lambda\sum_{t=T-l+1}^T(1+|Y_t|)-a(l-1)\right]_+ .
\]
Because $\sum_{i=1}^k n_i=T$, $\sum_{i=1}^k v_i^j=1$, and $v_b^j=v_j^j=c$, conservation implies $n_b,n_j\ge cT-(k-1)[O(1)+C_0\lambda Td+H_T]$. Since $K_j(n)\ge J\min(n_b,n_j)$ and $\beta_n>(k-1)J C_0\lambda$ for small $\lambda$, on this event
\[
 K_j(n)+\beta_nTd\ge cJT-O(1)-(k-1)J H_T .
\]

The terminal-gap term has absorbed the part of the count shortfall proportional to $Td$. To bound the remaining fluctuation, for any fixed $p>0$ let $M(u)$ be the supremum of $\E[e^{u(1+|Y|)}]$ over the compact range of tilted observation laws. It is finite near zero and tends to one as $u\downarrow0$. For each $l\in\{1,\ldots,T\}$, iterating conditional expectations step by step over the suffix $t=T-l+1,\ldots,T$ gives $\E_{R_{j,n}}[\exp(p C_1\lambda\sum_{t=T-l+1}^T(1+|Y_t|))]\le M(pC_1\lambda)^l$, so summing over $l$ yields
\[
 \E_{R_{j,n}}[e^{pH_T}]\le 1+\sum_{l=1}^T e^{-pa(l-1)}M(p C_1\lambda)^l=O(1),
\]
uniformly for small $\lambda$, since $e^{-pa}M(pC_1\lambda)\to e^{-p/k}<1$. Applying \eqref{dt:transfer} with $\gamma=cJ$ and $Z=(k-1)J H_T+O(1)$ (taking $p=(k-1)J$ in the moment bound just proved) gives probability at most $C e^{-cJT}$ for each good crossing. Writing $\pi$ for rule~\eqref{dt:rule} and summing over rivals and count vectors gives $\underline e_\pi(P)\ge cJ$. Together with the upper bound $B_{v_\lambda}(P)\le cJ$ from the perturbed pair and Lemma~\ref{lem:comparison},
\[
 cJ\le\underline e_\pi(P)\le\overline e_\pi(P)\le B_{v_\lambda}(P)\le cJ .
\]
Since $cJ=[1+(k/2-1)\lambda]\Gamma(w_{\rm EA},P)$, this proves \eqref{dt:exponent}. \Halmos\endproof

\proof{Proof of Theorem~\ref{dt:theorem} (post-pilot rule~\eqref{pv:block}).}
Write $c=a+\lambda/2$, and use $J_j$, $J=k\Gamma(w_{\rm EA},P)$, $R_{j,n}$, $d$, and $K_j(n)$ from the proof for rule~\eqref{dt:rule}.

\emph{Pilot floor.} Because every terminal count satisfies $m\le n_i\le m+L$, Lemma~\ref{lem:fixed-count-crossing} gives $K_j(n)\ge mJ_j$ for every post-pilot rule. Summing over count vectors and rivals gives a pilot floor of $(1-\lambda)\Gamma(w_{\rm EA},P)$ for the lower PICS exponent under pooled empirical-best selection (which also yields consistency by the strong law on each observation stream).

\emph{Consistency, tracking, and upper bound.} For any fixed $\lambda\in(0,1)$, each empirical mean converges to its true value uniformly over the second stage almost surely, since all prefixes contain at least $m$ observations per alternative and $m\to\infty$. At any unique-best configuration $Q\in\mathfrak D$, continuity of $\sbest$ implies that the estimated allocation likewise converges uniformly to $\sbest(Q)$ almost surely.

Under \eqref{pv:block}, the pre-draw deficits $s\widehat w_i-M_i(s-1)$ sum to one, so each selected alternative ends its draw with $M_i(s)\le s\widehat w_i+1$. Evaluating at each alternative's last draw gives $\limsup_{T\to\infty} M_i(L)/L\le\sbest_i(Q)$ almost surely (with the same bound if it is never drawn in the second stage). Conservation gives $M_i(L)/L\to\sbest_i(Q)$, and since $m/T\to(1-\lambda)/k$ and $L/T\to\lambda$, $N_i(T)/T=m/T+(L/T)(M_i(L)/L)\to v_{\lambda,i}(Q)$ almost surely. Thus \eqref{pv:block} tracks $v_\lambda$ throughout $\mathfrak D$ for every $\lambda\in(0,1)$ and has $B_{v_\lambda}(P)\le cJ$ by the two-mean perturbation in the proof for rule~\eqref{dt:rule}.

\emph{Inactive rivals and exit bound.} The pilot floor covers rivals with $J_j>J$ for small $\lambda$. For a runner-up, choose $U$ as in \eqref{dt:local-weights}. At every feasible $n$, the tilted mean vector is $x^j+O(\lambda)+O(1/T)$ coordinatewise. Every post-pilot decision uses prefixes of length at least $m$, so \eqref{dt:prefix} gives exit probability $O(e^{-hT})$ under $R_{j,n}$ for some $h>0$ independent of $n$ and small $\lambda$. Together with $K_j(n)\ge mJ$, Lemma~\ref{lem:fixed-count-crossing} yields exponent at least $mJ/T+h$, which strictly exceeds $cJ$ for small $\lambda$ (since $m/T=a+O(1/T)$ and $c-a=\lambda/2$).

\emph{Outsider counts on a good crossing.} For \eqref{pv:block}, set $S$ equal to the maximum absolute gap between the two decisive pooled means before any second-stage draw, truncated at a fixed $g_0$ larger than every such gap in $U$. Under $R_{j,n}$, the two streams share the population means $z_n$. A pre-draw gap of at least $x$ requires a prefix deviation of at least $x/2$ from $z_n$, so \eqref{dt:prefix} gives $\Pr_{R_{j,n}}(S\ge x)\le CT e^{-h'Tx^2}$ for $0<x\le g_0$, uniformly over feasible $n$ and small $\lambda$.

Call a history good when all its second-stage updates lie in $U$. Because every mean vector in $U$ has top two alternatives $\{b,j\}$, each outsider weight on a good history is $O(S^2)$ by \eqref{cls:outsider} (including at empirical ties between $b$ and $j$, where dividing weight equally between the two tied leaders assigns zero weight to every outsider). Applying the last-draw bound $M_i(L)\le L\cdot C S^2+1$ to each outsider $i\notin\{b,j\}$ shows that the total number $M_{\rm out}=\sum_{i\notin\{b,j\}}M_i(L)$ of second-stage draws spent on outsiders satisfies $M_{\rm out}\le C L S^2+k$. Since $L\le 2\lambda T$ for all large $T$ at fixed $\lambda$, for every fixed $p>0$ and sufficiently small $\lambda$ (so that $pCL<h'T$), substituting $y=x^2$ and integrating the tail on $[0,g_0^2]$ gives
\[
 \E_{R_{j,n}}[e^{pM_{\rm out}}\mathbf 1_{\rm good}]
 \le C\left(1+T\int_0^{g_0^2}pCL e^{(pCL-h'T)y}\,dy\right)=O(T).
\]

\emph{Decisive pair counts and completing the exponent.} For either member $i\in\{b,j\}$ last drawn on second-stage draw $s$, the pre-draw deficit bound and \eqref{dt:local-weights} give $M_i(L)=M_i(s)\le s\bigl(1/2+C_0|\bar X_b(km+s-1)-\bar X_j(km+s-1)|\bigr)+1$. If the final gap is $d\ge0$, \eqref{dt:pathwise-mean} gives $|\bar X_b(km+s-1)-\bar X_j(km+s-1)|\le d+(C/T)\sum_{r=km+s}^T(1+|Y_r|)$, where $Y_r$ is the observation on draw $r$. Since $s\le L=\lambda T+O(k)$, this count is at most $L/2+C\lambda Td+H+O(1)$ (including if the alternative is never drawn in the second stage), where
\[
 H=\max_{1\le l\le L}\left[C'\lambda\sum_{r=T-l+1}^T(1+|Y_r|)-(l-1)/2\right]_+ .
\]
Iterating conditional expectation over deterministic suffixes as in the proof for rule~\eqref{dt:rule} gives $\E_{R_{j,n}}[e^{pH}]=O(1)$ for every fixed $p>0$ and small $\lambda$.

Since $M_b(L)+M_j(L)+M_{\rm out}=L$ and $m+L/2=cT+O(k)$, subtracting the other pair member's second-stage count and $M_{\rm out}$ from $L$ and adding $m$ shows that both final counts $n_b,n_j$ exceed $cT-C\lambda Td-H-M_{\rm out}-O(1)$. Since $K_j(n)\ge J\min(n_b,n_j)$ and $\beta_n$ is bounded away from zero, \eqref{dt:transfer} applies on each good crossing with $\gamma=cJ$ and $Z=J(H+M_{\rm out})+O(1)$ for small $\lambda$. Writing $\pi$ for rule~\eqref{pv:block}, Cauchy--Schwarz and the two moment bounds with $p=2J$ give probability $O(T^{1/2})e^{-cJT}$. Summing over count vectors and rivals yields $\underline e_\pi(P)\ge cJ$, and combining this with $\underline e_\pi(P)\le\overline e_\pi(P)\le B_{v_\lambda}(P)\le cJ$ (by Lemma~\ref{lem:comparison} and the two-mean perturbation) proves \eqref{dt:exponent} for rule~\eqref{pv:block}. \Halmos\endproof

\subsection{Comparing Cumulative-Count and Post-Pilot Tracking}\label{pv:count-comparison}

Since $t=km+s$ and $N_i(t-1)=m+M_i(s-1)$, the cumulative-count score in \eqref{pv:total-clock-rule} decomposes as
\[
 t\widehat w_i-N_i(t-1)
 = \bigl[s\widehat w_i-M_i(s-1)\bigr] + km\left(\widehat w_i-\frac{1}{k}\right),
\]
where $\widehat w$ denotes the current estimated rate-optimal allocation. Under equal allocation, incorrect selection is governed by the active rivals $j\in\mathcal A$ whose sample mean meets that of alternative~$b$ at the pairwise crossing point $x_j$ solving \eqref{eq:cramer-oracle} with $w=w_{\rm EA}$.

In general, for one-parameter NEFs such as Poisson or exponential, when the sample means of $b$ and $j$ are both near $x_j$, the estimated rate-optimal ratios assign symmetric limiting weights ($1/2$ each) to $b$ and $j$, yet $b$ and $j$ generally have unequal per-observation rate costs $I_{P_b}(x_j)\ne I_{P_j}(x_j)$ at $x_j$. Retaining the $O(T)$ pilot offset $km(\widehat w_i-1/k)$ when the second stage has length $L\approx\lambda T$ would let small estimation fluctuations near $x_j$ tip $\widehat w_b$ against $\widehat w_j$ and steer all $L$ second-stage draws to a single alternative, potentially the one with the lower rate cost $I_{P_i}(x_j)$. Subtracting pilot counts in~\eqref{pv:block} removes this offset, preserves $v_\lambda$-tracking, and divides second-stage draws evenly between $b$ and $j$ to attain~\eqref{dt:exponent} exactly.

Under Gaussian sampling with unequal variances, by contrast, when the sample means of $b$ and $j$ are both near $x_j$, the estimated Gaussian ratios satisfy $\widehat w_i\approx\sigma_i/(\sigma_b+\sigma_j)$ for $i\in\{b,j\}$, so the higher-variance member of $\{b,j\}$ carries weight strictly above $1/2$ in $\widehat w$ and also yields the larger per-observation rate cost $I_{P_i}(x_j)$ at $x_j$. Keeping the cumulative offset in~\eqref{pv:total-clock-rule} therefore directs all second-stage draws near $x_j$ to that higher-variance alternative (rather than splitting them in proportion $\sigma_b:\sigma_j$ as post-pilot tracking of $\widehat w$ would), attaining the optimal first-order slope in Theorem~\ref{pv:gauss}.

\end{document}